\documentclass[10pt,a4paper,onecolumn]{article}

\usepackage[top=30mm, bottom=30mm, left=15mm, right=15mm]{geometry}
\makeatletter
\usepackage{indentfirst}

\usepackage{amssymb,amsmath,latexsym}
\usepackage{amsmath}
\usepackage{amsfonts}
\usepackage{amssymb}
\usepackage{mathrsfs}

\usepackage{graphicx}
\usepackage{array}
\usepackage{hhline}
\usepackage{booktabs}

\usepackage[colorlinks=true, linkcolor=red, urlcolor=blue, citecolor=cyan]{hyperref}

\usepackage{perpage} 
\MakePerPage{footnote}
\usepackage[font={small,it}]{caption}
\usepackage[T1]{fontenc}
\usepackage[utf8]{inputenc}
\usepackage{ulem}
\usepackage{xcolor}
\usepackage{float}
\usepackage{orcidlink}

\title{A class of $d$-dimensional regular black holes: Shadows, Thermodynamics and Gravitational collapse}

\author{A. Sadeghi \orcidlink{0000-0002-2437-0431},$^1$ F. Shojai \orcidlink{0000-0002-5070-057X},$^{1}$\\Department of Physics, University of Tehran, P.O. Box 14395-547\\Tehran, Iran.}

\begin{document}
\maketitle

\begin{abstract}
We investigate a general class of $d$-dimensional regular black holes characterized by a de Sitter core, which arises from the gravitational collapse of a polytropic star with an arbitrary polytropic index $n$. This framework generalizes the well-known Bardeen and Hayward black holes to higher dimensions and identifies nonlinear electrodynamics with a magnetic monopole charge as the physical source ensuring curvature regularity. We analyze the geometric structure and energy conditions, demonstrating that while the Weak and Null  Conditions are satisfied, the Strong Energy Condition is violated, a necessary feature for curvature singularity avoidance. Our study of optical properties reveals the existence of stable and unstable photon spheres, with shadows persisting only up to a critical magnetic charge limit; beyond this threshold, the object becomes a horizonless compact object. Numerical results indicate that the shadow size decreases as the dimension $d$, charge $q$, or index $n$ increases, allowing for constraints based on EHT observations of M87* and SgrA*. Thermodynamically, unlike the unstable Schwarzschild-Tangherlini solution, these regular black holes exhibit regions of local stability and phase transitions, with entropy deviating from the standard area law in higher dimensions. Finally, we generalize the Oppenheimer-Snyder-Datt collapse scenario to this background. We track the evolution of horizons, the nature of the trapping horizon, and derive a critical lower bound for the initial stellar radius required for physical black hole formation. Our results show that increasing dimensions and the polytropic index delay the collapse proper time, while magnetic charge facilitates the process by reducing the minimum initial radius. These findings provide new insights into the viability of curvature-regular black holes as possible curvature-singularity-free endpoints of gravitational collapse in higher-dimensional gravity.
\end{abstract}

\section{Introduction}

According to the Hawking-Penrose singularity theorems, the BH singularities are associated with geodesic incompleteness and it is generally believed that such singularities would not exist in a complete theory of quantum gravity \cite{34-35Mal}.  According to the weak cosmic censorship conjecture, if stellar matter satisfies the null energy condition (NEC), the BH singularity remains hidden behind the event horizon, thereby preserving the predictable structure of spacetime \cite{kh4-5}. In general relativity, BHs represent the ultimate state of classical gravitational collapse for massive stars. However, it has been proposed that quantum effects could produce a repulsive pressure that counteracts gravitational attraction and prevents the formation of a singularity at the end of collapse \cite{34-35Mal}.

Several approaches have been suggested to eliminate singularities in BHs. One method involves modifying the Einstein-Hilbert action by incorporating additional terms arising from vacuum polarization and particle creation \cite{Mukh}. 
Another approach  imposes an upper bound on spacetime curvature \cite{23-25Fro13,Fro16}. Early efforts were inspired by Einstein and Rosen's concept of wormhole geometry. Later, Sakharov proposed using the equation of state (EOS) $p= - \rho$ for high density regions in stars \cite{1DyCqg05}, while Gliner suggested that such conditions could represent the final state of gravitational collapse \cite{2DyCqg05}.

In this work, we focus on regular BHs with de Sitter cores. Hayward and Bardeen introduced two important solutions. The Bardeen solution arises from considering general relativity coupled to nonlinear electrodynamics (NLED) \cite{3RodJcap-9-21Kho,4-8DyCqg04}, while the Hayward solution is a modified Schwarzschild BH that incorporates a fundamental length scale \cite{54Fro16}. These models exhibit de Sitter behavior near the center and asymptotically approach Schwarzschild spacetime at larger distances. The terminology "regular BH" must be used with care, since regularity of the spacetime requires geodesic completeness, and curvature regularity alone is not sufficient. It has been shown that, unlike the Bardeen metric, the Hayward metric is geodesically incomplete at a negative radius \cite{Zhou:2022yio}. A recent paper \cite{our1} presents a general regular solution with a de Sitter core within a gravitational collapse scenario.  The stellar EOS is shown to be of the polytropic type, resulting from the smooth matching of the interior and exterior geometries of the star, without  imposing any a priori assumptions about the EOS of the collapsing matter. The end product of gravitational collapse is a general static regular BH. This solution is characterized by the polytropic index of stellar matter; the Bardeen and Hayward BHs are special cases that correspond to specific polytropic indices \cite{our1}.

In recent years, the study of BHs has moved beyond the traditional $4$-dimensional framework to explore the intriguing realm of higher-dimensional spacetimes. This is due to theoretical advances in string theory and the potential for new physics at the TeV scale, suggesting the existence of large extra dimensions \cite{vafa,malda1}. 

In the context of general relativity, Tangherlini was the first to obtain a $d$-dimensional solution, which is a generalization of the Schwarzschild solution to higher dimensions \cite{tang}. Myers and Perry found exact solutions for a rotating BH in $d$ dimensions \cite{Myers:1986un}. Other studies of BHs in higher dimensions can be found in \cite{Emparan:2008eg}. The generalization to a $d$-dimensional Bardeen BH has been studied in \cite{ali}.

The shadow of a BH has been an active topic of interest in recent years due to the observations of M$87^*$ and  the supermassive BH in  the Milky Way galaxy Sgr A$^*$ \cite{EventHorizonTelescope:2019dse, EventHorizonTelescope:2022wkp}. Therefore, by studying shadows theoretically and comparing them with the observations, we can test the theoretical models. Shadows can result from the presence of a massive compact object or a BH and are formed due to a strong gravitational field. Studies on the shadows of compact objects can be found in \cite{Cardoso:2019rvt,Shaikh:2022ivr,Abdikamalov:2019ztb}. The shadow of the $d$-dimensional Schwarzschild-Tangherlini (ST) BH is studied  in \cite{Singh:2017vfr} and other interesting topics regarding BHs' shadow in higher dimensions can be found in \cite{Roy:2023ine, Papnoi:2014aaa}. 

Bardeen, Carter, and Hawking established BHs as thermodynamic systems. They related changes in mass to the horizon area through a factor proportional to the surface gravity, which they called the first law of BH thermodynamics. This law is analogous to the first law of ordinary thermodynamics, with mass, surface gravity and horizon area playing the roles of energy, temperature and entropy, respectively \cite{bch, Dymnikova:2015yma}. Hawking solidified the analogy between temperature and surface gravity by showing that the creation of quantum particles leads to BH radiation \cite{Hawkingtemp}.

Considering the BH as a thermodynamic system characterized by temperature and entropy makes studying its heat capacity crucial. Heat capacity indicates stability (instability) for positive (negative) values, and divergence signals a possible phase transition \cite{bonn1, Hut}. If a BH were to absorb mass, negative heat capacity would mean its temperature decreases because the absorption rate would exceed the emission rate. Thus, the BH would grow, and the system would become unstable. However, the presence of other factors, such as magnetic charge, can stabilize the system \cite{Pavn}. It should be noted that local thermodynamic stability, diagnosed for example by a positive heat capacity at fixed charge, is distinct from linear dynamical stability of the coupled metric and NLED perturbations. In \cite{DeFelice:2024seu}, it is shown that, in four-dimensional general relativity coupled to NLED, nonsingular BHs with regular centers are subject to an angular Laplacian instability near the center: a vector-field perturbation develops a negative squared angular propagation speed and drives the growth of an even-parity gravitational perturbation. Therefore, the existence of a positive-heat-capacity branch does not establish dynamical viability. In the present work, our stability statements refer only to local thermodynamic stability.

Alongside studying the BH itself, an important question is whether such objects could result from stellar collapse. The classical model of collapse, developed by Oppenheimer, Snyder and Datt (OSD) \cite{Oppenheimer:1939ue, Datt:1938uwc}, smoothly connects the interior Friedmann--Lema\^itre--Robertson--Walker (FLRW) metric to  the Schwarzschild metric, ultimately resulting in a singularity. Various proposals attempt to modify this scenario. One approach is to replace the Schwarzschild exterior with a regular BH containing a de Sitter core, which prevents the formation of a central singularity due to repulsive effects related to the violation of the strong energy condition (SEC) \cite{our1}. Certain OSD model modifications introduce a bounce, thereby eliminating the singularity \cite{Bambi:2014, Bambi:2013}. 
In \cite{Zhang:2015}, the authors suggest that quantum effects may replace classical singularities with a bounce.

In this paper, we introduce a general class of regular $d$-dimensional BHs with a de Sitter core. This generalizes the four-dimensional regular BHs proposed in \cite{our1} to arbitrary $d$ dimensions. We achieve this generalization by coupling the Einstein-Hilbert action to an appropriate NLED theory in which the magnetic charge acts as the free parameter ensuring regularity in each dimension. This extends the work of \cite{Ayon}, which was focused on the Bardeen BH. Next, we study the energy conditions (ECs) of this model, find the circular orbits, and identify the photon spheres and shadows. We perform an exact analysis of the photon spheres and a numerical analysis of the shadow size for various  values of the magnetic charge and spacetime dimensions up to 11. Within the model, we study both BH and horizonless compact object to determine the limits of photon sphere and shadow existence. Using a numerical approach, we also study the thermodynamic properties of the solution, including temperature, entropy, and heat capacity.  Then, we examine the stability of the model and possible phase transitions. Finally, we generalize the OSD collapse scenario by smoothly joining a $d$-dimensional flat FLRW interior to a $d$-dimensional regular BH. We then study the evolution of the surface of the star, the apparent horizon and the event horizon. We investigate the nature of the trapping horizon and show that increasing the number of spacetime dimensions and the polytropic index delays the collapse proper time, whereas the presence of magnetic charge facilitates the collapse scenario by reducing the minimum initial radius.

While the mathematical generalization of four-dimensional regular BHs to arbitrary dimensions may appear formal, the physical consequences of this extension are profoundly non-trivial and yield several novel insights that have no direct analog in $d=4$. First, the dimensional dependence fundamentally alters the effective EOS governing the Oppenheimer–Snyder collapse, introducing a dimension- and matter-dependent effective polytropic index $n_{\text{eff}}=2 n (d-1)/[3(d-2)]$ which reduces to the original polytropic index $n$ in four dimensions. Second, we demonstrate that the deviation from the standard law for entropy systematically increases with dimension, while the critical thresholds for energy condition violations (DEC and SEC) become explicitly tied to $d$, revealing new regimes of exotic matter required to sustain regular cores. Finally, the collapse dynamics exhibit a critical lower bound on the initial stellar radius $R_{0,\min}(d,q,n)$, which is sensitive to both dimension and magnetic charge, and for $d>5$, the apparent horizon can undergo a change of causal character, becoming null	at a critical radius. These findings demonstrate that the $d$-dimensional framework is not merely a technical extension, but an essential tool for uncovering universal scaling laws.

The outline of this paper is as follows: Section \ref{sec2}  studies a general $d$-dimensional static spherically symmetric geometry and derives the corresponding Einstein equations. Section \ref{sec3} introduces a  family of $d$-dimensional regular BHs with a de Sitter core using an appropriate Lagrangian. We also compute the scalar curvature, horizons, extremal conditions, and energy conditions. Section \ref{sec4} analyzes photon spheres using an exact approach and shadows using a numerical approach.   Section \ref{sec5} delves into the thermodynamics of BHs by studying surface gravity, temperature, and entropy using analytical and numerical approaches. Additionally, we assess the thermodynamic stability of BHs by identifying potential phase transitions through heat capacity analysis. Section \ref{SecCol} presents the scenario of a star collapsing into a $d$-dimensional regular BH and studies the evolution of the star's surface, as well as the apparent and event horizons. Section \ref{seccon} provides a summary and conclusion. 

Throughout this paper, we assume the signature of the metric tensor is $(-, +, +, +,...)$. We use natural units in which $G = c = 1$.  To simplify some calculations, we scale the variables by the mass parameter $m$ and denote them with tildes, i.e., $\tilde r= r/m$.
\section{$d$-dimensional static spherically symmetric geometry}
\label{sec2}
We will consider a general static spherically symmetric spacetime in $d$ dimensions with $g_{tt}=-1/g_{rr}$ as
\begin{align}\label{genmet}
ds^2 = -A_d(r) dt^2 + A_d^{-1}(r) dr^2 + r^2 d \Omega^2_{d-2},
\end{align}
where
\begin{align}
d\Omega^2_{d-2}=  \sum_{i=1}^{d-2} \left[ \prod_{j=2}^{i} \sin^2 \theta_{j-1}\right] d\theta^2_i,
\end{align}
is the line element of a $(d-2)$-dimensional sphere \cite{myers,dian}.
 The Einstein equations then give \cite{Tang, Yin}
\begin{align}
\label{err}
G^{t }{}_{t }&=G^{r }{}_{r }=-\frac{(d-3) (d-2)}{2 r^2}+\frac{(d-3) (d-2)}{2 r^2} A_d(r)+\frac{(d-2)}{2 r} A_d'(r)=8\pi T^{t }{}_{t }= -8\pi \rho(r) =T^{r }{}_{r } = 8\pi p_\text{r} (r),\\
\label{epp}
G^{\theta_i}{}_{\theta_i }&=-\frac{(d-3) (d-4)}{2 r^2}+\frac{(d-3) (d-4)}{2 r^2} A_d(r)+\frac{(d-3) }{r}A_d'(r)+\frac{1}{2} A_d''(r)=8\pi T^{\theta_{i} }{}_{\theta_{i} } = 8\pi p_\text{t}(r),
\end{align}
where $i\geq1$ and a prime denotes differentiation with
respect to the radial coordinate. Considering an anisotropic matter source with the energy-momentum tensor (EMT) $T^\mu_{\,\,\,\nu} = \text{diag}[-\rho(r) , p_\text{r} (r) , p_\text{t}(r) ,..., p_\text{t}(r)]$, Eq. \eqref{err} can be integrated to
\begin{align}
\label{firstar}
A(r) = 1 - \frac{m^{d-3}}{r^{d-3}} - \frac{2}{(d-2) r^{d-3}} \int \rho(r) r^{d-2} dr,
\end{align}
where $m$ is an integration constant chosen so that $A(r)=1$ when $\rho (r)=0$ and $r \rightarrow \infty$. A comparison with the ST BH gives us \cite{tang}
\begin{align}
\label{mfirst}
m^{d-3} = \frac{16 \pi M}{(d-2) \Omega_{d-2}},
\end{align}
where $M$ is the Arnowitt–Deser–Misner (ADM) mass and $\Omega_{d-2}$ is the volume of a unit $(d-2)$-dimensional sphere
\begin{align}
	\Omega_{d-2} = \frac{2 \pi^{\frac{d-1}{2}}}{\Gamma\left(\frac{d-1}{2}\right)}.
\end{align}
 Combining  \eqref{err} and \eqref{epp} yields the relation between $p_\text{t}(r)$ and $\rho(r)$
\begin{align}
\label{rhop}
-\rho(r) - \frac{1}{d-2} r \rho^\prime (r) = p_\text{t}(r) \hspace{0.7cm}\Rightarrow \hspace{0.7cm}\rho(r) r^{d-2}= - (d-2)\int r^{d-3} p_\text{t}(r) \, dr + C_1,
\end{align}
where $C_1$ is an arbitrary integration constant.  Therefore, the relevant EMT must have the form  $T^t{}_t=T^r{}_r$ and $T^{\theta_{1}}{}_{\theta_{1}}=T^{\theta_{2}}{}_{\theta_{2}}$ from \eqref{err} and \eqref{epp}. Furthermore, given the energy density function $\rho(r)$, the transverse pressures are given by \eqref{rhop}.

In the following section, we introduce the appropriate nonlinear magnetic monopole Lagrangian and derive the exact form of the metric by applying \eqref{rhop}. 

\section{A class of $d$-dimensional regular BHs}
\label{sec3}
Our goal is to study a metric that describes a general class of regular $d$-dimensional BHs  introduced in \cite{our1}. This class of geometries is parameterized by the free parameters $m$, $q$, and $n$. The latter is the polytropic index of stellar matter in the OSD collapse \cite{our1}. The metric in  $4$ dimensions is
\begin{align}
	\label{met4}
	ds^2 = -\big( 1 - \frac{m}{r(1+ m^{\frac{3}{n} -2}q^2 r^{-\frac{3}{n}})^n}\big) dt^2 + \big( 1 - \frac{m}{r(1+ m^{\frac{3}{n}  -2}q^2 r^{-\frac{3}{n}})^n}\big) ^{-1} dr^2 + r^2 d\Omega_2^2,
\end{align}
where using \eqref{mfirst}, $m = 2M$ in $4$ dimensions. This geometry describes a de Sitter core and as $r\to \infty$, the Schwarzschild solution is recovered. Known solutions for regular BHs with de Sitter cores include the Hayward and Bardeen metrics. Both metrics belong to the aforementioned class of spacetimes when the polytropic index takes the specific values $n=1$ and $n=3/2$, respectively. In addition to generalizing the metric \eqref{met4} to $d$ dimensions, we are interested in interpreting the sources of these geometries via NLED. Previous studies on the Bardeen BH have employed NLED to incorporate magnetic charge as a free parameter of the metric \cite{Ayon}\footnote{There is also a study that considers an electric source for the Bardeen BH, but that is not our focus here \cite{Rodrigues:2018bdc}.}. The Hayward BH contains a free parameter that acts as a fundamental length scale.  For more information on studying the Hayward metric with NLED see \cite{Malaf, Malaf2}. As mentioned above, both BHs can be subsumed under a more general metric \cite{our1} and identified by specific values of the polytropic index.

Here, we consider the general class of regular BHs described by Eq. \eqref{met4} and  demonstrate that this class has a de Sitter core with NLED as its source. We also generalize the solution to a class of $d$-dimensional regular BHs with de Sitter cores.
We begin with the action for NLED coupled to GR in $d$ dimensions
\begin{align}
\label{action}
\mathcal{A} = \frac{1}{16\pi} \int d^d x \sqrt{-g}\left[{\mathcal R} - 4 \mathcal{L}(F)\right],
\end{align}
where ${\mathcal R} $ is the Ricci scalar and $F$ is the Maxwell scalar defined by contracting the Maxwell tensor 
\begin{align}
F= \frac{1}{4} F_{\mu\nu}F^{\mu\nu},\quad\quad\quad F_{\mu\nu} = 2 \nabla_{[\mu}A_{\nu]}.
\end{align}
The variation of the action \eqref{action} with respect to the metric yields
\begin{align}
{\mathcal R} _{\mu\nu} - \frac{1}{2} g_{\mu\nu} {\mathcal R} = 8\pi T_{\mu\nu},\quad\quad\quad\nabla_\mu (\mathcal{L}_F F^{\mu\nu})=0,\quad\quad\quad \nabla_{[\mu}F_{\nu\rho]}=0,
\end{align}
where $\mathcal{L}_F = \partial\mathcal{L}/\partial F$. The EMT is
\begin{align}
\label{EMT11}
T_{\mu\nu}= \frac{1}{4\pi}(\mathcal{L}_F F_{\mu\alpha}F^{\nu\alpha} - g_{\mu\nu} \mathcal{L}).
\end{align}
In $4$ dimensions, substituting \eqref{met4} into \eqref{err} yields the  Lagrangian as a function of $r$
\begin{align}
	\mathcal L(r)\big|_{d=4} =\frac{3m^{\frac{3}{n}-1}}{2q^{2 n}}   \left(\frac{q^2 r^{-3/n}}{1+m^{\frac{3}{n}-2} q^2  r^{-3/n}}\right)^{n+1}.
\end{align}
To find the Lagrangian in terms of $F$, we must use an ansatz. Assuming  that the only non-zero component of the Maxwell tensor is $F_{\theta_1\theta_2}$, as in \cite{Ayon}, the free parameter $q$ plays the role of a magnetic charge. Thus, the Maxwell scalar would be $F = q^2/2 r^4$. Consequently, we obtain the Lagrangian density in terms of the Maxwell scalar
\begin{align}
		\label{LFd4}
	\mathcal{L}(F)\big|_{d=4} =\frac{3 m^{\frac{3}{n}-1}}{2 q^{2 n}}\left(\frac{\left(2 F q^{2 \left(\frac{4 n}{3}-1\right)}\right)^{\frac{3}{4 n}}}{1+m^{\frac{3}{n}-2} \left(2 F q^{2 \left(\frac{4 n}{3}-1\right)}\right)^{\frac{3}{4 n}}}\right)^{n+1}.
\end{align}
This result reduces to the Bardeen Lagrangian for $n=3/2$ \cite{Ayon, Rodrigues:2018bdc}\footnote{Using \eqref{EMT11} and the metric \eqref{met4}, the Lagrangian \eqref{LFd4} satisfies both \eqref{err} and \eqref{epp} in $4$ dimensions.}. However, the Lagrangian \eqref{LFd4} does not reduce to Maxwell's Lagrangian, $\mathcal{L}(F)=F$, and therefore, the Reissner–Nordström metric is not recovered in any limit. This is the main result of the above ansatz in which the monopole charge of a self–gravitating magnetic field is used instead of the Coulomb charge \cite{Ayon, Rodrigues:2018bdc}.

To generalize the $4$-dimensional Lagrangian \eqref{LFd4} to higher dimensions,  we must start by suggesting the Lagrangian density. Our suggestion for the appropriate form of the Lagrangian density in $d$ dimensions is as follows 
\begin{align}
	\label{LFf}
	\mathcal{L}(F)=\frac{(d-1) (d-2) m^{\frac{3 (d-2)}{2 n}-1}}{4 q^{\frac{2n}{3} (d-1) }}\left[\frac{\left(2 F q^{2(\frac{2 n}{3} (d-2)-(d-3)) }\right)^{\frac{3}{4 n}}}{1+m^{\left(\frac{d-2}{2}\right)  \left(\frac{3}{n}-2\right)} \left(2 F q^{2(\frac{2 n}{3} (d-2)-(d-3))}\right)^{\frac{3}{4 n}}}\right]^{\frac{2n}{3}\frac{d-1 }{d-2}+1},
\end{align}
where $n$ is an arbitrary positive parameter that plays the role of the polytropic index in the aforementioned collapse scenario and $m$ is the parameter introduced in \eqref{mfirst} that is related to the ADM mass. The cases $n = 1$ and $n = 3/2$ correspond to the nonlinear Lagrangians  sourcing the Hayward and Bardeen BHs. 

\begin{figure}[h]
		\begin{minipage}[b]{0.5\linewidth}
			\includegraphics[width=1\linewidth]{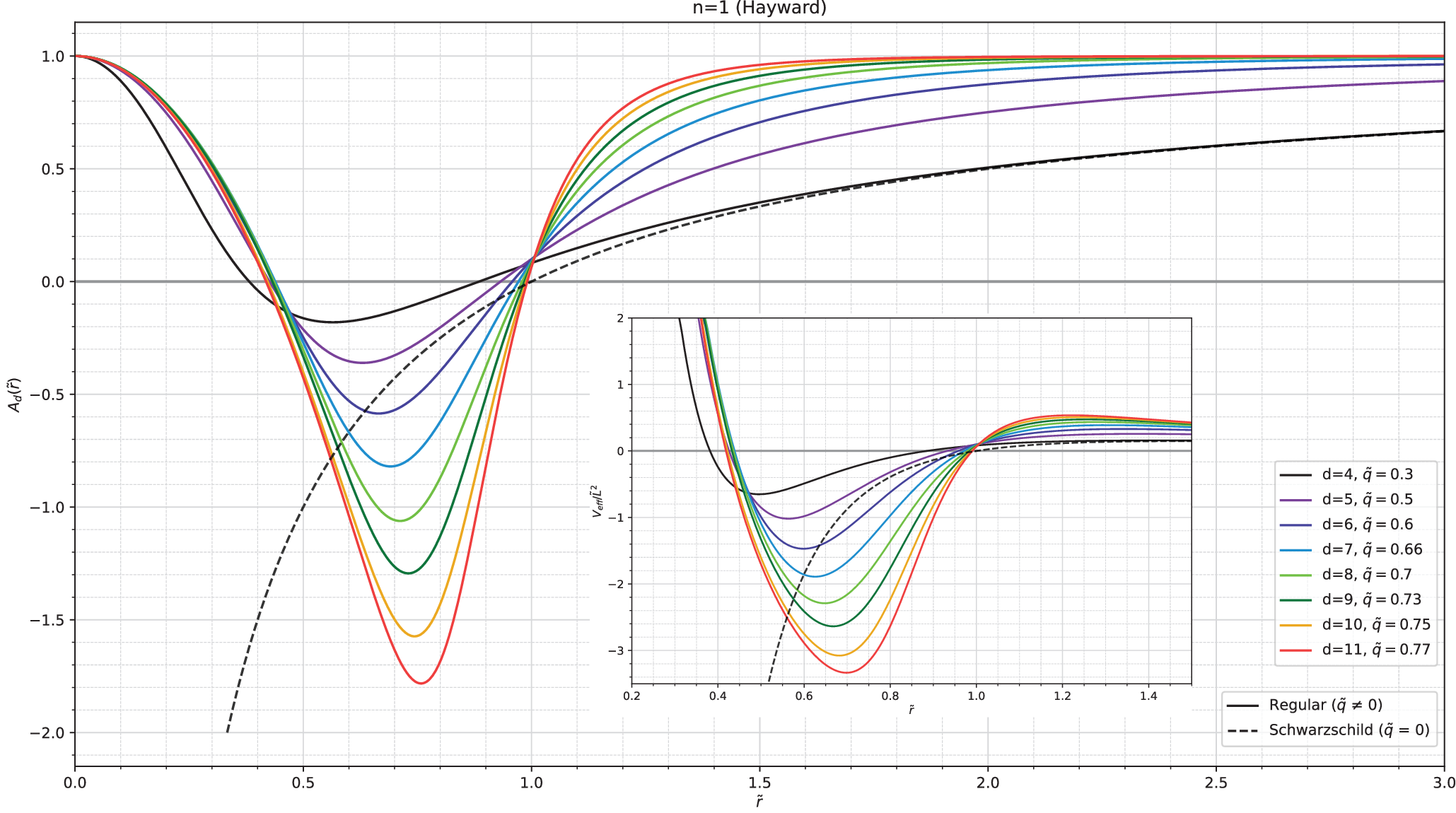} 
		\end{minipage}  \vspace{0.5cm}
		\begin{minipage}[b]{0.5\linewidth}
			\includegraphics[width=1\linewidth]{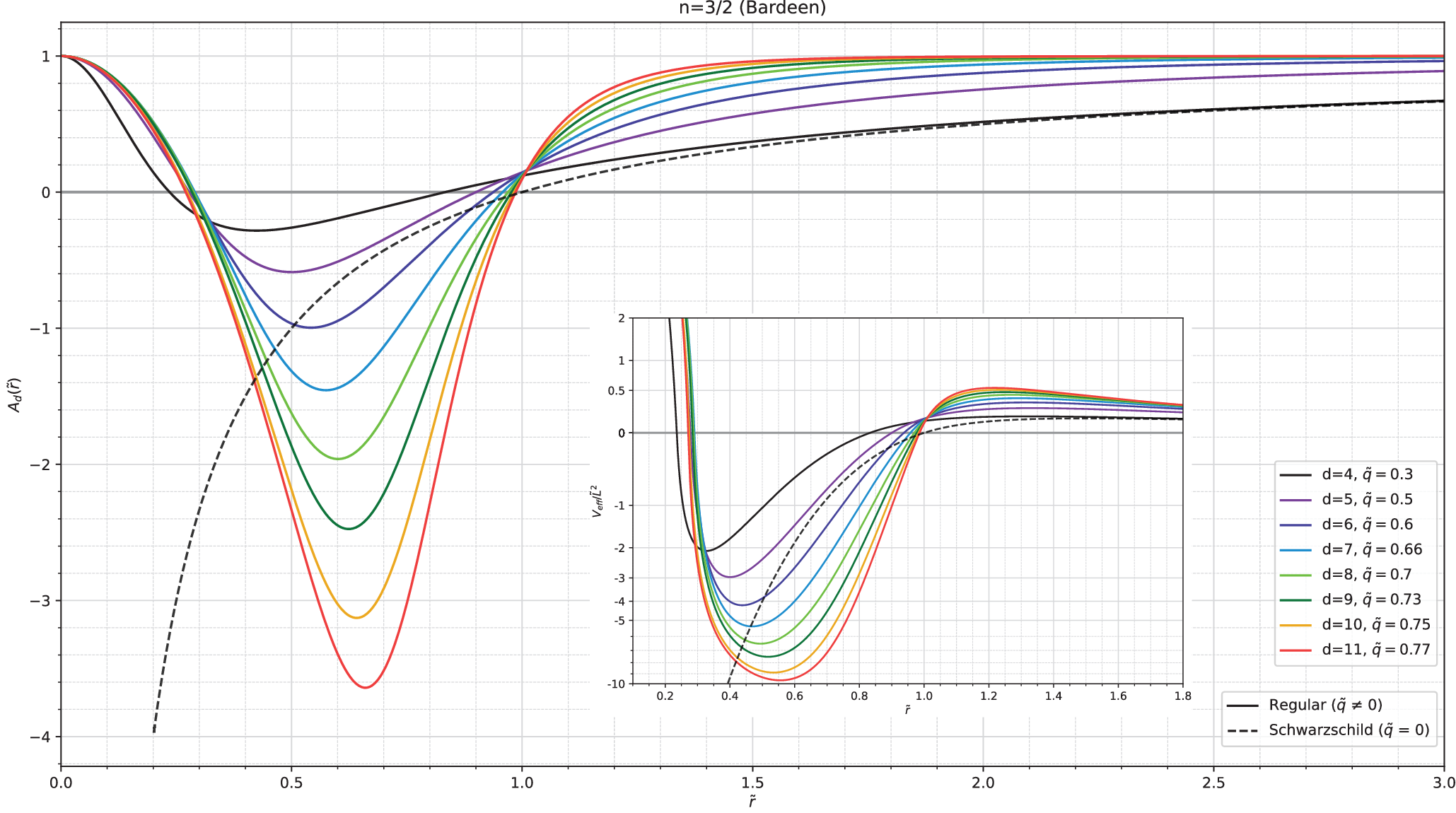} 
		\end{minipage} \\
		\begin{minipage}[b]{0.5\linewidth}
			\includegraphics[width=1\linewidth]{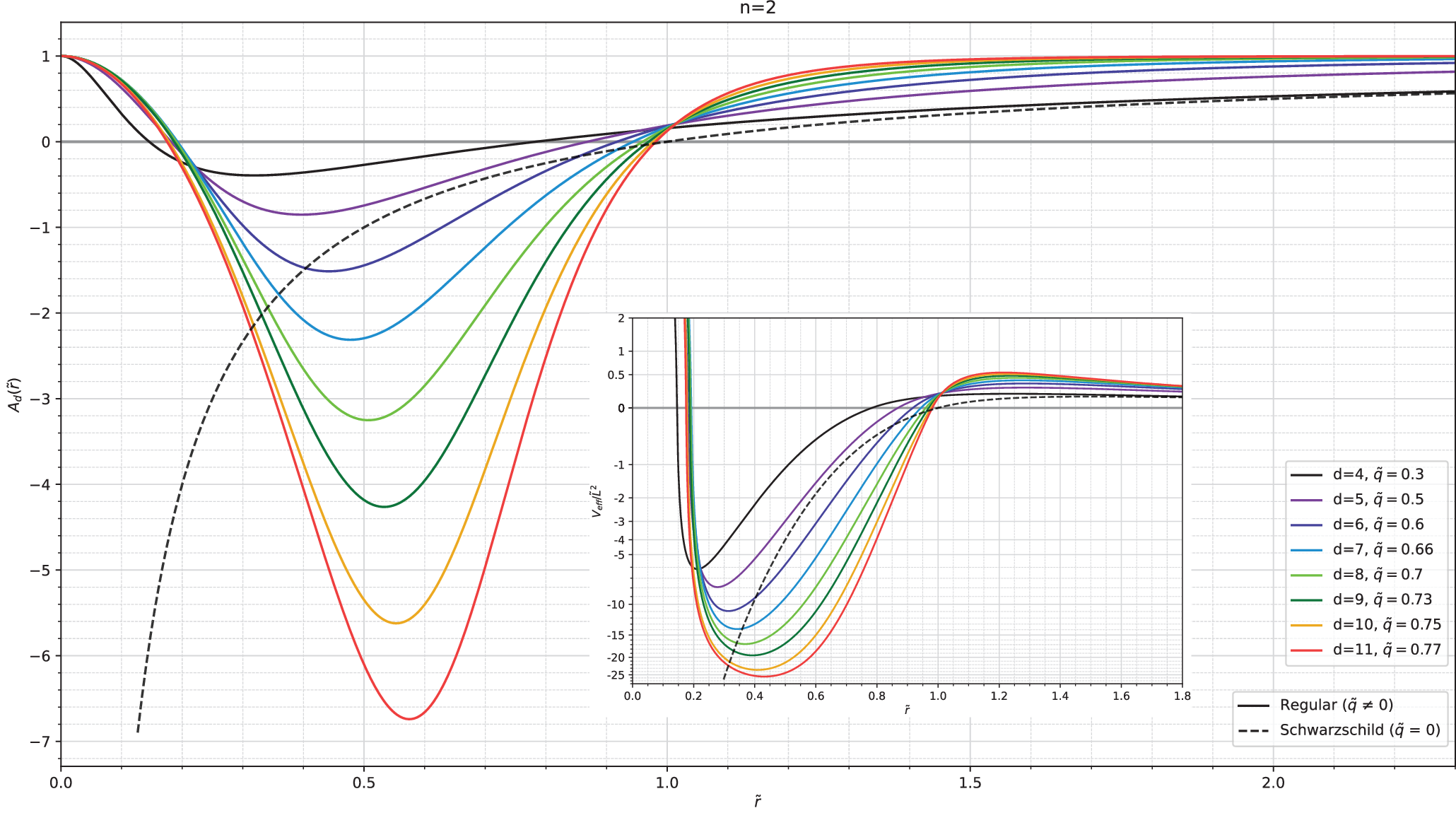} 
		\end{minipage}
		\begin{minipage}[b]{0.5\linewidth}
			\includegraphics[width=1\linewidth]{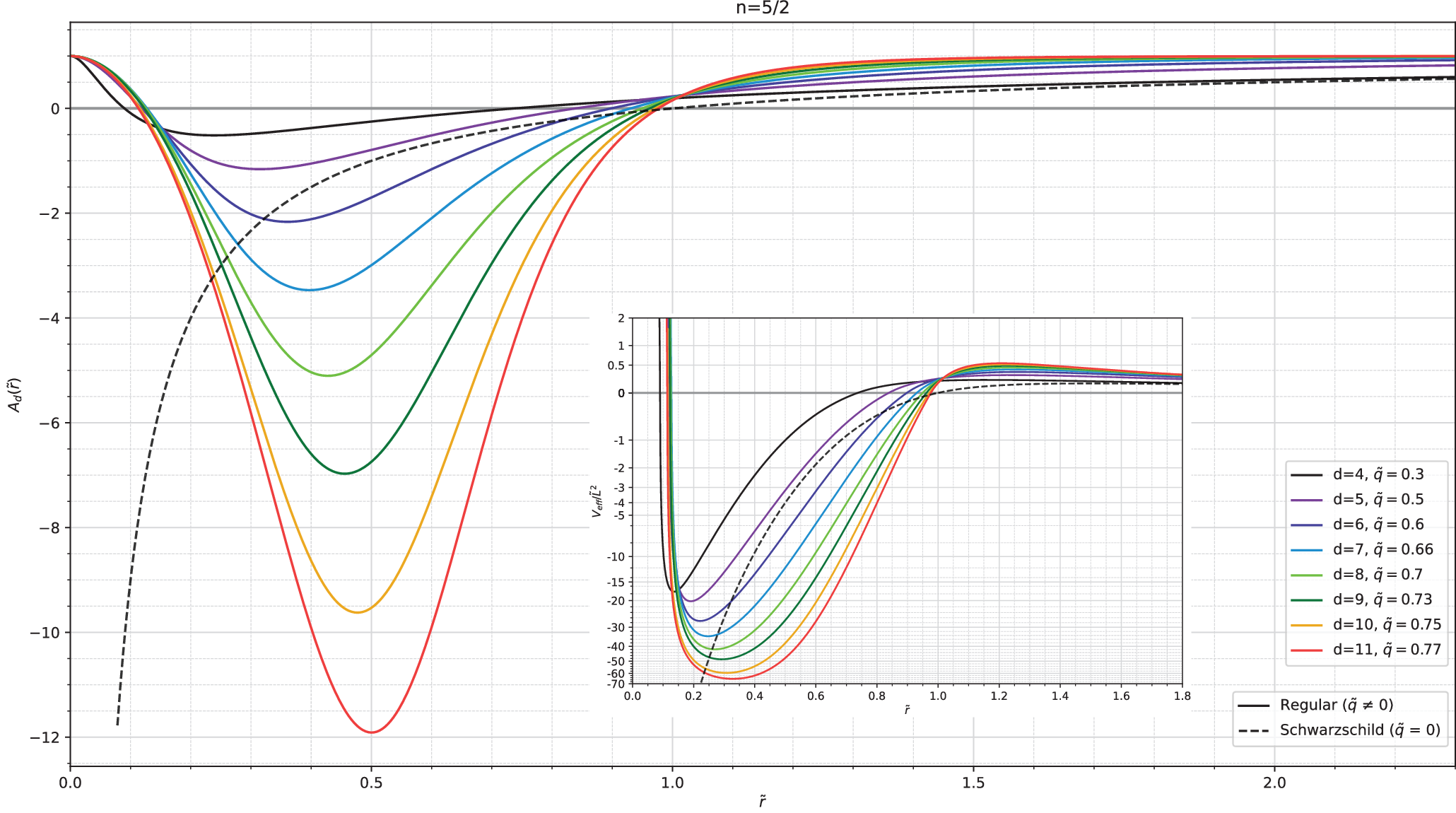} 
		\end{minipage} 
		\caption{ The plot of $A_d( \tilde r)$ and the corresponding effective potential for a massless particle as a function of radius ($\tilde r$ is scaled by $m$ as $\tilde r = r/m$) for different dimensions and for the cases $n = 1$ (i.e., Hayward), $n = 3/2$ (i.e., Bardeen), $n = 2$ and $n = 5/2$. The magnetic charge $q$ is set to different values in each dimension to make the plots comparable. The value of the rescaled magnetic charge $\tilde q$ is chosen so that the BH has two distinct roots. This ensures that the BH is neither extremal nor horizonless. The dashed line represents the Schwarzschild BH in $4$ dimensions. We have  omitted other dimensions for the ST BH plot due to the indistinguishability of the curves. The same $\tilde q$ values are used for the effective potential plot.}
		\label{figfr}
	\end{figure}

Substituting Eqs. \eqref{EMT11} and \eqref{LFf} into Eq. \eqref{rhop} yields
\begin{align}
	\label{Ffirst}
	4F + \frac{2r}{d-2} F' = 0 \quad \rightarrow \quad F= \frac{C_2}{r^{2(d-2)}},
\end{align}
where $C_2$ is the integration constant. On the other hand, the Maxwell field tensor describing a magnetic charge in $d$ dimensions is \cite{ali}
\begin{align}
	F_{\mu\nu} = 2 \delta_{[\mu}^{\theta_{d-3}}\delta_{\nu]}^{\theta_{d-2}} \frac{q^{d-3}}{r^{d-4}} \sin \theta_{d-3} \left[ \prod_{j=1}^{d-4}\sin^2 \theta_j\right],
\end{align}
which gives the Maxwell scalar as
\begin{align}
	\label{Fseco}
	F = \frac{q^{2(d-3)}}{2r^{2(d-2)}}.
\end{align}
Comparing this with \eqref{Ffirst} yields the constant $C_2$. We can then substitute \eqref{Fseco} into \eqref{LFf} to obtain the radial dependence of the Lagrangian 
\begin{align}
	\label{Lrf}
	\mathcal{L}(r)=\frac{(d-1) (d-2) q^{d-2} m^{\frac{3 (d-2)}{2 n}-1}}{4 \left(r^{\frac{3 (d-2)}{2 n}}+m^{\left(\frac{d-2}{2} \right) \left(\frac{3}{n}-2\right)}q^{d-2} \right)^{\frac{2n}{3}\frac{ d-1 }{ d-2}+1}}.
\end{align}
Using this expression, the metric component $A_d(r)$ can be found from \eqref{EMT11}, and \eqref{firstar} as
\begin{align}
	\label{ad1}
	A_d(r)=1-\frac{ m^{d-3}}{r^{d-3} } \left(1+\frac{ m^{\left(\frac{d-2}{2}\right)\left(\frac{3}{n}-2\right)}q^{d-2}}{r^{\frac{3 (d-2)}{2 n}}}  \right)^{\frac{-2 n}{3}\frac{(d-1) }{ (d-2)}},
\end{align}
where the integration constant resulting from \eqref{firstar} is chosen so that the above solution agrees with the ST  solution for $q=0$. Eqs. \eqref{Lrf} and \eqref{ad1} are consistent with $d$-dimensional Bardeen BH spacetime (i.e., $n=3/2$) \cite{ali} and also with $4$-dimensional Hayward (i.e., $n=1$) \cite{54Fro16}. \\
The physical interpretation of the NLED source can be clarified by examining the limiting behavior of the Lagrangian. Eq. \eqref{LFf} has a saturating nonlinear structure. For the magnetic configuration \eqref{Fseco}, $F=q^{2(d-3)}/[2r^{2(d-2)}]$, and hence $F \to\infty $ as $r\to 0$. In this strong-field limit, $\mathcal{L}(F)$ approaches a finite constant while $F\mathcal{L}_F \to 0$. Consequently, Eq. \eqref{EMT11} imply that the effective source approaches $p_\text{r} =p_\text{t}=-\rho$, producing the vacuum-like EOS associated with the de Sitter core. Thus, curvature regularity is achieved not because the electromagnetic invariant $F$ remains finite, but because the nonlinear response of the source saturates sufficiently rapidly to keep the stress-energy tensor finite. At large radius, Eq. \eqref{Lrf} behaves as $\mathcal{L}(r)\propto r^{-[d-1+3(d-2)/(2n)]}$, so that the NLED contribution rapidly decays and the ST asymptotics are recovered.\\
The behavior of the metric component $A_d(r)$ with respect to the radius is plotted in Fig. \ref{figfr}. As the dimension increases, the minima of $A_d(r)$ become more separated, while the maxima converge.  Increasing the polytropic index enhances this behavior.  From metric component \eqref{ad1}, we see that $A_d(r)$ asymptotically approaches
\begin{align}
		\label{adrinf}
		A_d(r)\approx&1-\frac{ m^{d-3}}{r^{d-3}}\left[1-\left(\frac{2n}{3}\left(\frac{d-1}{d-2}\right)\frac{m^{\left(\frac{d-2}{2} \right) \left(\frac{3}{n}-2\right)}q^{d-2}}{ r^{\frac{3 (d-2)}{2 n}}}\right)+\text{terms of higher orders $\frac{1}{r}$}\right],\hspace{1.5cm} r\to \infty,\\
		\label{adrzero}
		A_d(r)\approx&1-\frac{ m^{d-3} r^2}{\left(m^{\left(\frac{d-2}{2}\right) \left(\frac{3}{n}-2\right)} q^{d-2} \right)^{\frac{2n}{3}\frac{d-1}{d-2}}}\left[1-\frac{2n}{3}\frac{d-1}{d-2}\frac{  r^{\frac{3}{2n} (d-2)}}{m^{\left(\frac{d-2}{2}\right) \left(\frac{3}{n}-2\right)} q^{d-2} }+\text{terms of higher orders $r$}\right],\hspace{0.5cm} r\to 0.
\end{align}
Thus, despite the presence of the magnetic charge $q$, Eq. \eqref{adrinf} and Fig. \ref{figfr} show that the solution increasingly resembles the ST solution as the radius increases. Conversely, \eqref{adrzero} shows that the magnetic charge has a greater influence on smaller radii.
	
For the line element \eqref{genmet} with the metric component \eqref{ad1}, the Ricci scalar is given by 
\begin{align}
	{\mathcal R} &=\frac{1}{r^2}\left[(d-3) (d-2) (1-A_d(r)) -2 (d-2) r A_d'(r) -r^2 A_d''(r)\right],\\
	&= \frac{(d-1) m^{\frac{3 (d-2)}{2 n}-1} q^{d-2}  }{\left(m^{\left(\frac{d-2}{2}\right)\left(\frac{3}{n}-2\right)} q^{d-2}+r^{\frac{3 (d-2)}{2 n}}\right)^{\frac{2n (d-1) }{3 (d-2)}+2}}\left( m^{\left(\frac{d-2}{2}\right)\left(\frac{3}{n}-2\right)} d q^{d-2} -\frac{3 d-2 (n+3) }{2 n} r^{\frac{3 (d-2)}{2 n}}\right).
\end{align}
For  $q\neq0$, the scalar curvature is obviously finite at $r=0$.
  \subsection{Horizons}
 The horizons are the roots of $A_d(r)$, so from \eqref{ad1}, we have  
\begin{align}
	\label{firsthor}
	\mathfrak{R}_\text{h}^{3-d} +\mathfrak{R}^2_\text{h} m^{\left(\frac{d-2}{2}\right)\left(\frac{3}{n}-2\right)} q^{d-2} -{m}^{\frac{3}{2n}\frac{(d-2)(d-3)}{(d-1)}}=0,
\end{align}
where  $ \mathfrak{R}_\text{h} =  r_\text{h}^{-\frac{3 (d-2)}{2n(d-1)}}$ and $ r_\text{h}$ is the radius of the horizon(s). The order of the algebraic equation \eqref{firsthor} depends on $d$. This emphasizes the importance of specifying the dimension in order to analytically find the horizon radius. Mathematically, for odd dimensions up to $d = 9$, Eq. \eqref{firsthor} is analytically solvable for $r_\text{h}$. Some exact solutions are examined in Appendix \ref{appa}. For the general dimension, Eq. \eqref{firsthor} can have two real roots,  one real root, or no real roots. This can be seen in the zero points of $A_d(r)$ in  Fig. \ref{figfr}.

Although the horizon equation cannot be solved for $r_{\text{h}}$ in any dimension, it can be solved analytically for $q$ as
\begin{align}
	\label{qhorzionntilde}
	q= m^{1-\frac{3}{2 n}} r_{\text{h}}^{\frac{3}{2 n}} \bigg(\big(\frac{m^{d-3}}{r_{\text{h}}^{d-3}}\big)^{\frac{3 (d-2)}{2 (d-1) n}}-1\bigg)^{\frac{1}{d-2}},
\end{align}
which can be written in terms of dimensionless quantities, i.e., $\tilde r = r/m$ and $\tilde q = q/m$, as
	\begin{align}
		\label{qhorizon}
		\tilde q= \tilde r_{\text{h}}^{\frac{3}{2 n}} \left(\left(\tilde r_{\text{h}}^{d-3}\right)^{-\frac{3 (d-2)}{2n (d-1) }}-1\right)^{\frac{1}{d-2}}.
	\end{align}
	The extremal case with only one real root can be found analytically using the condition $ A_d(\tilde r_{\text{ext}})= A'_d(\tilde r_{\text{ext}})=0$.  These equations give the extremality condition and also the radius of the extremal horizon, respectively as
	\begin{align}\label{qext}
		\tilde q_{\text{ext}}&=\left(\frac{d-3}{2}\right)^{\frac{1}{d-2}} \left(\frac{2}{d-1}\right)^{\frac{d-1}{(d-3) (d-2)}},\\
		\label{rext}
		\tilde r_{\text{ext}}&=\left(\frac{2}{d-1}\right)^{\frac{2 n(d-1) }{3 (d-2) (d-3)}}.
	\end{align}
	The above equations show that the value of  $\tilde q$ for extremality  is independent of the polytropic index $n$, while the extremal radius depends on it. According to  Eqs. \eqref{qext} and \eqref{rext}, the scaled radius of the extremal horizon and the corresponding scaled charge approach unity as $d$ approaches infinity. This shows that the rate of change of $\tilde q_{\text{ext}}$ and $\tilde r_{\text{ext}}$ decreases with increasing dimensions. In each dimension, $\tilde q_{\text{ext}}$ is the maximum charge value. For larger values, the horizon equation \eqref{firsthor} has no roots, thus, the metric \eqref{ad1} represents a compact object without a horizon.
	
		\subsection{Energy conditions}
		\label{SSecEC}

\begin{figure}[!h]\centering
		\begin{minipage}[b]{0.49\linewidth}
			\includegraphics[width=1\linewidth]{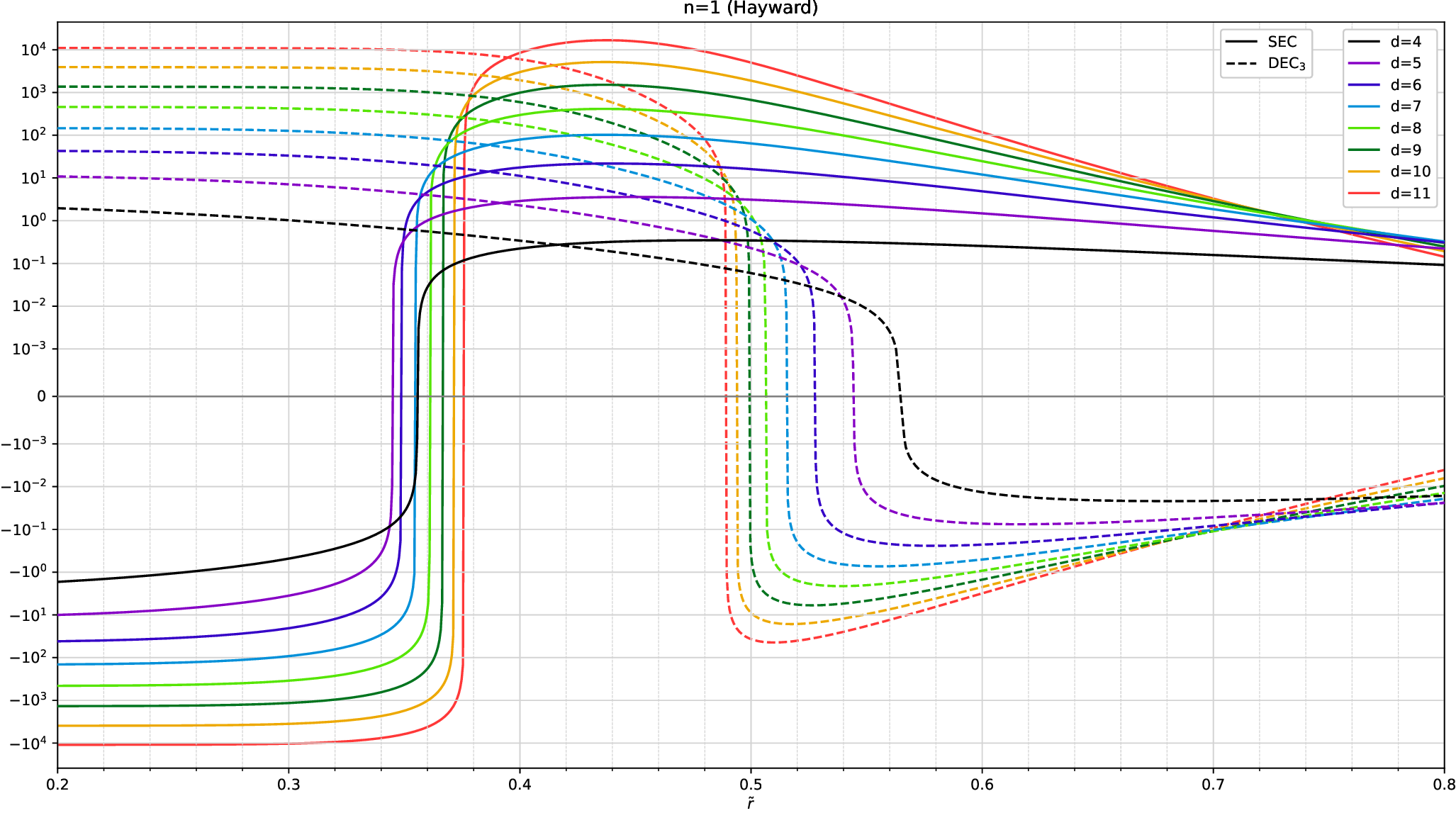}
		\end{minipage}  
		\begin{minipage}[b]{0.49\linewidth}
			\includegraphics[width=1\linewidth]{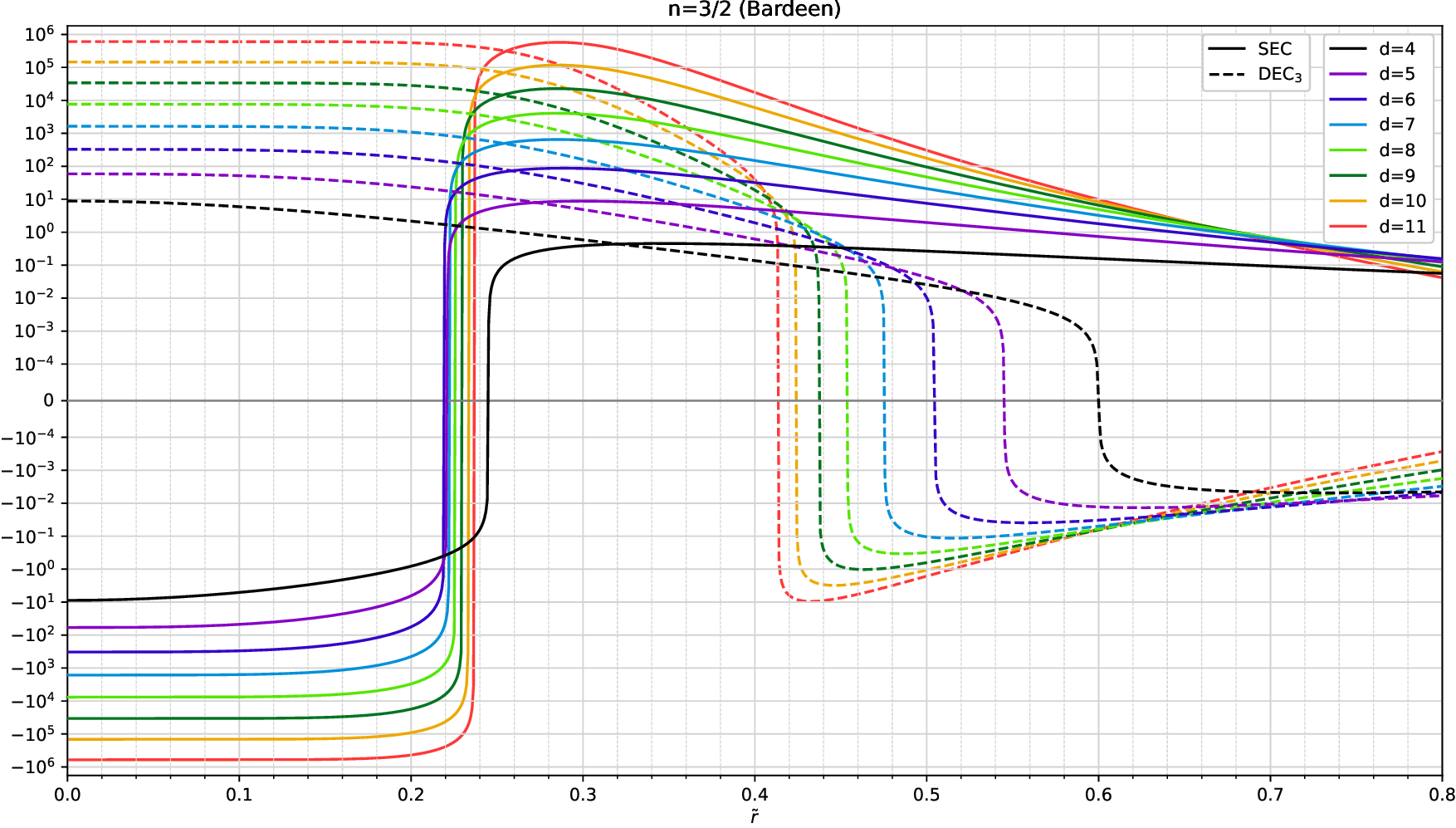}
		\end{minipage}\\
		\begin{minipage}[b]{0.49\linewidth}
			\includegraphics[width=1\linewidth]{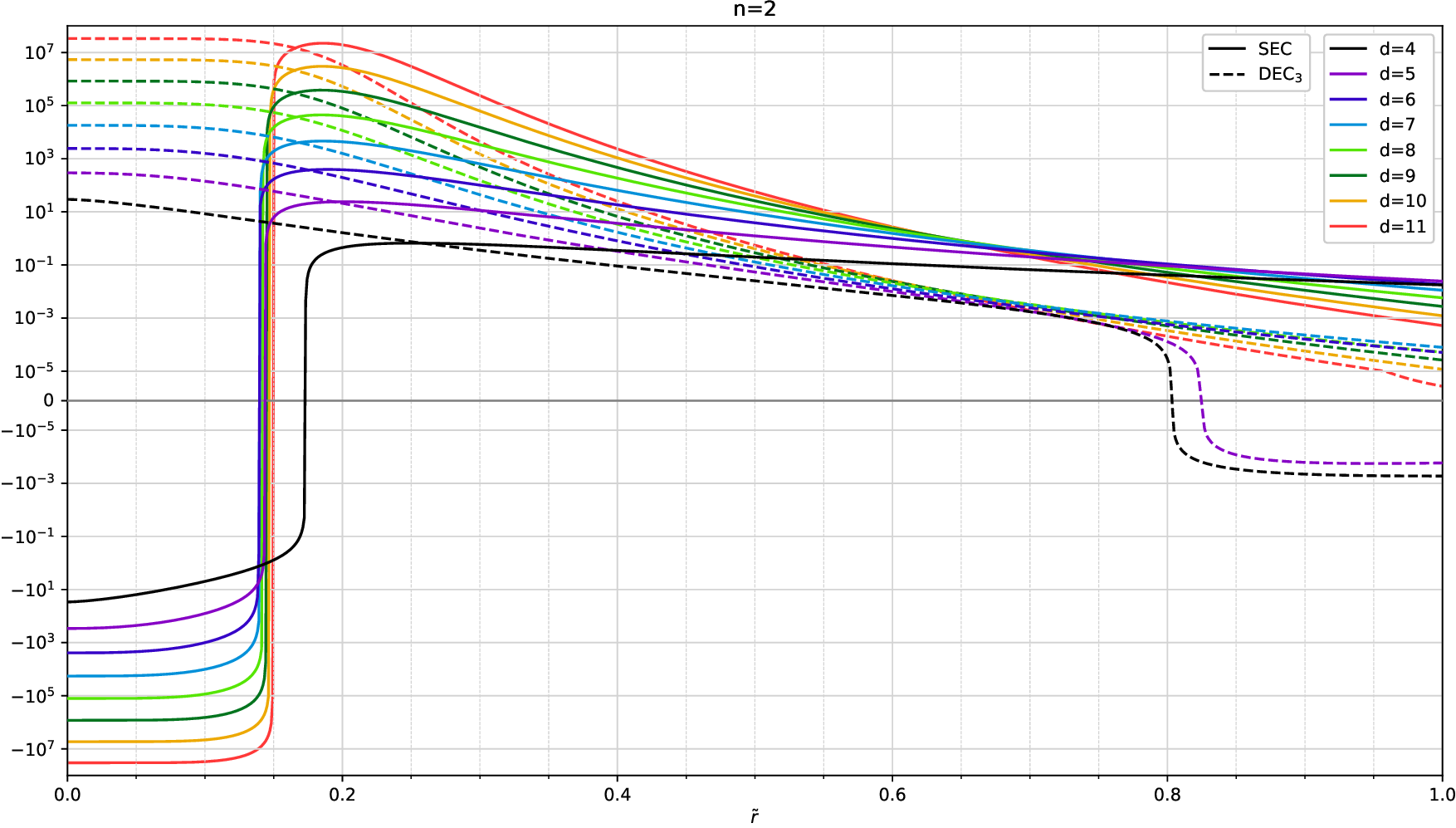}
		\end{minipage}  
		\begin{minipage}[b]{0.49\linewidth}
			\includegraphics[width=1\linewidth]{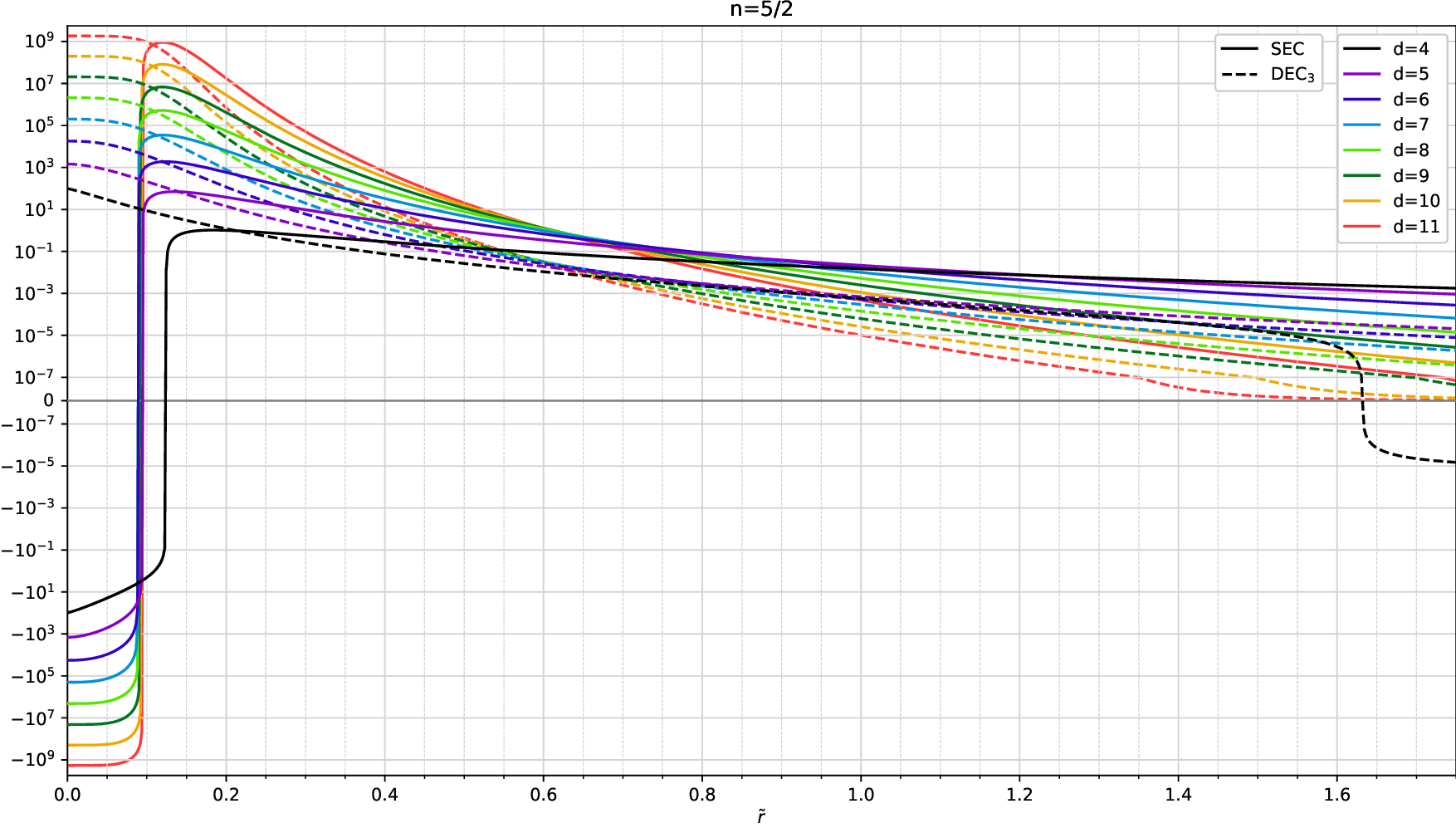}
		\end{minipage}	\hfill
		\caption{SEC and DEC$_3$ as a function of the dimensionless radius for different polytropic indices and dimensions. The solid and dashed lines represent SEC and DEC$_3$, respectively.  All other energy conditions are satisfied.}
		\label{figEC} 
	\end{figure}

		According to Einstein’s equations \eqref{err} and \eqref{epp}, we discuss the energy conditions using the EMT in the form of $T^\mu{}_{\nu} = \text{diag}[-\rho(r) , p_\text{r}(r) , p_\text{t}(r) ,..., p_\text{t}(r)]$. For this form of the EMT, weak energy condition (WEC), null energy condition (NEC), dominant energy condition (DEC), and strong energy condition (SEC) are expressed as \cite{Guo:2022ghl}
	\begin{itemize}
		\item[(i)] WEC\({}_{1,2}(r) \)= {NEC}\({}_{1,2}(r) \):  \( \rho + p_i \geq 0 \),
		\item[(ii)] WEC\({}_{3}(r) \) = {DEC}\({}_{1}(r) \) : \( \rho \geq 0 \),
		\item[(iii)]SEC\((r) \)= \( (d - 3) \rho + p_\text{r} + (d-2) p_\text{t} \geq 0 \)
		\item[(iv)] DEC\({}_{2,3}(r) \) : \( \rho + p_i \geq 0 \), and  \( \rho - p_i \geq 0 \)  
	\end{itemize}
	where $ i = r, t $  denotes the radial and tangential pressures. Using Einstein's equations \eqref{err} and \eqref{epp} 
		\begin{align}
		8 \tilde \pi \tilde \rho (\tilde r) &= - 8\pi \tilde p_\text{r}  (\tilde r) = \frac{(d-1) (d-2)}{2\tilde r^{d-1}} \tilde q^{d-2} \tilde r^{\frac{-3 (d-2)}{2 n}}\left(\tilde q^{d-2}\tilde  r^{-\frac{3 (d-2)}{2 n}}+1\right)^{-\frac{2 (d-1) n}{3 (d-2)}-1},\\
		8\pi \tilde p_\text{t} (\tilde r) &= \frac{(d-1) }{4 n \tilde r^{d-1}}\tilde q^{d-2} \tilde r^{-\frac{3 (d-2)}{2 n}}\left(\tilde q^{d-2} \tilde r^{-\frac{3 (d-2)}{2 n}}+1\right)^{-\frac{2 (d-1) n}{3 (d-2)}-2}\left((3 d+2 n-6)-2 (d-2) n \tilde q^{d-2} \tilde r^{\frac{-3(d-2)}{2n}}\right).
	\end{align}
It is evident that the WEC\({}_{1,2}\) and NEC\({}_{1,2}\) are satisfied everywhere because $\tilde \rho + \tilde{p}_\text{r} =0$. Also, since $\tilde q$ and $n$ are positive, $\rho\geq0$ and therefore, the WEC\({}_{3}\) and DEC\({}_{1} \) are satisfied as well.  Additionally,
	\begin{align}
		\tilde \rho + \tilde p_\text{t} = \frac{(d-1) (2 (d-1) n+3 (d-2))}{32 \pi n \tilde r^{d-1}}\tilde q^{d-2} \tilde r^{-\frac{3 (d-2)}{2 n}}\left(\tilde q^{d-2} \tilde r^{-\frac{3 (d-2)}{2 n}}+1\right)^{-\frac{2 (d-1) n}{3 (d-2)}-2},
		\end{align}
is always positive. Thus, the WEC${}_{1,2}$ and consequently, the NEC${}_{1,2}$ hold everywhere. Since $\tilde\rho + \tilde{p}_i \geq 0$, violating  DEC$_{2}$, means violating $\tilde\rho - \tilde{p}_\text{r} \geq 0$, which is satisfied everywhere since both $\tilde\rho$ and $-\tilde{p}_\text{r} $ are positive. Therefore, only the tangential component must be checked. A‌ simple calculation also reveals that 
\begin{align}
	\label{DEC23}
	\text{DEC}_{3}(\tilde r) = \tilde\rho - \tilde p_\text{t} &= \frac{(d-1) \tilde q^{d-2} \tilde r^{-\frac{3 (d-2)}{2 n}}}{32 \pi n \tilde r^{d-1}} \left(\tilde q^{d-2} \tilde r^{-\frac{3 (d-2)}{2 n}}+1\right)^{-\frac{2 (d-1) n}{3 (d-2)}-2}\nonumber\\
	&\times \left(4 (d-2) n \tilde q^{d-2} \tilde r^{-\frac{3 (d-2)}{2 n}}+(d (2 n-3)-6 n+6)\right) \geq 0
\end{align}
Only the last parenthesis can be negative. It is positive if 
\begin{align}
	\label{DECnd}
	n\geq\frac{3 (d-2)}{2 (d-3)}
\end{align}
In this case,  DEC$_{3}$ is satisfied everywhere. Otherwise, it is only satisfied for 
\begin{align}
	\label{vioDEC}
	\tilde r \leq \left(\frac{4 n(d-2)  \tilde q^{d-2}}{3 (d-2)-2n (d-3) }\right)^{\frac{2 n}{3 (d-2)}}\equiv \tilde r_{\text{DEC}}.
\end{align}
To satisfy the SEC, it is required that
		\begin{align}
		 \frac{(d-2) (d-1)}{32 \pi n  \tilde r^{d-1}} \tilde q^{d-2}\tilde r^{-\frac{3 (d-2)}{2 n}}\left(\tilde q^{d-2} \tilde r^{-\frac{3 (d-2)}{2 n}}+1\right)^{-\frac{2 (d-1) n}{3 (d-2)}-2}\nonumber\\
		 \times\left(2 (d-3) n+3 (d-2)-4 n \tilde q^{d-2} \tilde r^{-\frac{3 (d-2)}{2 n}}\right) \geq 0,
		\end{align}
holds. Only the last expression in parentheses can be negative. Therefore, the SEC is satisfied if
	\begin{align}
		\label{vioSEC}
		\tilde r\geq \left(\frac{4 n \tilde q^{d-2}}{2n(d-3)+3 (d-2)}\right)^{\frac{2 n}{3 (d-2)}} \equiv \tilde r_{\text{SEC}}.
		\end{align}
From \eqref{vioDEC} and \eqref{vioSEC}, we get
	\begin{align}
		\frac{\tilde r_{\text{DEC}}}{\tilde r_{\text{SEC}}}=\left(\frac{(d-2) (3 (d-2)+2 n(d-3))}{3 (d-2)-2 n(d-3) }\right)^{\frac{2 n}{3 (d-2)}}.
		\end{align}
	For a given $d$ and $n$,  it is clear that $\tilde r_{\text{DEC}}\geq \tilde r_{\text{SEC}}$. The SEC and DEC$_{3}$ are plotted  as a function of radius in Fig. \ref{figEC}. According to \eqref{vioDEC} and \eqref{vioSEC}, 
	these energy conditions can be violated at certain radii.  Additionally, if the polytropic indices and dimensions satisfy \eqref{DECnd}, then there is no DEC$_{3}$ violation.
\section{Geodesics}
\label{sec4}
In this section, we will find the null circular orbits of the metric component \eqref{ad1} and check geodesic completeness. The first step is to write the Lagrangian as
	\begin{align}
		\label{Lagpho1}
		\mathscr{L}=\frac{1}{2}g_{\alpha\beta} \dot x ^\alpha \dot x^\beta =\frac{1}{2}\left[- A_d(r) \dot t^2+ A_d(r)^{-1}\dot r^2 +r^2 \sum_{i=1}^{d-2} \left[ \prod_{j=2}^{i} \sin^2 \theta_{j-1}\right] \dot \theta_i^2 \right],
	\end{align}
	where the dot indicates differentiation with respect to the affine parameter. The canonically conjugate momentum $P_{\mu} = \partial \mathcal{L}/\partial \dot x^\mu$ can be calculated as \cite{Singh:2017vfr}
	\begin{align}
     P_\text{t} &= - A_d(r) \dot t,\\
     P_\text{r}  &= A_d(r)^{-1} \dot r,\\
     P_{\theta_i} &= r^2 \sum_{i=1}^{d-3} \prod_{j=1}^{i-1} \sin^2 \theta_j \dot{\theta}_{d-3},\hspace{1cm} i = 1,2,\dots,d-3,
     \\
     P_{\theta_{d-2}} &= r^2 \prod_{i=1}^{d-3} \sin^2 \theta_i \dot \theta_{d-2},
	\end{align}
where $\theta_{d-2}$ is the last tangent coordinate of spacetime. Due to the spherical symmetry of the gravitational field,  we can restrict ourselves to $\theta_i = \pi/2$ (except for $\theta_{d-2}$), without loss of generality. The conserved quantities  along the geodesic, $E$ and $L$, are given by the following Euler-Lagrange equations
		\begin{align}
			\label{ELt}
		t:\hspace{0.1cm} & - A_d(r) \dot t = E,\\
		\label{ELphi}
		\theta_{{d-2}}:\hspace{0.1cm} & r^2 \prod_{i=1}^{d-3} \sin^2 \theta_i \dot \theta_{d-2} = r^2 \dot \theta_{d-2}= L,
	\end{align}
which allows us to rewrite the Lagrangian \eqref{Lagpho1} as
\begin{align}
	\label{Lagpho2}
	\mathscr{L} = \frac{\epsilon}{2} = -\frac{E^2}{2 A_d(r) } + \frac{\dot r^2}{2 A_d(r) }+ \frac{L^2}{2 r^2}.
\end{align}
The parameter $\epsilon$ for the timelike and null geodesics is set to  $-1$ and $0$, respectively.  Multiplying \eqref{Lagpho2} by $A_d(r)$ and rearranging the terms yields
\begin{align}
	\label{LagEep}
E^2+\epsilon=\dot r^2-\epsilon(A_d(r)-1)+ \frac{A_d(r) L^2}{r^2}.
\end{align}
For null geodesics (i.e., $\epsilon =0$)
\begin{align}
\label{EffVeff}
E^2 = \dot r^2+V_{\text{eff}},\,\ \hspace{1cm}\text{where} \hspace{1cm}V_{\text{eff}} = \frac{A_d(r) L^2}{ r^2}.
\end{align}
The effective potential, normalized to $L^2$, is plotted in Fig. \ref{figfr} for different values of dimension and $\tilde q$. This potential tends to infinity as $r\to0$ and approaches zero for large values of $r$.
\subsection{Geodesic completeness}
	To determine whether the spacetime is fully regular, geodesic completeness must be investigated in addition to curvature regularity. 
	For a radial timelike geodesic, $L=0$ and $\epsilon=-1$, Eq. \eqref{LagEep} gives $\dot r^2=E^2-A_d(r)$. 	Therefore, the proper time (affine parameter) required to reach the center is	
	\begin{align}
		\lambda(r)-\lambda(r_i)=\int_{r_i}^{r}\frac{dr}{\sqrt{E^2-A_d(r)}}.
	\end{align} 
	Near the regular center, $A_d(r)=1+O(r^2)$. Hence, for $E^2>1$, the integrand approaches $1/\sqrt{E^2-1}$, and the geodesic
	reaches $r=0$ in a finite proper time. Curvature regularity at $r=0$ is therefore not by itself sufficient to establish geodesic completeness; the extendibility of the spacetime beyond the center must also be
	examined \cite{Zhou:2022yio}.
One finds that when $\frac{3(d-2)}{2n}$ is an odd integer, the metric function \eqref{ad1} diverges at a negative radius
\begin{align}
	r=- m^{1-\frac{2n}{3}}q^{\frac{2n}{3}}
\end{align}
For odd-integer values of $s=\frac{3(d-2)}{2n}$, the real extension to $r<0$ develops an additional negative-radius singularity, $r=-m^{1-2n/3}q^{2n/3}$. For even-integer $s$, this particular negative-radius singularity is absent, and the real extension may be chosen symmetric under $r \to -r$. For non-integer $s$, however, the continuation of $r^s$ to negative values of $r$ is branch dependent and is not automatically a real single-valued metric. Therefore, the absence of a real negative-radius zero in this expression should not be interpreted as a proof of full geodesic
completeness. In the present work, regularity is understood as curvature regularity on the physical domain $r\ge0$, while full geodesic completeness of the maximally extended spacetime requires a separate global analysis.\\
For the Hayward metric ($n=1$) in 4 dimensions, $\frac{3(d-2)}{2n}$ is an odd integer and the metric is geodesically incomplete, with the singularity occurring at $r= - m^{1/3}q^{2/3}$ which lies in the $r<0$ region. The extended spacetime is geodesically complete for the Bardeen metric ($n=3/2$) in 4 dimensions.\\
In summary, the metric introduced here is curvature-regular for $r \geq0$, but geodesic completeness of the maximally extended spacetime is not guaranteed. Therefore, throughout this work we use the term “regular” in the standard curvature-regular sense, while explicitly distinguishing it from full geodesic completeness.

\subsection{Photon spheres}
\label{SSPS}
   Here we focus on photon spheres.  For a null circular geodesic, $\dot r = 0$ which yields $E^2=V_{\text{eff}}$ using Eq. \eqref{EffVeff}. To determine the radius of the photon sphere, we require $\ddot r =0$  which is equivalent to ${d V_{\text{eff}}}/{d r}=0$. The two conditions mentioned above give the following equations, which describe the photon spheres of the geometry \eqref{ad1} 
    \begin{align}
    	\label{rdoteq0}
    	\dot{\tilde r} &= 0 \hspace{1cm} \Rightarrow \hspace{1cm} \frac{\tilde E^2}{\tilde L^2}= \frac{A_d(\tilde r)}{\tilde r^2} =  \frac{1}{\tilde r^2}\left(1-\frac{1}{\tilde r^{d-3}}\left(1+\frac{\tilde q^{d-2}}{\tilde r^{\frac{3 (d-2)}{2 n}}}\right)^{-\frac{(d-1) (2 n)}{3 (d-2)}}\right), \\
    	\label{dvdr}
    	    \frac{\tilde V_{\text{eff}}^\prime}{L^2}&=\frac{A_d^\prime(\tilde r)}{\tilde r^2}- \frac{2A_d(\tilde r)}{\tilde r^3}=0\hspace{1cm} \Rightarrow \hspace{1cm}\frac{2 \tilde r^{d-3}}{d-1} = \left(1+\frac{\tilde q^{d-2}}{ \tilde r^{\frac{3}{2n} (d-2)}}\right)^{-\frac{2n (d-1) }{3 (d-2)}-1} .
    \end{align}
   Combining these equations leads to the following equation for the radius of the photon sphere, $\tilde r_{\text{PS}}$
        \begin{align}
        	\label{rps}
         \tilde r_{\text{PS}} = \left(\frac{2}{d-1}\right)^{\frac{2n (d-1)}{3(d-2)(d-3)}} \left(1 - \frac{E^2 \tilde r^2_{\text{PS}}}{L^2}\right)^{-\frac{2n (d-1)}{(d-2)(d-3)}- \frac{1}{d-3}}.
    \end{align}
    The right-hand side of the above equation is the product of two factors. The first factor is the radius of the extremal horizon, as given by \eqref{rext} and the second factor is greater than one. From this, we deduce that $\tilde r_{\text{PS}} \geqslant \tilde r_{\text{ext}}$ , i.e., the photon sphere is always outside the outer horizon, unless 
    	\begin{align}
    		\frac{\tilde E^2}{\tilde L^2}=\frac{1}{\tilde r^2}-\frac{1}{\tilde r^{d-3}}\bigg(1+\big(\frac{2}{d-1}\big)^{\frac{d-1}{d-3}}\tilde r^{\frac{-3 (d-2)}{2 n}}\bigg)^{-\frac{2 (d-1) n}{3 (d-2)}},
    	\end{align}
         in which case the photon sphere coincides with the extremal horizon.
         
          Now, we want to find the radius of the photon sphere for which the extreme value of $\tilde q$ occurs. According to Eq. \eqref{dvdr}, on the photon sphere,  $A_d^\prime = 2A_d/r$. This gives $\tilde q=\tilde q({\tilde r_{\text{ph}}})$, and then we can find its extremum. To do so, we write Eq. \eqref{dvdr} as
             	\begin{align}\label{ph}
            A^\prime \left(\tilde r_{\text{PS}}, \tilde q(\tilde r_{\text{PS}})\right) = \frac{2}{\tilde r_{\text{PS}}} A\left(\tilde r_{\text{PS}},\tilde q(\tilde r_{\text{PS}})\right).
         \end{align}
        Calculating the total derivative of the above equation with respect to the radius of the photon sphere yields
         \begin{align}
         	\label{pqpr}
         	\frac{\partial \tilde{q}}{\partial \tilde{r}} = -\tilde{V}^{\prime\prime}_{\text{eff}}\left(A\left(\tilde{r}_{\text{PS}}, \tilde{q}(\tilde{r}_{\text{PS}})\right) - \frac{2}{\tilde{r}} \frac{\partial A\left(\tilde{r}_{\text{PS}}, \tilde{q}(\tilde{r}_{\text{PS}})\right)}{\partial \tilde{q}}\right)^{-1},
         \end{align}
         and setting $\partial \tilde q/\partial \tilde r_{\text{PS}} = 0$, we obtain
         \begin{align}\label{ee}
         	A_d^{\prime\prime} (\tilde r_{\text{PS}}, \tilde q(\tilde r_{\text{PS}}))= \frac{2 A_d^\prime(\tilde r_{\text{PS}}, \tilde q(\tilde r_{\text{PS}}))}{\tilde r_{\text{PS}}}- \frac{2 A_d(\tilde r_{\text{PS}}, \tilde q(\tilde r_{\text{PS}})) }{\tilde r^2_{\text{PS}}} = \frac{2 A_d(\tilde r_{\text{PS}}, \tilde q(\tilde r_{\text{PS}}))}{\tilde r_{\text{PS}}^2}.
         \end{align}
               
               Eq. \eqref{ee} is actually $\tilde V^{\prime\prime}_{\text{eff}}=0$, which has been evaluated at the photon sphere\footnote{This can easily be seen by calculating the second derivative of Eq. \eqref{EffVeff} on the photon sphere.}  and  yields the extremum of $\tilde q_{\text{PS}}$ in Eq. \eqref{dvdr}. For the corresponding  radius,  the two photon spheres coincide, forming a degenerate photon sphere. Its radius and charge are given by
         \begin{align}
         	\label{maxqph}
         	\tilde q_{\text{PS,\,deg}}&=\left(\frac{d-1}{2}\right)^{\frac{3}{2n (d-3)}} \left(\frac{3 d+4 n-6}{2 n(d-3) }\right)^{\frac{1}{2-d}} \left(\frac{2 n(d-3) }{3 d+4 n-6}+1\right)^{\frac{2 n(1-d) +3 (2-d)}{2 n(d-3) (d-2) }},\\
         	\label{rdis}
         	\tilde r_{\text{PS,\,deg}}&=\left(\frac{d-1}{2}\right)^{\frac{1}{d-3}} \left(\frac{2n (d-3) }{3 d+4 n-6}+1\right)^{\frac{2n (1-d) +3 (2-d)}{3 (d-2) (d-3)}}.
         \end{align}
         To understand how $\tilde r_{\text{PS,\,deg}}$ behaves with respect to $n$, it is sufficient to consider 
         \begin{align}
         \frac{d \tilde r_{\text{PS,\,deg}}}{d n}=-\tilde r_{\text{PS,\,deg}} \left(\frac{2}{3 d+4 n-6}+\frac{2 (d-1) \ln \left(\frac{2 (d-3) n}{3 d+4 n-6}+1\right)}{3 (d-3) (d-2)}\right).
         \end{align}
Since  $d>4$ and $n\geq0$, the expression in parentheses is always positive, so ${d \tilde r_{\text{PS,\,deg}}}/{d n}<0$. In other words, an increase in the polytropic index causes a decrease in $\tilde r_{\text{PS,\,deg}}$.
         For  $\tilde q = \tilde q_{\text{PS,\,deg}}$, Eq. \eqref{dvdr} has only one positive real root, which corresponds to a horizonless compact object with a degenerate photon sphere. For any value of $\tilde q > \tilde q_{\text{PS,\,deg}}$, Eq. \eqref{dvdr} has no acceptable roots, and thus no photon spheres. Conversely, for $\tilde q < \tilde q_{\text{PS,\,deg}}$, this equation has at most two real positive roots. In both cases, as previously mentioned, the photon spheres are always outside the extremal horizon. See Fig. \ref{figfr} for clarification.
         
         We can analyze the stability of photon spheres by examining $\tilde V^{\prime\prime}_{\text{eff}}$ on the photon sphere. Negative and positive values indicate unstable and stable photon spheres, respectively. Substituting \eqref{ad1} into \eqref{pqpr} reveals that the expression in parentheses is always negative. Therefore $\partial \tilde q /\partial \tilde r<0$ (or equivalently $\tilde r_{\text{PS}} > \tilde r_{\text{PS,\,deg}}$) corresponds to $\tilde V^{\prime\prime}_{\text{eff}}<0$  meaning the photon sphere is unstable, and $\partial \tilde q /\partial \tilde r>0$ (or equivalently, $\tilde r_{\text{PS}} < \tilde r_{\text{PS,\,deg}}$) corresponds to $\tilde V^{\prime\prime}_{\text{eff}}>0$, meaning the photon sphere is stable. When $\partial \tilde q /\partial \tilde r=0$, i.e., $\tilde q = \tilde q_{\text{PS,\,deg}}$, there is a marginally unstable photon sphere at $\tilde r_{\text{PS,\,deg}}$.
         
For $ \tilde q=0$, $\tilde V^\prime_{\text{eff}}$ has a single root at $\tilde r =\tilde r_{\text{PS,\,ST}}$. This corresponds to the single photon sphere of a ST BH. The radius of the photon sphere, together with the corresponding effective potential, can be calculated as
\begin{align}
	\label{rpsst}
\tilde r_{\text{PS,\,ST}}= \left(\frac{d-1}{2}\right)^{\frac{1}{d-3}}, \hspace{2cm}\tilde V_{\text{eff,\,ST}}\big|_{\tilde r_{PS,\,ST}}= \tilde L^2\left(\frac{ d-3 }{d-1}\right)\left(\frac{2}{d-1}\right)^{\frac{2}{d-3}},
\end{align}
where $\tilde V^{\prime\prime}_{\text{eff}}|_{\tilde r_{\text{PS,\,ST}}}<0$ indicating an unstable photon sphere, as expected.
At $d=4$ and $\tilde q =0$, we recover the Schwarzschild maximum of the effective potential $\tilde V_{\text{eff,\, S}}|_{\tilde r_{\text{PS,\,S}}}=4 \tilde L^2/27$ with a radius $\tilde r_{\text{PS,\,S}}=3/2$\footnote{Note that, in $4$ dimensions, $V_{\text{eff}}|_{\tilde r_{\text{PS}}}$ and $r_{\text{PS,\,S}}$ for Schwarzschild spacetime  are $L^2/27 M^2$ and $3M$, respectively. These values can be related to the tilde notation using \eqref{mfirst}  by substituting $\tilde m = 2 M$.}. 

\begin{figure}
	\begin{minipage}[h]{1\linewidth}
		\includegraphics[width=1\linewidth]{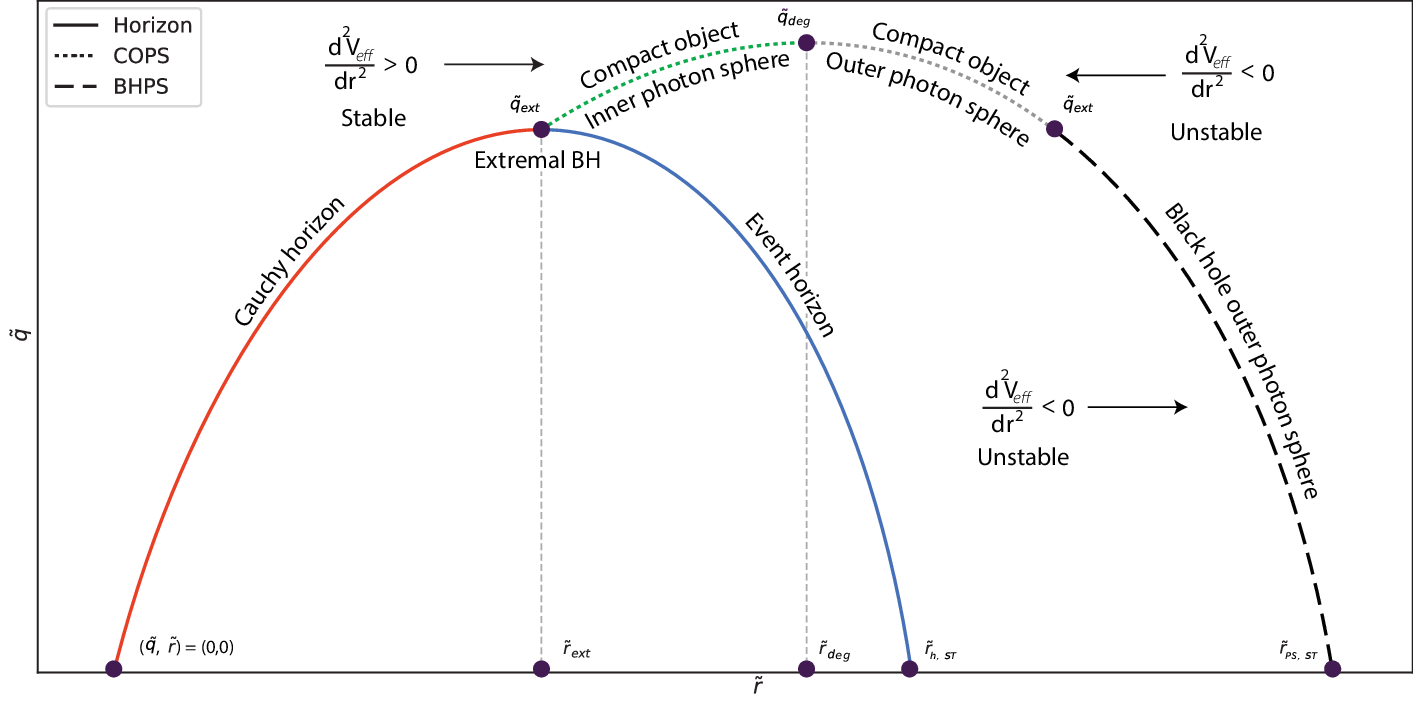}
	\end{minipage}
	\caption{Schematic illustration of the horizons, the BH photon sphere (BHPS), and the compact object photon sphere (COPS), as well as the stability of the photon spheres.}\vspace{0.2cm}
	\label{figguide}
	\begin{minipage}[b]{0.5\linewidth}
		\includegraphics[width=1\linewidth]{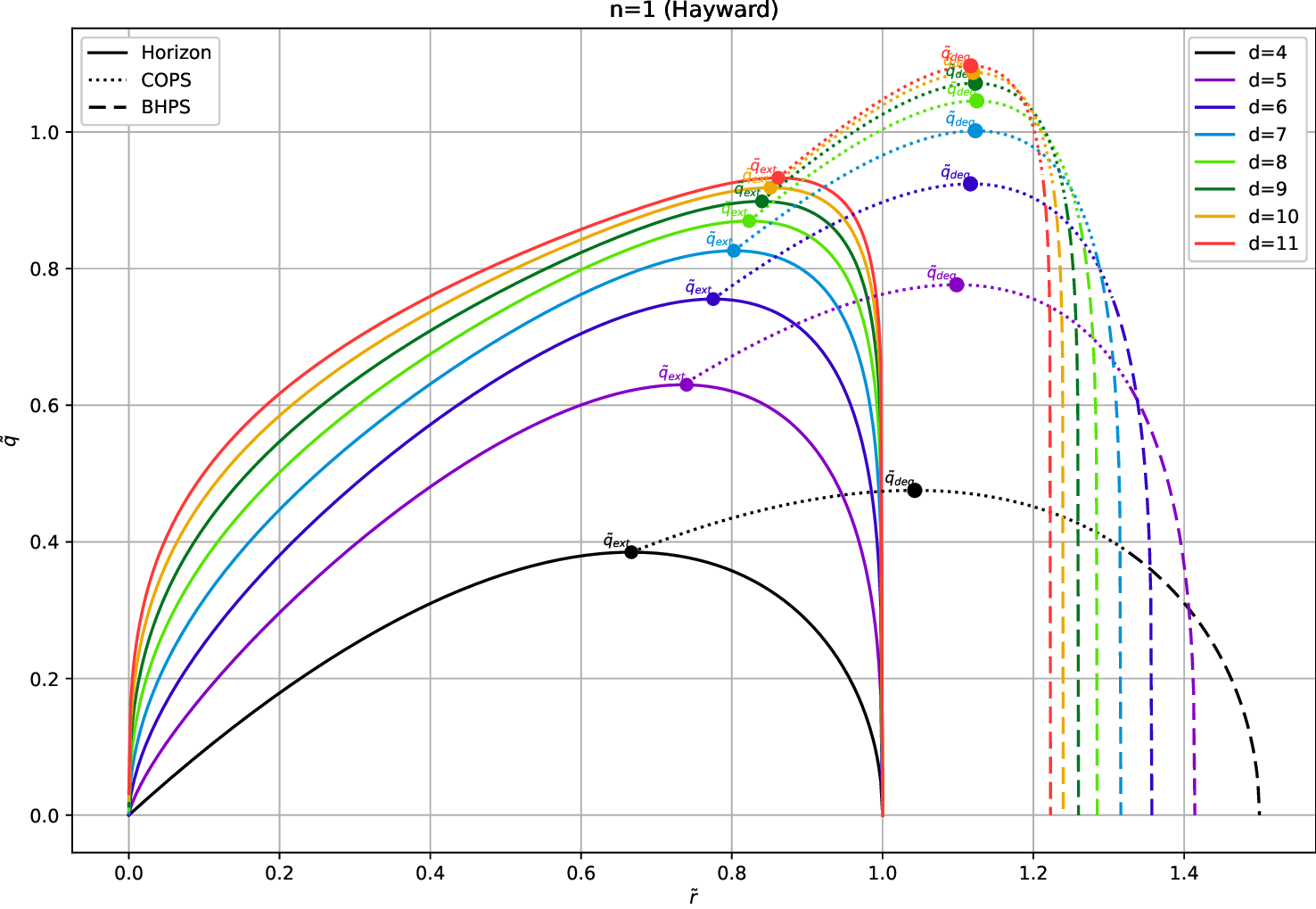} 
	\end{minipage} \vspace{0.2cm}
	\begin{minipage}[b]{0.5\linewidth}
		\includegraphics[width=1\linewidth]{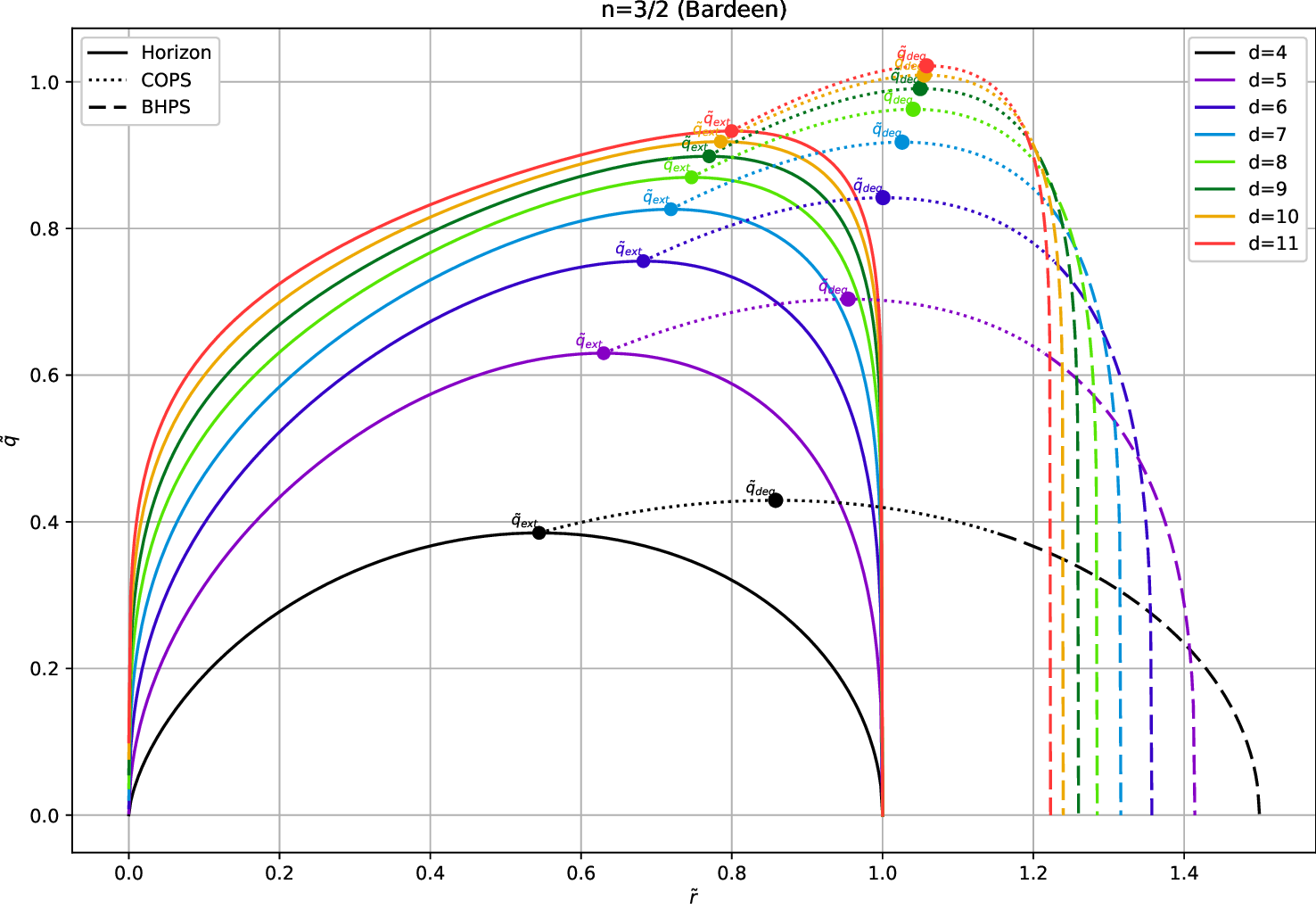} 
	\end{minipage} \\
	\begin{minipage}[b]{0.5\linewidth}
		\includegraphics[width=1\linewidth]{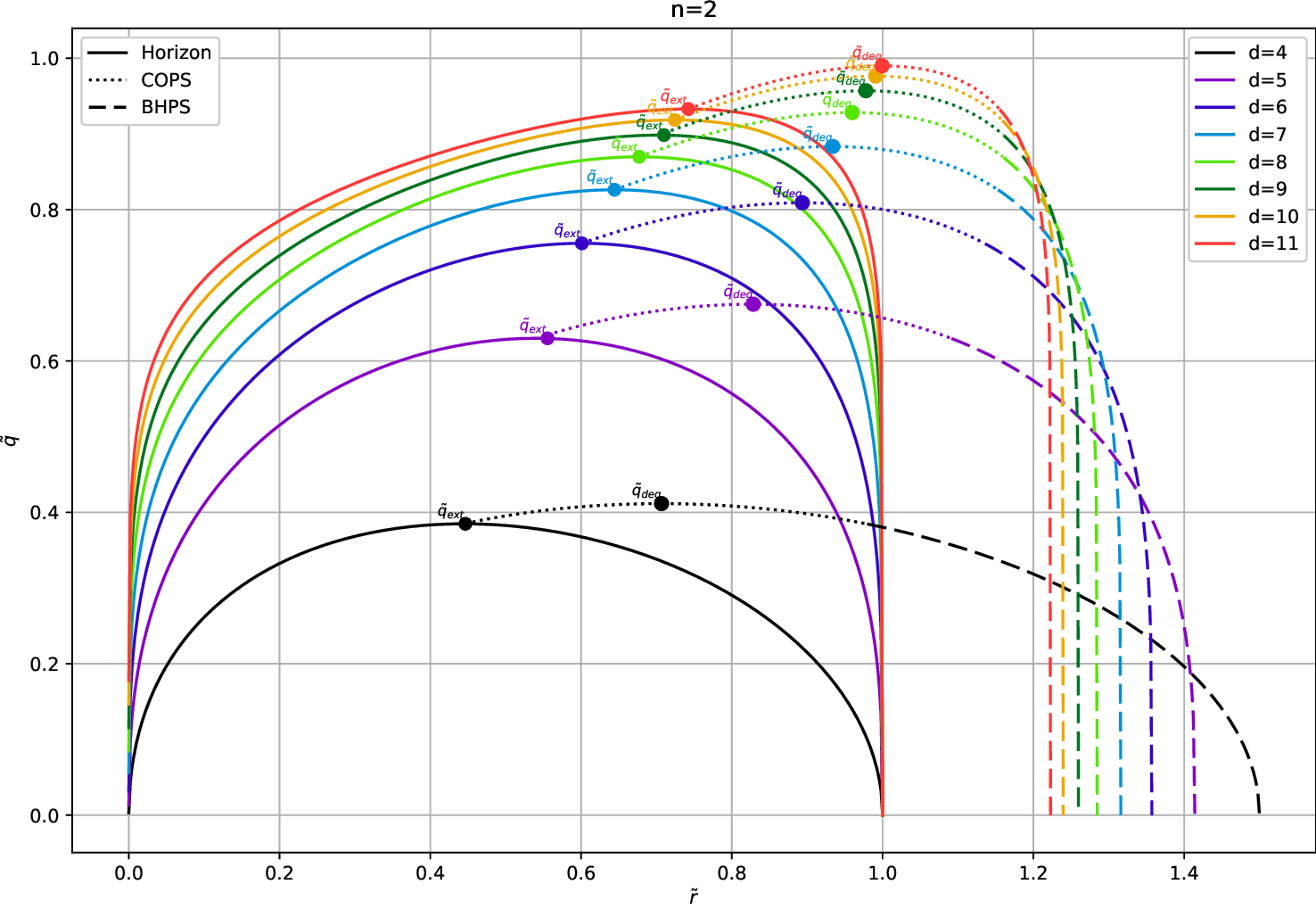} 
	\end{minipage}
	\begin{minipage}[b]{0.5\linewidth}
		\includegraphics[width=1\linewidth]{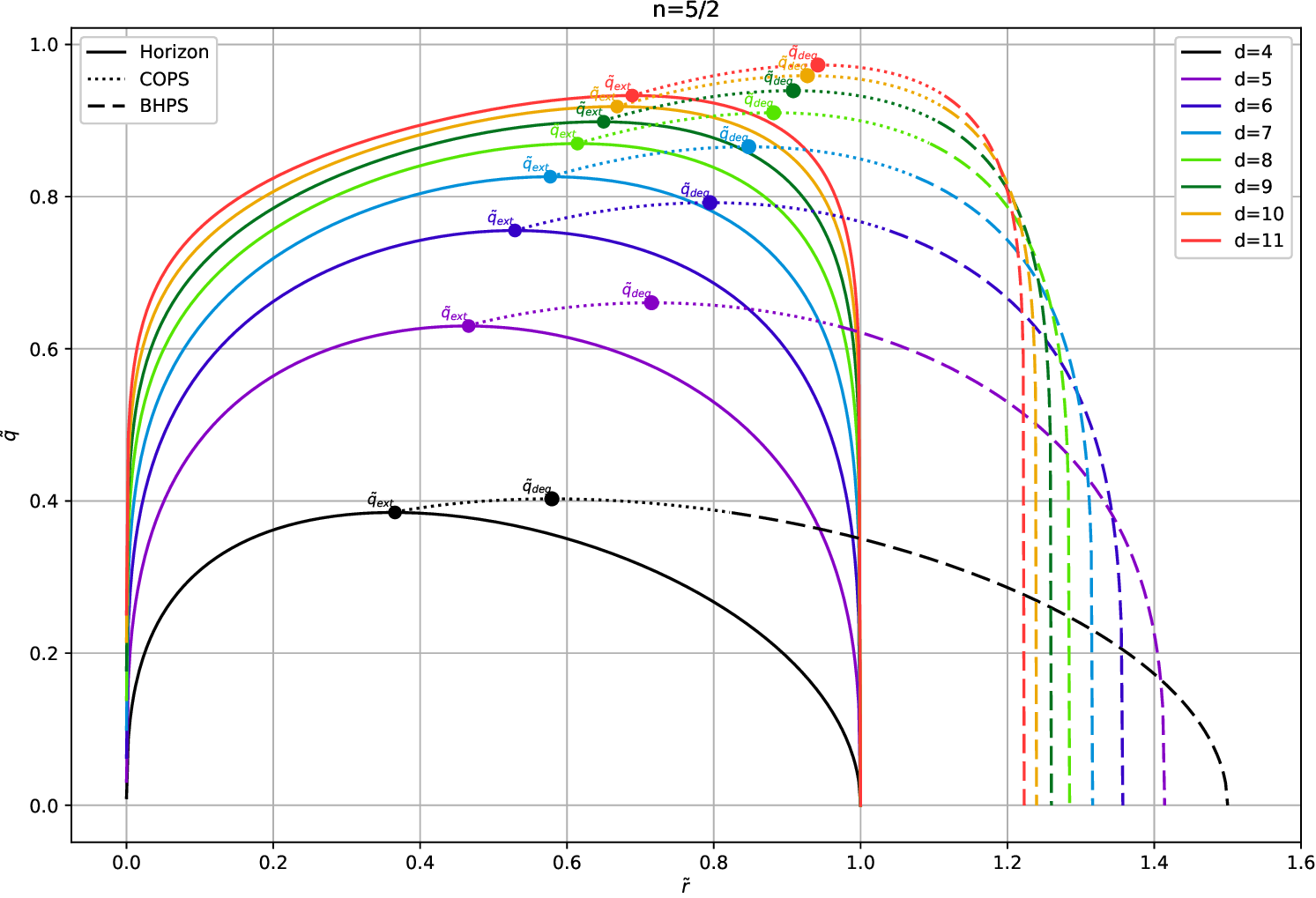} 
	\end{minipage} 
	\caption{The inner and outer horizons, and photon spheres as a function of $\tilde q$ for different dimensions and for $n=1$ (i.e., Hayward BH), $n=3/2$ (i.e., Bardeen BH), $n=2$ and $n=5/2$ are presented.  $\tilde q =0$ represents ST BH.}
	\label{figqr}
\end{figure}

Fig. \ref{figguide} shows the schematic behavior of the horizons, photon spheres and their stability.
Using Fig. \ref{figguide} and noting Eq. \eqref{rps}, Fig. \ref{figqr} plots the BH photon spheres (BHPS) and compact object photon spheres (COPS) as well as the outer and inner horizons \eqref{qhorizon}, for different dimensions and polytropic indices.

For the BH case, whether extremal or non-extremal, there is an unstable outer photon sphere. For a compact object, there are two photon spheres: the outer one is unstable  and the inner one is stable. For a given $d$ and $n$, $\tilde q_{\text{PS,\,deg}}> \tilde q_{\text{ext}}$ and $\tilde r_{\text{PS,\,deg}}> \tilde r_{\text{ext}}$, and for $\tilde q = \tilde q_{\text{PS,\,deg}}$, there is one degenerate photon sphere. For $\tilde q > \tilde q_{\text{PS,\,deg}}$, there are no photon spheres.
\subsection{Shadows}
\label{SecShad}
\begin{figure}[!h]\centering
		\begin{minipage}[b]{0.32\linewidth}
			\includegraphics[width=1\linewidth]{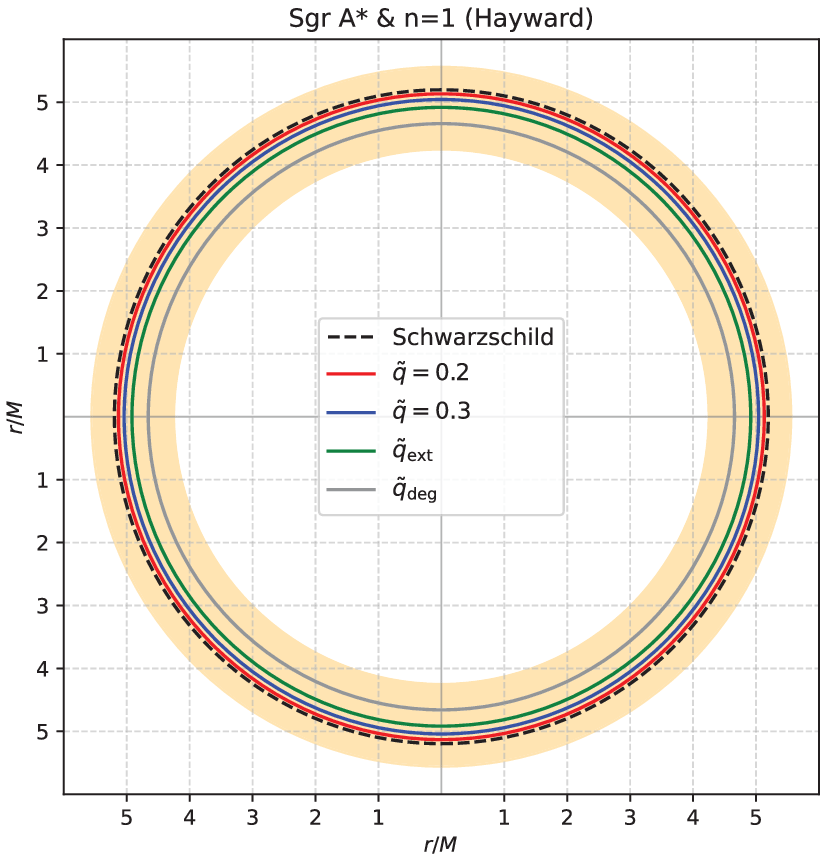}
		\end{minipage}  
		\begin{minipage}[b]{0.32\linewidth}
			\includegraphics[width=1\linewidth]{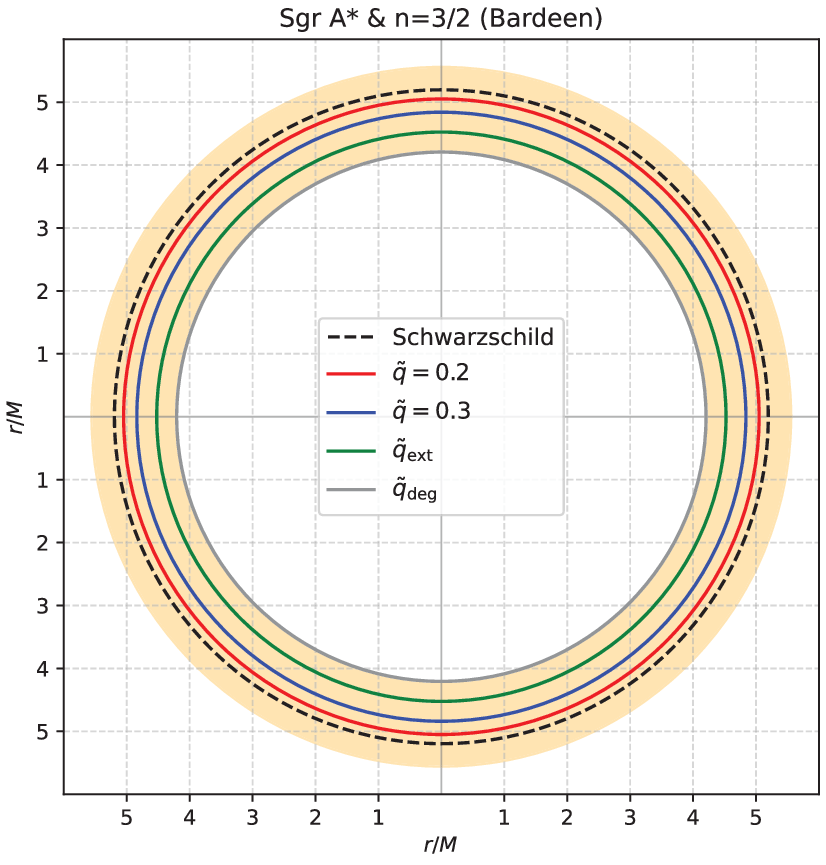}
		\end{minipage} 
		\begin{minipage}[b]{0.32\linewidth}
			\includegraphics[width=1\linewidth]{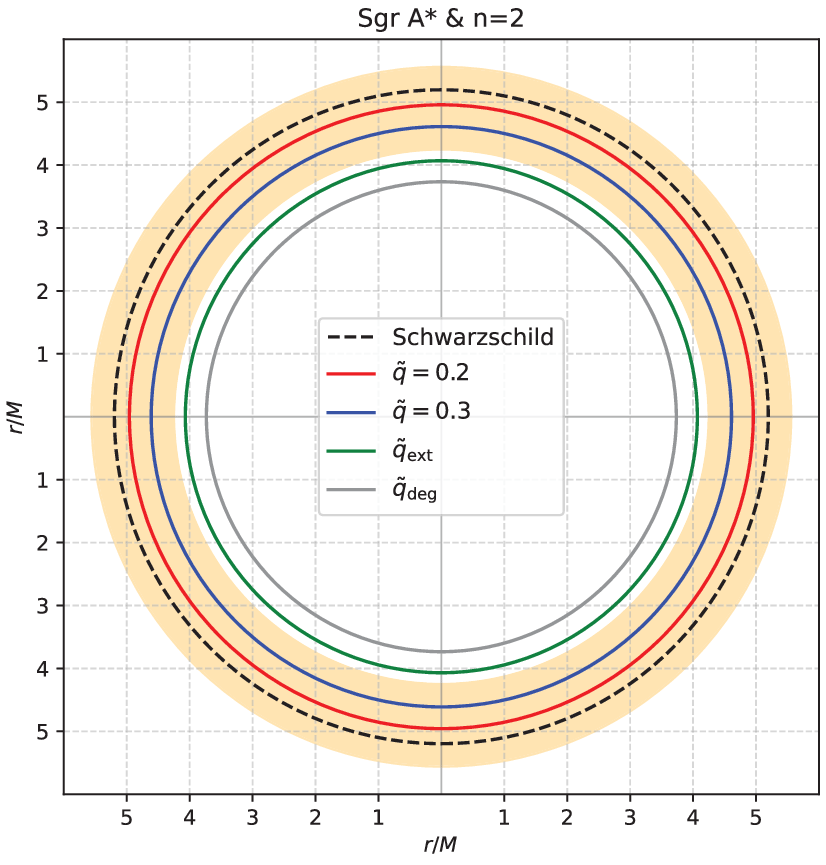}
		\end{minipage}
		\begin{minipage}[b]{0.32\linewidth}
			\includegraphics[width=1\linewidth]{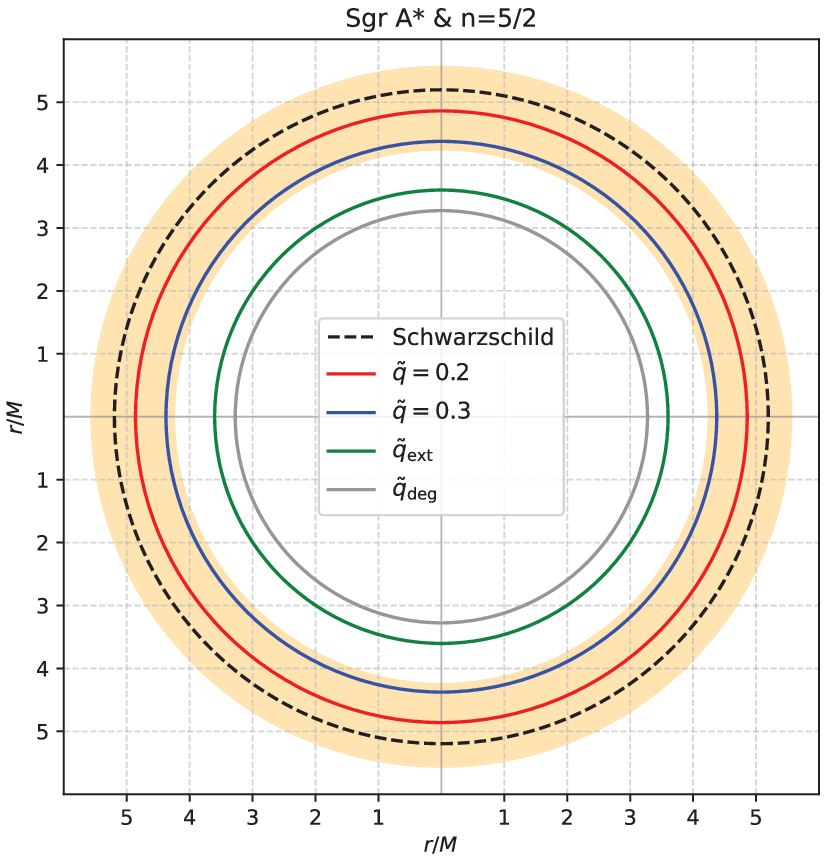}
		\end{minipage} 
		\begin{minipage}[b]{0.32\linewidth}
			\includegraphics[width=1\linewidth]{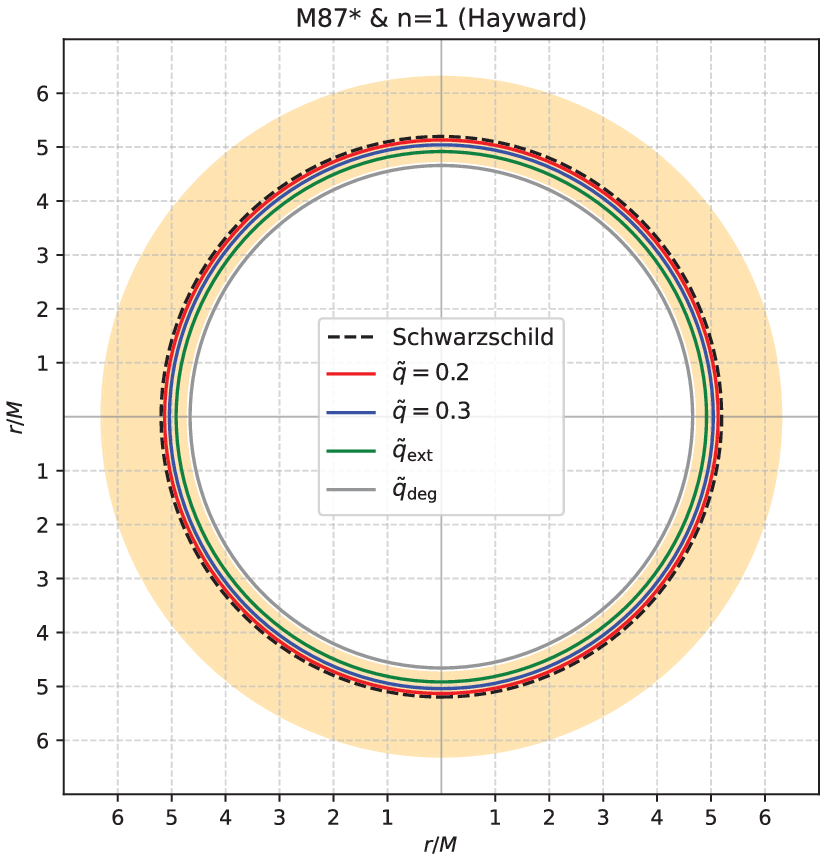}
		\end{minipage}  
		\begin{minipage}[b]{0.32\linewidth}
			\includegraphics[width=1\linewidth]{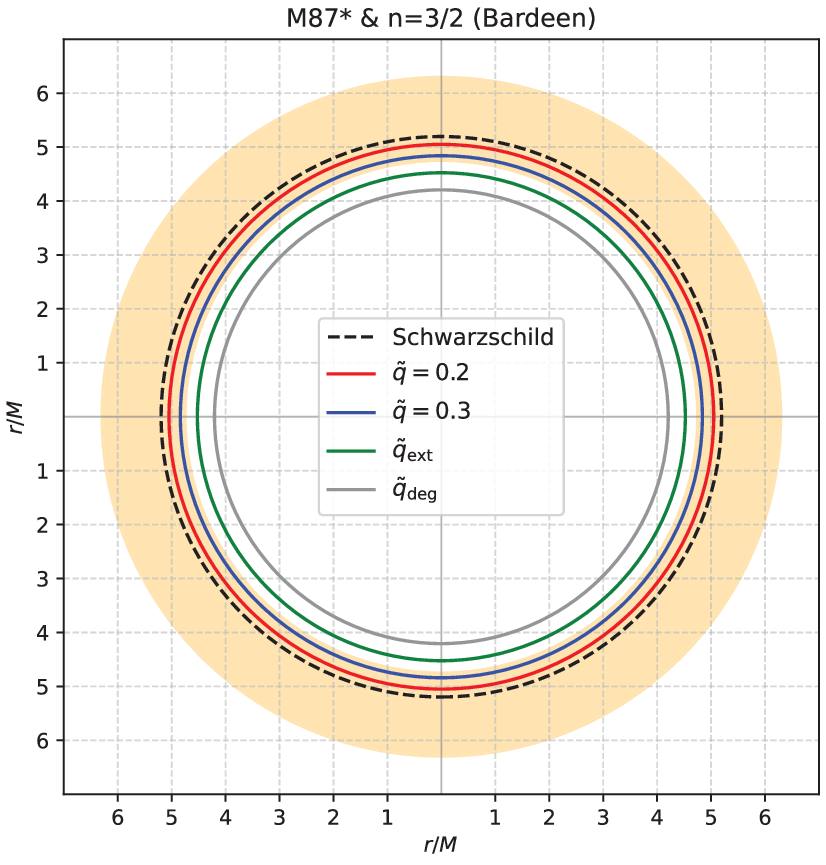}
		\end{minipage} \\
		\begin{minipage}[b]{0.32\linewidth}
			\includegraphics[width=1\linewidth]{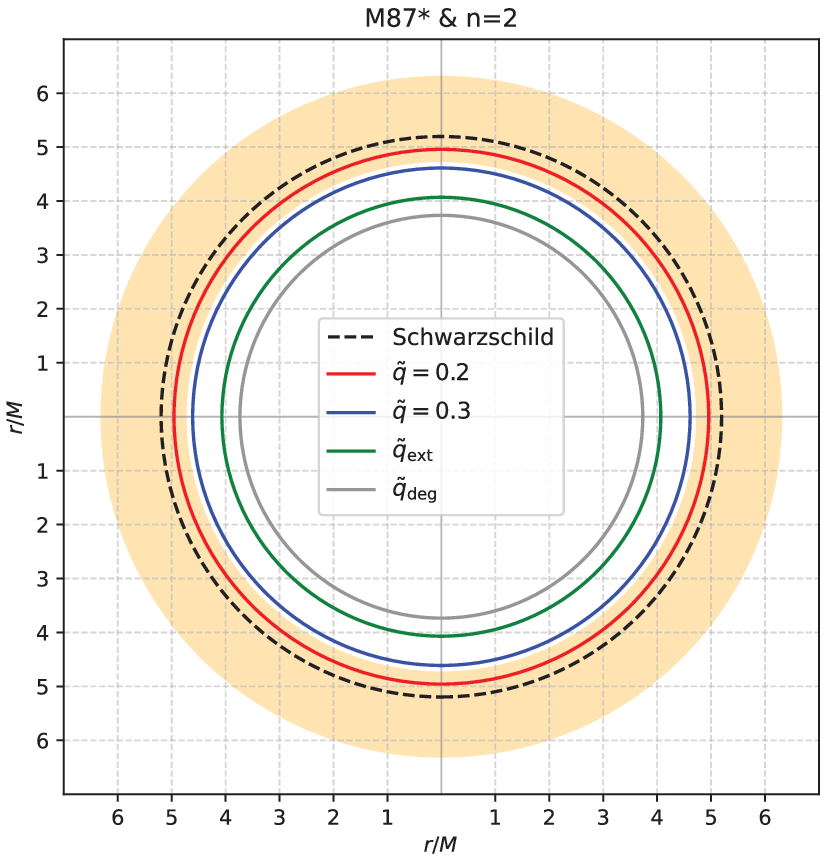}
		\end{minipage}
		\begin{minipage}[b]{0.32\linewidth}
			\includegraphics[width=1\linewidth]{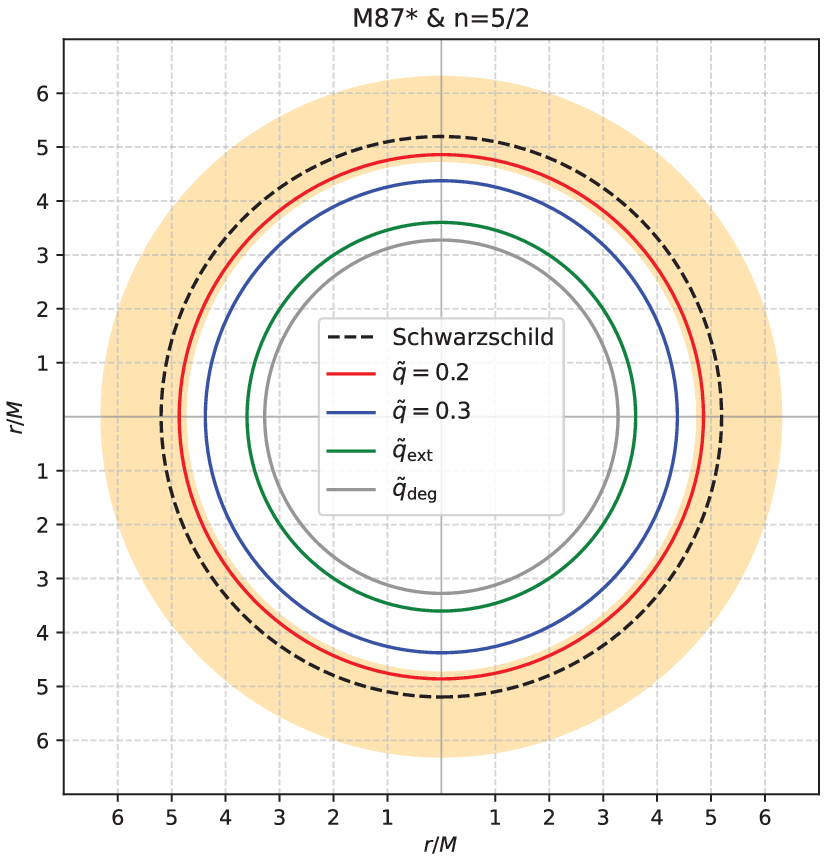}
		\end{minipage} 		\hfill
		\caption{Shadow of the regular $4$-dimensional metric \eqref{met4} compared to the shadow boundaries of $Sgr A^*$ and $M87^*$ for various polytropic indices. The dashed line corresponds to the Schwarzschild BH.}
		\label{figsha} 
	\end{figure}

{A photon emitted from a source toward a spherically symmetric BH or compact object can have two fates depending on its energy and angular momentum. The photon may be captured by the BH's or compact object's gravitational field or it can escape and go to infinity. The shadow corresponds to the region where the light rays are captured by the BH or compact object. The boundary of the shadow consists of light rays that neither escape to infinity nor fall into the BH. Instead, they approach a circular orbit, i.e., the photon sphere. It is the locus of the trapped photons and is defined by the roots of $\dot r=\ddot r=0$, as discussed in \ref{SSPS}. 

	To determine the shadow's size \cite{Synge:1966okc}, we assume a static observer at a given radius sends light rays into the past. Then, we calculate the angle between  the light rays' trajectories that asymptotically approach the photon sphere and the radial direction. A straightforward calculation yields \cite{Perlick:2021aok}
	\begin{align}
		\sin \alpha_{\text{Sh}} = \frac{r_{\text{PS}}}{r_\text{o}} \sqrt{\frac{A_d(r_\text{o})}{A_d(r_{\text{PS}})}},
	\end{align}
where $r_\text{o}$ corresponds to the observer's position. For an observer taken asymptotically far from the compact object, the angular size itself vanishes as $1/r_\text{o}$. The finite
quantity characterizing the shadow on the observer’s image plane is the critical impact
parameter,\footnote{It should be mentioned that the shadow of the $d$-dimensional BH is defined on the 2-dimensional surface corresponding to the observer. Accordingly, the shadow of the 5-dimensional rotating Myers–Perry BH has also been studied on a 2-dimensional surface, as discussed in \cite{Papnoi:2014aaa}.}
\begin{align}
	\label{sinalphaend}
	\tilde r_\text{o} \sin \alpha_{\text{Sh}}\big|_{\tilde r_\text{o}\to \infty} =  \frac{\tilde r_{\text{PS}}}{\sqrt {A_d(\tilde r_{\text{PS}})}} = \frac{\left(\frac{2}{d-1}\right)^{\frac{2n (d-1)}{3(d-2)(d-3)}}}{ \left(1 - b^{-2} \tilde r^2_{\text{PS}}\right)^{\frac{2n (d-1)}{(d-2)(d-3)}+\frac{1}{d-3}}} \left(1-\frac{ 1}{\tilde r_{\text{PS}}^{d-3} } \left(1+\frac{\tilde q^{d-2}}{\tilde r_{\text{PS}}^{\frac{3 (d-2)}{2 n}}}  \right)^{\frac{-2 n}{3}\frac{(d-1) }{ (d-2)}}\right)^{-1/2},
\end{align}
where $ b =  L/ E$ is the impact parameter and Eqs. \eqref{ad1} and \eqref{rps} are used.
 For the case of $\tilde q =0$, i.e., ST BH, using \eqref{rpsst} and \eqref{sinalphaend} yields the radius of the shadow
 \begin{align}
 	\tilde r_{\text{Sh,\,ST}}=\left(\frac{d-1}{2}\right)^{\frac{1}{d-3}} \sqrt{\frac{d-1}{d-3}}.
 \end{align}

Note that according  to \eqref{maxqph}, the shadow  exists only for  $0\leq\tilde q\leq\tilde q_{\text{deg}}$. At $\tilde q=\tilde q_{\text{deg}}$, the stable and unstable photon spheres merge into a single degenerate photon sphere, and for  $\tilde q >\tilde q_{\text{deg}}$,  the photon sphere and thus, the shadow disappears\footnote{Alternatively, consider a photon arriving from infinity, which  will only be deflected if $\tilde E^2<\tilde V_{\text{eff}}|_{\tilde r_{PS}}$. Otherwise, it would be captured by the BH or compact object. For a photon to be deflected rather than captured, its impact parameter must satisfy $ b > \tilde L/(\tilde V_{\text{eff}}|_{\tilde r_{PS}})^{1/2}$. For $\tilde q >\tilde q_{\text{deg}}$, there are no extrema in $\tilde V_{\text{eff}}$ and hence no $\tilde r_{\text{PS}}$.}.

The size of the shadow for different values of $\tilde q$, up to the maximum  $\tilde q_{\text{deg}}$ that shadows can exist, is calculated numerically  and listed in table \ref{tabsh}. As $q$, the dimension $d$, or the polytropic index $n$ increases, the shadow's size decreases. Therefore, for a given $d$ and $n$, $\tilde q=\tilde q_{\text{deg}}$ gives the minimum shadow radius and for a Schwarzschild BH in $4$ dimensions, the BH shadow would be maximally sized.

To compare the $4$-dimensional results with the observations, the shadow radii of  $M87^*$ and $\text{Sgr} A^*$ are used here. These are given by $r_{\text{Sh}}/M \approx 5.5 \pm 0.8$ and $4.21\lesssim r_{\text{Sh}}/M \lesssim 5.56$, respectively \cite{Bambi:2019tjh, Vagnozzi:2022moj}.  Note that the results in table \ref{tabsh}  should be written in terms of the ADM mass using Eq. \eqref{mfirst}. In $4$ dimensions, this means multiplying the values in table \ref{tabsh} by two. Figure \ref{figsha} plots the shadow radius of the regular class of BH spacetime \eqref{ad1} in $4$ dimensions (i.e., Eq. \eqref{met4})} compared to that of $\text{Sgr} A^*$ and $M87^*$. Solutions with lower $\tilde q$ and $n$ are more consistent with observational data than those with higher $\tilde q$ and $n$.
\begin{table}[h]
						\centering
		\caption{Numerical values for the size of the shadow of the BH using \eqref{sinalphaend}. There are no shadows for $\tilde q > \tilde q_{\text{deg}}$. To observe variations in shadow sizes due to the convergence of radii at lower values of $\tilde q$, appropriate values of $\tilde q$ are selected.}\vspace{0.1in}
		\centering
		\begin{tabular}{c|c@{\hskip 0.18in}c@{\hskip 0.18in}c@{\hskip 0.18in}c|c@{\hskip 0.18in}c@{\hskip 0.18in}c@{\hskip 0.18in}c}
			\toprule
			\multicolumn{1}{c}{} & \multicolumn{4}{c}{$d=4$} & \multicolumn{4}{c}{$d=5$} \\
			\midrule
			$n$ & $\tilde q=0.2$ & $\tilde q=0.3$ & $\tilde q=\tilde q_{\text{ext}}$ & $\tilde q= \tilde q_{\text{deg}}$ & $\tilde q=0.3$ & $\tilde q=0.5$  & $\tilde q=\tilde q_{\text{ext}}$ &$\tilde q= \tilde q_{\text{deg}}$\\
			\midrule
			$n=1$ & $\tilde{r} = 2.5659$ & $\tilde{r} = 2.5209$ & $\tilde{r} = 2.4585$ & $\tilde{r} =2.3292$ & $\tilde{r} = 1.9949$ & $\tilde{r} = 1.9750$ & $\tilde{r} =1.9455$ & $\tilde{r} = 1.8657$ \\
			$n=\frac{3}{2}$ & $\tilde{r} = 2.5250$ & $\tilde{r} = 2.4199$ & $\tilde{r} = 2.2620$ & $\tilde{r} = 2.1033$ & $\tilde{r} =1.9870$ & $\tilde{r} = 1.9351$ & $\tilde{r} = 1.8512$ & $\tilde{r} =1.7437$ \\
			$n=2$ & $\tilde{r} = 2.4787$ & $\tilde{r} = 2.3057$ & $\tilde{r} = 2.0344$ & $\tilde{r} = 1.8673$ & $\tilde{r} = 1.9775$ & $\tilde{r} = 1.8864$ & $\tilde{r} = 1.7307$ & $\tilde{r} =1.6088$ \\
			$n=\frac{5}{2}$ & $\tilde{r} =2.4302$ & $\tilde{r} =2.1881$ & $\tilde{r} = 1.8014$ & $\tilde{r} = 1.6379$ & $\tilde{r} =1.9672$ & $\tilde{r} = 1.8336$ & $\tilde{r} = 1.5976$ & $\tilde{r} = 1.4703$ \\
		\end{tabular}
		\centering
		\begin{tabular}{c|c@{\hskip 0.18in}c@{\hskip 0.18in}c@{\hskip 0.18in}c|c@{\hskip 0.18in}c@{\hskip 0.18in}c@{\hskip 0.18in}c}
			\toprule
			\multicolumn{1}{c}{} & \multicolumn{4}{c}{$d=6$} & \multicolumn{4}{c}{$d=7$} \\
			\midrule
			$n$ & $\tilde q=0.4$ & $\tilde q=0.6$ & $\tilde q=\tilde q_{\text{ext}}$ & $\tilde q=\tilde q_{\text{deg}}$ & $\tilde q=0.5$ & $\tilde q=0.66$ & $\tilde q=\tilde q_{\text{ext}}$& $\tilde q=\tilde q_{\text{deg}}$ \\
			\midrule
			$n=1$ & $\tilde r = 1.7501$ & $\tilde r =  1.7416$ & $\tilde r = 1.7234$ & $\tilde r = 1.6654$ & $\tilde r = 1.6106$ & $\tilde r =1.6065$ & $\tilde r =1.5943$ & $\tilde r =1.5488$ \\
			$n=\frac{3}{2}$ & $\tilde r = 1.7466$ & $\tilde r = 1.7223$ & $\tilde r = 1.6662$ & $\tilde r = 1.5832$ & $\tilde r = 1.6080$ & $\tilde r = 1.5957$ & $\tilde r = 1.5553$ & $\tilde r = 1.4874$ \\
			$n=2$ & $\tilde r = 1.7420$ & $\tilde r = 1.6975$ & $\tilde r =  1.5880$ & $\tilde r = 1.4896$ & $\tilde r = 1.6046$ & $\tilde r =1.5812$ & $\tilde r = 1.4993$ & $\tilde r = 1.4160$ \\
			$n=\frac{5}{2}$ & $\tilde r =1.7370$ & $\tilde r = 1.6696$ & $\tilde r = 1.4973$ & $\tilde r = 1.3904$ & $\tilde r =1.6006$ & $\tilde r = 1.5645$ & $\tilde r = 1.4320$ & $\tilde r = 1.3390$ \\
		\end{tabular}
		\centering
		\begin{tabular}{c|c@{\hskip 0.18in}c@{\hskip 0.18in}c@{\hskip 0.18in}c|c@{\hskip 0.18in}c@{\hskip 0.18in}c@{\hskip 0.18in}c}
			\toprule
			\multicolumn{1}{c}{} & \multicolumn{4}{c}{$d=8$} & \multicolumn{4}{c}{$d=9$} \\
			\midrule
			$n$ & $\tilde{q}=0.5$ & $\tilde{q}=0.7$ & $\tilde{q}=\tilde{q}_{\text{ext}}$ & $\tilde{q}=\tilde{q}_{\text{deg}}$ & $\tilde{q}=0.6$ & $\tilde{q}=0.73$ & $\tilde{q}=\tilde{q}_{\text{ext}}$ & $\tilde{q}=\tilde{q}_{\text{deg}}$ \\
			\midrule
			$n=1$ & $\tilde{r}=1.5197$ & $\tilde{r}=1.5171$ & $\tilde{r}=1.5084$ & $\tilde{r}=1.4709$ & $\tilde{r}=1.4544$ & $\tilde{r}=1.4530$ & $\tilde{r}=1.4465$ & $\tilde{r}=1.4147$ \\
			$n=\frac{3}{2}$ & $\tilde{r}=1.5189$ & $\tilde{r}=1.5105$ & $\tilde{r}=1.4799$ & $\tilde{r}=1.4223$ & $\tilde{r}=1.4533$ & $\tilde{r}=1.4486$ & $\tilde{r}=1.4247$ & $\tilde{r}=1.3745$ \\
			$n=2$ & $\tilde{r}=1.5177$ & $\tilde{r}=1.5012$ & $\tilde{r}=1.4372$ & $\tilde{r}=1.3648$ & $\tilde{r}=1.4517$ & $\tilde{r}=1.4422$ & $\tilde{r}=1.3908$ & $\tilde{r}=1.3266$ \\
			$n=\frac{5}{2}$ & $\tilde{r}=1.5163$ & $\tilde{r}=1.4903$ & $\tilde{r}=1.3846$ & $\tilde{r}=1.3021$ & $\tilde{r}=1.4499$ & $\tilde{r}=1.4346$ & $\tilde{r}=1.3481$ & $\tilde{r}=1.2737$ \\
		\end{tabular}
		\centering
		\begin{tabular}{c|c@{\hskip 0.18in}c@{\hskip 0.18in}c@{\hskip 0.18in}c|c@{\hskip 0.18in}c@{\hskip 0.18in}c@{\hskip 0.18in}c}
			\toprule
			\multicolumn{1}{c}{} & \multicolumn{4}{c}{$d=10$} & \multicolumn{4}{c}{$d=11$} \\
			\midrule
			$n$ & $\tilde q=0.65$ & $\tilde q=0.75$ & $\tilde q=\tilde q_{\text{ext}}$ & $\tilde q=\tilde q_{\text{deg}}$ & $\tilde q=0.77$ & $\tilde q=0.85$ & $\tilde q=\tilde q_{\text{ext}}$  & $\tilde q=\tilde q_{\text{deg}}$ \\
			\midrule
			$n=1$ & $\tilde r = 1.4053$ & $\tilde r = 1.4045$ & $\tilde r =1.3995$ & $\tilde r = 1.3720$ & $\tilde r =1.3664$ & $\tilde r = 1.3652$ & $\tilde r =1.3624$ & $\tilde r =  1.3381$ \\
			$n=\frac{3}{2}$ & $\tilde r =1.4044$ & $\tilde r = 1.4015$ & $\tilde r =1.3822$ & $\tilde r =1.3378$ & $\tilde r =1.3642$ & $\tilde r = 1.3596$ & $\tilde r =1.3483$ & $\tilde r = 1.3085$ \\
			$n=2$ & $\tilde r = 1.4030$ & $\tilde r = 1.3971$ & $\tilde r = 1.3545$ & $\tilde r = 1.2968$ & $\tilde r = 1.3608$ & $\tilde r = 1.3510$ & $\tilde r = 1.3252$ & $\tilde r = 1.2726$ \\
			$n=\frac{5}{2}$ & $\tilde r = 1.4014$ & $\tilde r =1.3917$ & $\tilde r =1.3189$ & $\tilde r =1.2511$ & $\tilde r = 1.3567$ & $\tilde r = 1.3403$ & $\tilde r = 1.2949$ & $\tilde r = 1.2325$ \\
			\bottomrule
		\end{tabular}
								\label{tabsh}
		\end{table}

As the number of dimensions increases, magnetic charge-dependent shadows also tend to converge. The same is true for the position of horizons, BHPSs, and COPSs. As the number of dimensions increases, the differences between the radius of the photon sphere and the radius of the shadow decrease. As these differences become negligible in higher dimensions, studying horizons and photon spheres in detail becomes less important. Consequently, complex mathematical calculations for higher dimensions seem unnecessary.

We adopt the standard dimensional-reduction viewpoint for connecting the $d$-dimensional framework to observable 4-dimensional physics: the four-dimensional spacetime corresponds to a hypersurface of constant extra coordinates in the full $d$-dimensional manifold, with all fields independent of the extra dimensions (cylindrical symmetry). Under this minimal embedding assumption, the induced metric on the 4-dimensional slice is precisely the $d=4$ limit of Eq. \eqref{ad1}. Consequently, photon trajectories confined to this hypersurface obey the standard four-dimensional null geodesic equations, and all observational comparisons (e.g., shadow sizes, photon-sphere radii) are performed using the $d=4$ sector of the theory. 
\section{BH Thermodynamics}
\label{sec5}
To study the thermodynamics of the static spherically symmetric BH spacetime \eqref{genmet}, it is necessary to find its surface
gravity. Since the spacetime is static, the event horizon is also a Killing horizon, which is generated by the Killing vector $\xi^\mu$. Using the metric component \eqref{ad1}, the surface gravity is defined as $\kappa =\sqrt{-\xi_{\mu;\nu}  \xi^{\mu;\nu}/2}\big|_{r=r_\text{h}}$, which is  evaluated at the event horizon, $r=r_\text{h}$ and can be simplified to
\begin{align}
	\label{SG1}
\kappa &= \frac{1}{2} \frac{d A_d(r)}{dr}\big|_{r=r_\text{h}}=\left[\frac{1}{2 r}\frac{(d-3)-2 m^{\left(\frac{d-2}{2}\right)  \left(\frac{3}{n}-2\right)}  q^{d-2} r^{-\frac{3 (d-2)}{2 n}}}{1+ m^{\left(\frac{d-2}{2}\right)\left(\frac{3}{n}-2\right)} q^{d-2} r^{-\frac{3 (d-2)}{2 n}}}\right]_{r=r_\text{h}}.
\end{align}
This can be expressed in terms of the dimensionless variables obtained by normalizing $m$, i.e., the tilde variables in the previous sections, as
\begin{align}
	\label{SGnor}
	\bar\kappa =\left[\frac{1}{2 \tilde r}\frac{(d-3)-2 \tilde  q^{d-2} \tilde r^{-\frac{3 (d-2)}{2 n}}}{1+  \tilde q^{d-2} \tilde r^{-\frac{3 (d-2)}{2 n}}}\right]_{\tilde r=\tilde r_\text{h}}.
\end{align}
where $\bar \kappa=m \kappa $. For small values of $\tilde q$, it is clear that the surface gravity tends to that of the ST spacetime, which is $(d-3)/2 \tilde r_h$. As expected, it can also be seen that in the extremal case, $\tilde q = \tilde q_{\text{ext}}$, the surface gravity becomes zero.

\begin{figure}[t]
	\begin{minipage}[b]{0.5\linewidth}
		\includegraphics[width=1\linewidth]{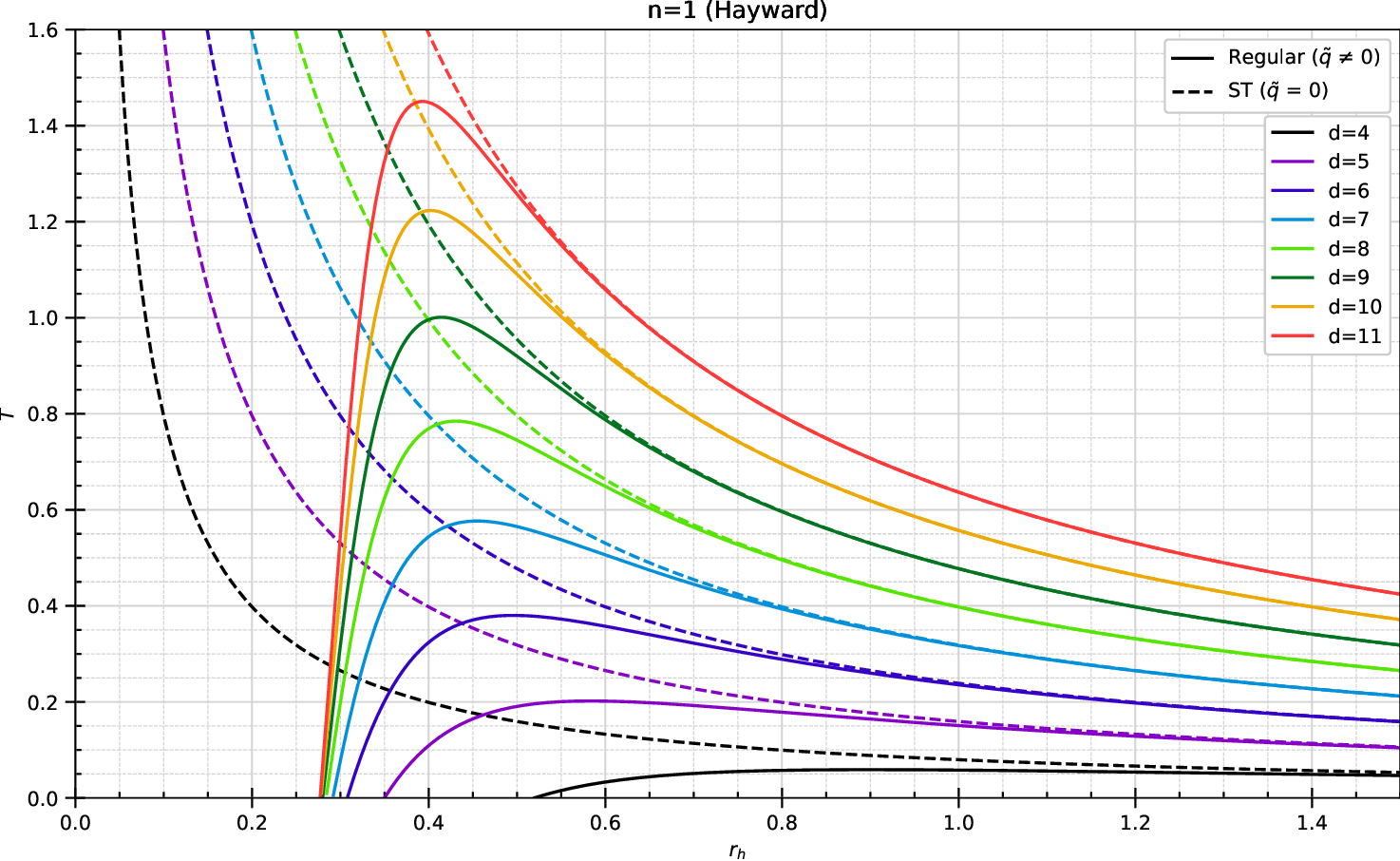}
	\end{minipage}  \vspace{0.5cm}
	\begin{minipage}[b]{0.5\linewidth}
		\includegraphics[width=1\linewidth]{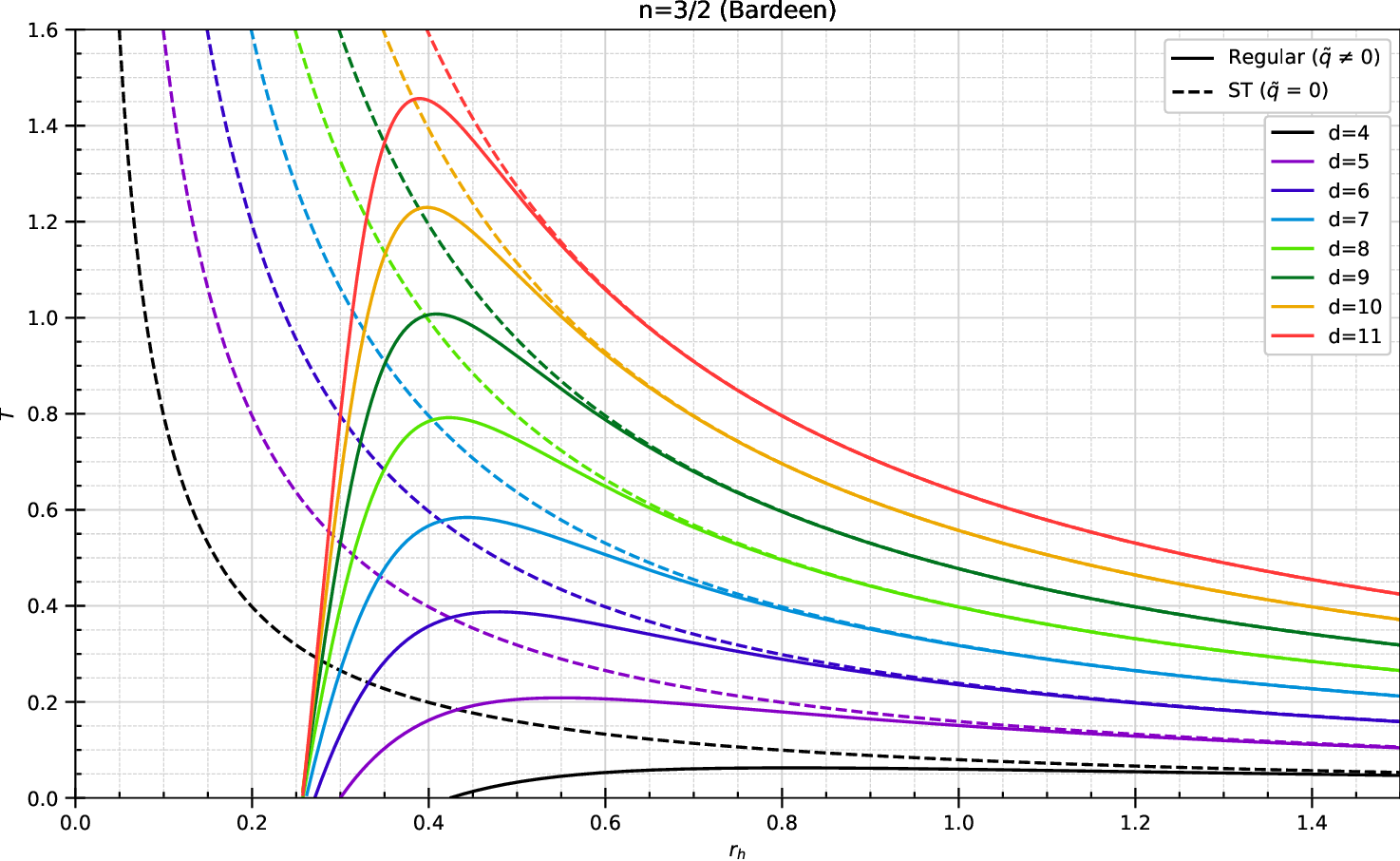}
	\end{minipage} 
	\begin{minipage}[b]{0.5\linewidth}
		\includegraphics[width=1\linewidth]{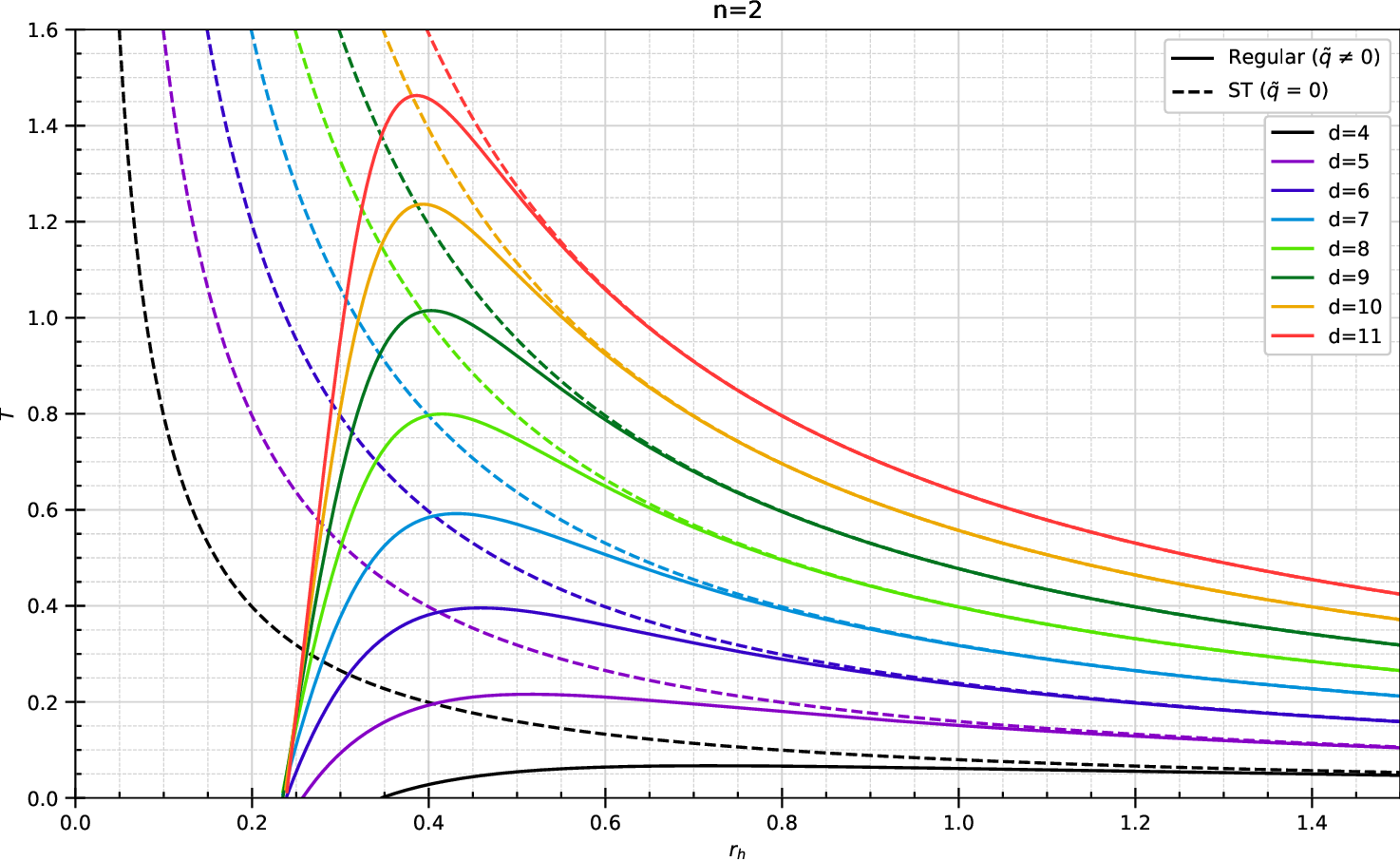}
	\end{minipage}
	\hfill
	\begin{minipage}[b]{0.5\linewidth}
		\includegraphics[width=1\linewidth]{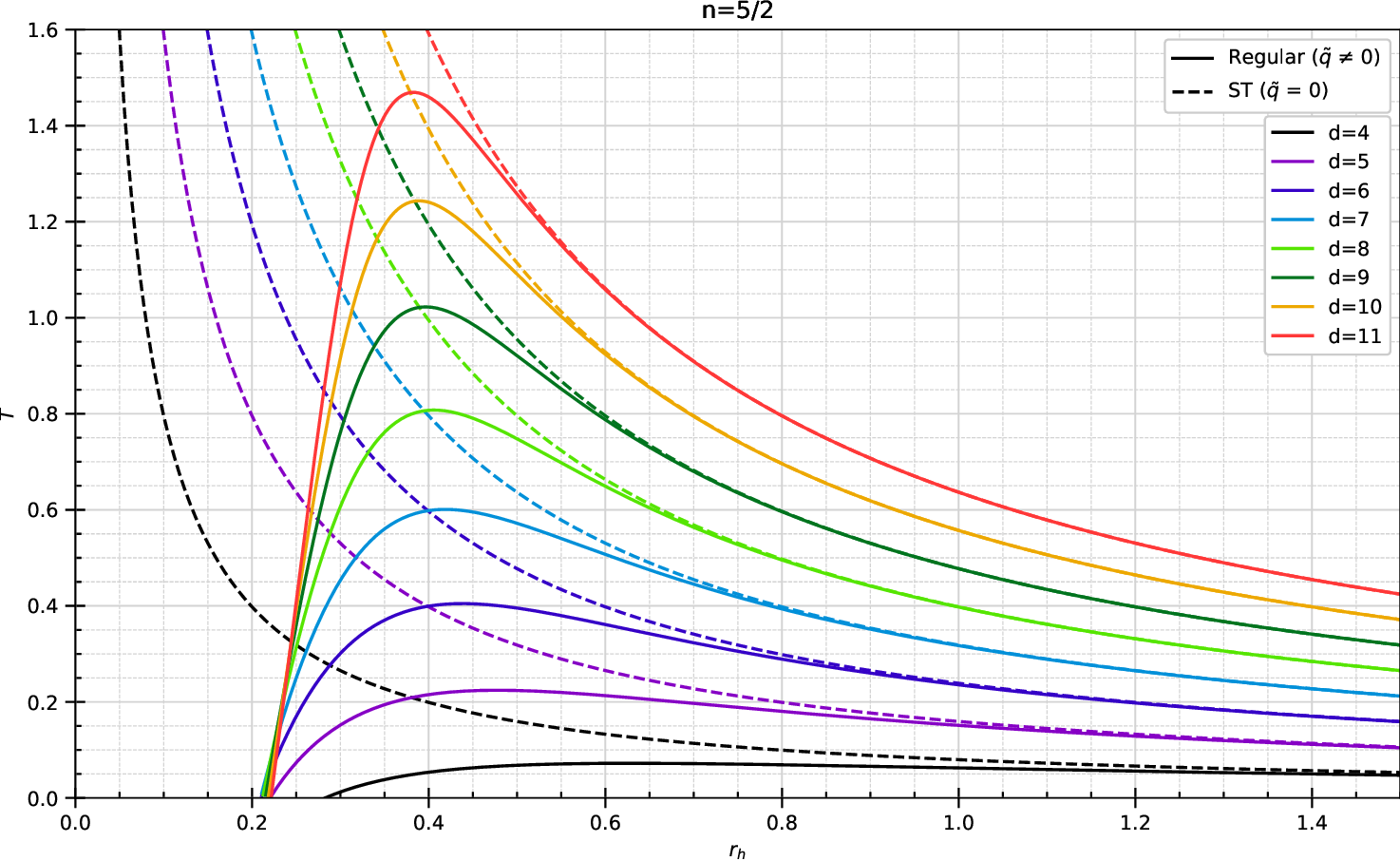}
	\end{minipage} 
\caption{The numerical result of the horizon temperature as a function of the horizon radius for different polytropic indices values and dimensions with $q = 0.3$. 
The solid and dashed lines represent the regular BH \eqref{ad1} and the ST BH, respectively. The horizon temperature vanishes at the extremal radius, increases to a maximum, and then decreases toward the ST behavior at larger horizon	radius.}
\label{figtemp} 
\end{figure}

We can plot the Hawking temperature $ T=  \kappa/2\pi$ as a function of the horizon radius by numerically computing the roots of $A_d(r)=0$ and then substituting them into the surface gravity \eqref{SG1}. See Fig. \ref{figtemp}. The temperature starts from zero, at the extremal horizon, increases up to a maximum, and then decreases toward the ST behavior at large horizon radius corresponding to $m \gg q$.\\
The vanishing of the Hawking temperature at the extremal radius marks the boundary of the BH branch. For fixed $d$ and $n$, when the magnetic charge exceeds the extremal value, the equation $A_d(r)=0$ has no real positive root and the spacetime describes a horizonless compact object. In this regime, the Hawking temperature and heat capacity are no longer BH thermodynamic quantities, since there is no event horizon. The energy conditions of this horizonless branch are governed by the same effective matter source discussed in Sec. \ref{SSecEC}. Therefore, the WEC and NEC remain satisfied everywhere. The SEC is necessarily violated in the central region, reflecting the de Sitter core required to avoid the curvature singularity. The DEC is satisfied everywhere when condition \eqref{DECnd} holds; otherwise its tangential component is violated outside the radius given in Eq. \eqref{vioDEC}. Thus, the horizonless or “planet-like” branch should not be interpreted as an ordinary matter planet, but rather as a compact object supported by the NLED source whose central region violates the SEC.
	
\subsection{Entropy}

The thermodynamic properties \cite{bch, Hawkingtemp, Beke} of a BH, which is described by mass and charge parameters, are defined by 
 \begin{align}
	\label{thermo}
T=\frac{\partial M}  {\partial S}  =\frac{\kappa}{2\pi }, \hspace{2cm}  \Phi  =\frac{\partial M}  {\partial q} ,\hspace{2cm} S= \int \frac{1}{T} \left(d M - \Phi d q\right)
\end{align}
where $T$ is the temperature, $\Phi $ is the electric or magnetic potential, $S$ is the Bekenstein-Hawking entropy, and $M$ is the ADM mass introduced in \eqref{mfirst}. The first law is given by  \cite{bch, Smarr}
\begin{align}
\delta M =  T \delta S +  \Phi  \delta q.                           
\end{align}
Here, $\delta $ denotes variations of the BH's parameters.
With constant charge, the entropy can then be calculated as
\begin{align}
	\label{firstS}
S = \int_{r_{\text{ext}}}^{ r_{\text{h}}} \frac{1}{ T}\left(\frac{\partial M}{\partial  r_{\text{h}}}\right)_q d  r_{\text{h}}.
\end{align}

As shown in Fig. \ref{figtemp}, for $r_\text{h}<r_{\text{ext}},$ no regular BH horizon exists and the formal continuation of the surface gravity is not interpreted as a physical Hawking temperature. The extremal horizon radius $r_{\text{ext}} $ is defined as the radius at which the temperature vanishes. This radius can be obtained by Eqs. \eqref{qext} and  \eqref{rext}  as
\begin{align}
	\label{rh0}
	r_{\text{ext}}
	&=q_{\text{ext}} \left(\frac{2}{d-3}\right)^{\frac{1}{d-2}}  \left(\frac{2}{d-1}\right)^{\frac{(d-1) (2 n-3)}{3 (d-3) (d-2)}}.
\end{align}
 Now, we consider the special case of a $d$-dimensional Bardeen BH ($n=3/2$), for which the entropy can be derived  analytically.
Using \eqref{ad1}
\begin{align}
	M_{\text{Bardeen},\,d}=\frac{(d-2) \Omega_{d-2}  }{16 \pi  r_{\text{h}}^2}\left(q^{d-2}+r_{\text{h}}^{d-2}\right)^{\frac{d-1}{d-2}}.
\end{align}
whose derivative with respect to the horizon radius is given by
\begin{align}
	\label{dmdrhB}
	\frac{\partial M_{\text{Bardeen},\,d}}{\partial  r_{\text{h}}}=\frac{(d-2) \Omega_{d-2}  }{16 \pi  r_{\text{h}}^3}\left(q^{d-2}+r_{\text{h}}^{d-2}\right)^{\frac{1}{d-2}} \left((d-3) r_{\text{h}}^{d-2}-2 q^{d-2}\right).
\end{align}
Therefore, using Eqs. \eqref{SG1} and \eqref{firstS},  the entropy of a $d$-dimensional Bardeen BH becomes
\begin{align}
S_{\text{Bardeen},\,d} &= \int_{r_{\text{ext}}}^{ r_{\text{h}}}\frac{(d-2) \Omega_{d-2}}{4 r_{\text{h}}^2}  \left(q^{d-2}+r_{\text{h}}^{d-2}\right)^{\frac{d-1}{d-2}} dr_{\text{h}},\\
&=-\frac{(d-2) \Omega_{d-2} }{4 r_{\text{h}} q^{d-2}} \left(q^{d-2}+r_{\text{h}}^{d-2}\right)^{\frac{2 d-3}{d-2}}\, {}_2F_1\left[1,2,\frac{d-3}{d-2},-\frac{r_{\text{h}}^{d-2}}{q^{d-2}} \right] \nonumber\\
\label{entdB}
&+\frac{(d-2)\Omega_{d-2} }{4} q^{d-2} \left(\frac{d-1}{d-3}\right)^{\frac{2 d-3}{d-2}} \left(\frac{d-3}{2}\right)^{\frac{1}{d-2}} \, {}_2F_1\left(1,2,\frac{d-3}{d-2},-\frac{2}{d-3}\right),
\end{align} 
where ${}_2 F_1$ is the hypergeometric function, and $r_{\text{ext}}$ is given by \eqref{rh0}. An analytical derivation of the Hayward BH ($n=1$) entropy 
is also possible in $4$ dimensions. A simple calculation using Eqs. \eqref{firstS} and \eqref{entdB} yields\footnote{The entropy of a $4$-dimensional Bardeen spacetime is reported in  \cite{ali, Man:2013hza, Singh:2019tgw, Sadeghi:2023aii, Sharif:2010pj}. However, the authors did not use the lower bound of the integral. Instead, they set the integration constant to zero (i.e., $q = 0$), which is undesirable. Therefore, our results differ from those in the aforementioned references by a constant.}
\begin{align}
	\label{ent4H}
	S_{\text{Hayward , $d=4$}}&=\pi  \left(2 q^2 \log \left(\frac{r_{\text{h}}^2-q^2}{2 q^2}\right)+\frac{q^4}{q^2-r_{\text{h}}^2}+r_{\text{h}}^2\right)-\frac{5 \pi  q^2}{2},\\
	\label{ent4B}
	S_{\text{Bardeen , $d=4$}}&=\frac{\pi}{ r_{\text{h}}}  \left( r_{\text{h}}^2-2  q^2\right) \sqrt{ q^2+ r_{\text{h}}^2}+3 \pi   q^2 \ln \left(\sqrt{  q^2+ r_{\text{h}}^2}+ r_{\text{h}}\right)-3 \pi   q^2 \ln \left( q(\sqrt{2}+\sqrt{3}) \right).
\end{align}
 For $q=0$ the result of \eqref{firstS} reduces to the entropy of the ST, which obeys the area law \cite{Singh:2017vfr}. 
 \begin{align}
 	S_{\text{ST}}= \frac{\Omega_{d-2}}{4}r_{\text{h}}^{d-2}.
 \end{align}

\begin{figure}[h]
	\begin{minipage}[b]{0.5\linewidth}
	\includegraphics[width=1\linewidth]{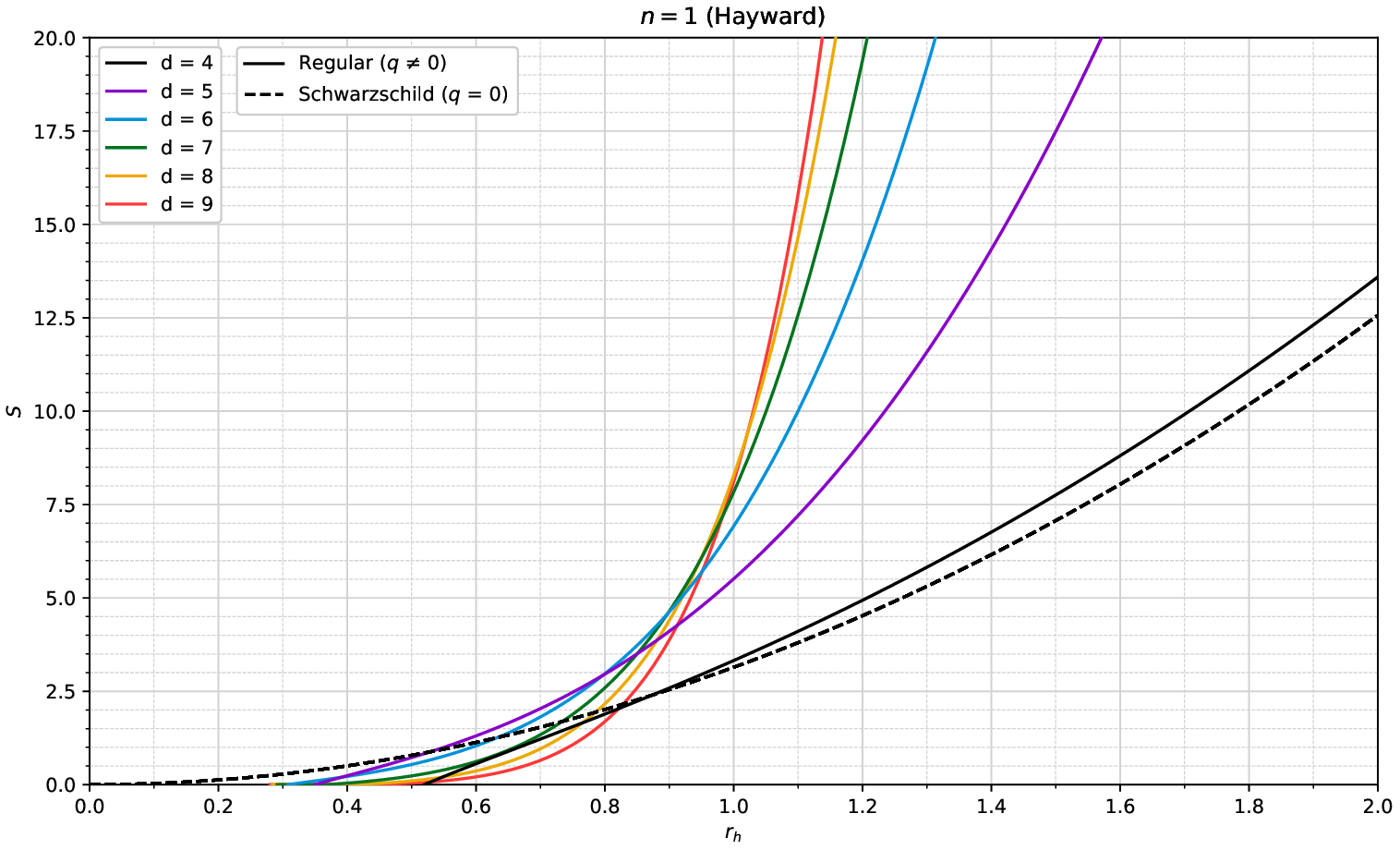}
	\end{minipage}  
	\begin{minipage}[b]{0.5\linewidth}
		\includegraphics[width=1\linewidth]{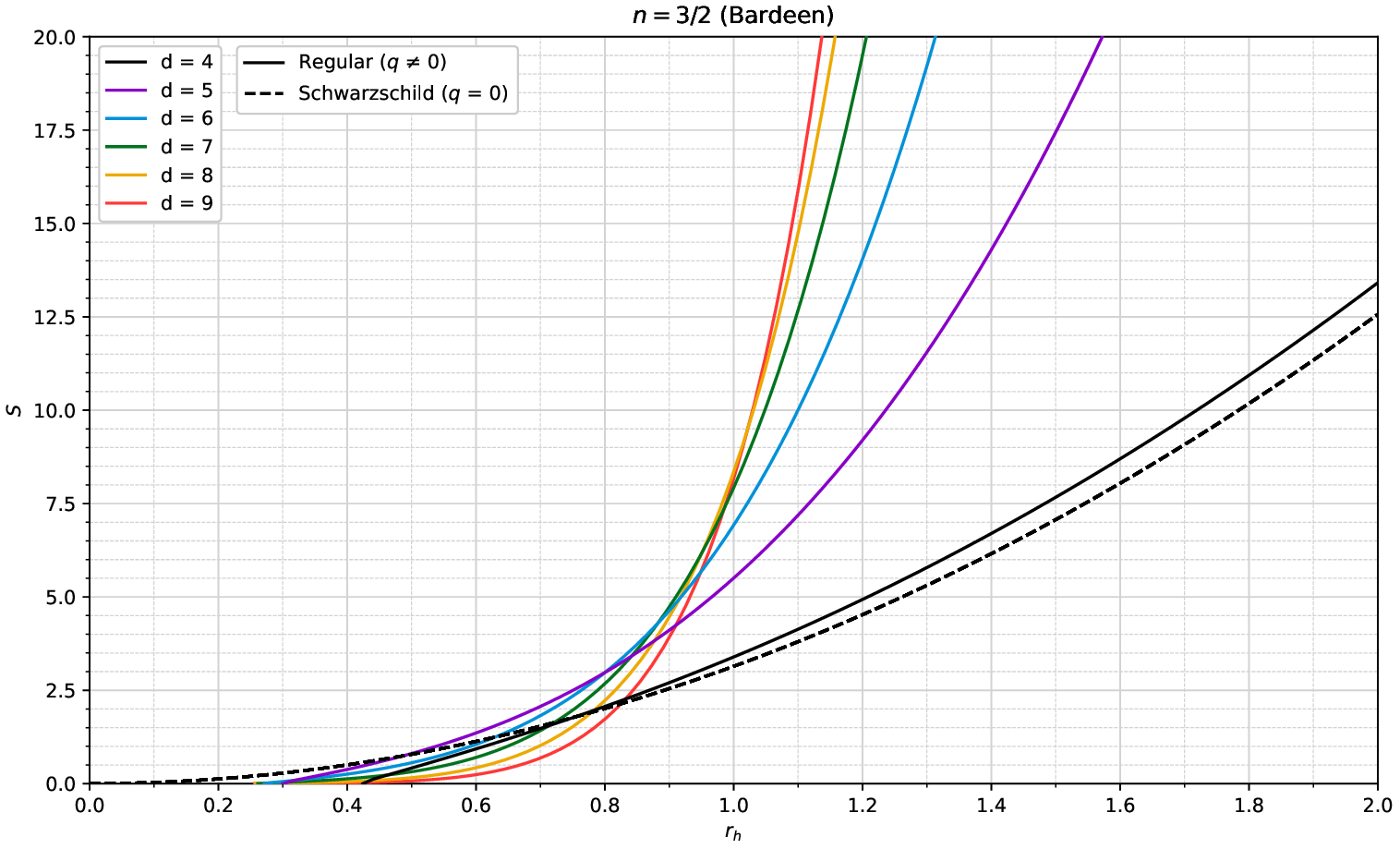}
	\end{minipage} \\
	\begin{minipage}[b]{0.5\linewidth}
		\includegraphics[width=1\linewidth]{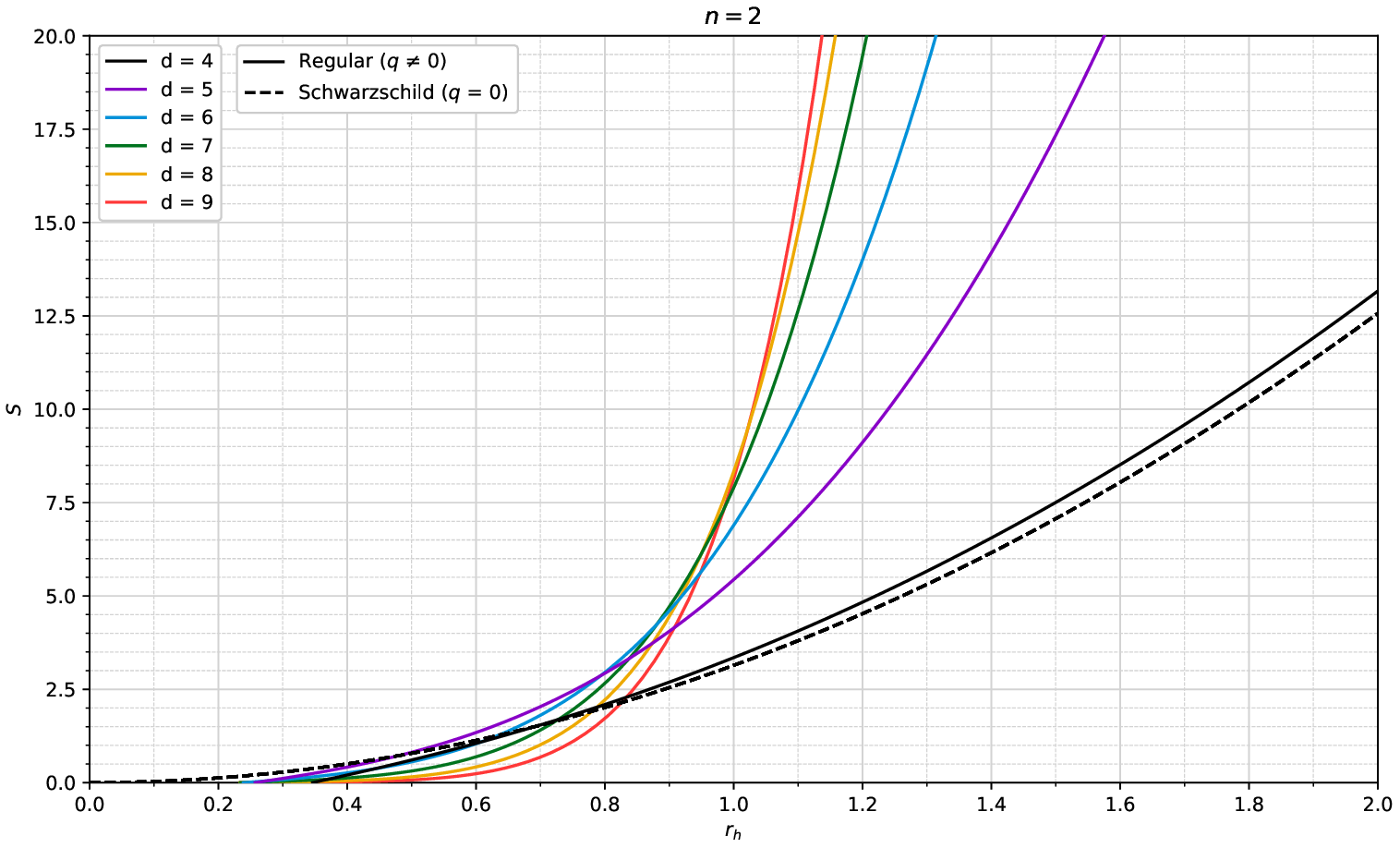}
	\end{minipage}
	\begin{minipage}[b]{0.5\linewidth}
		\includegraphics[width=1\linewidth]{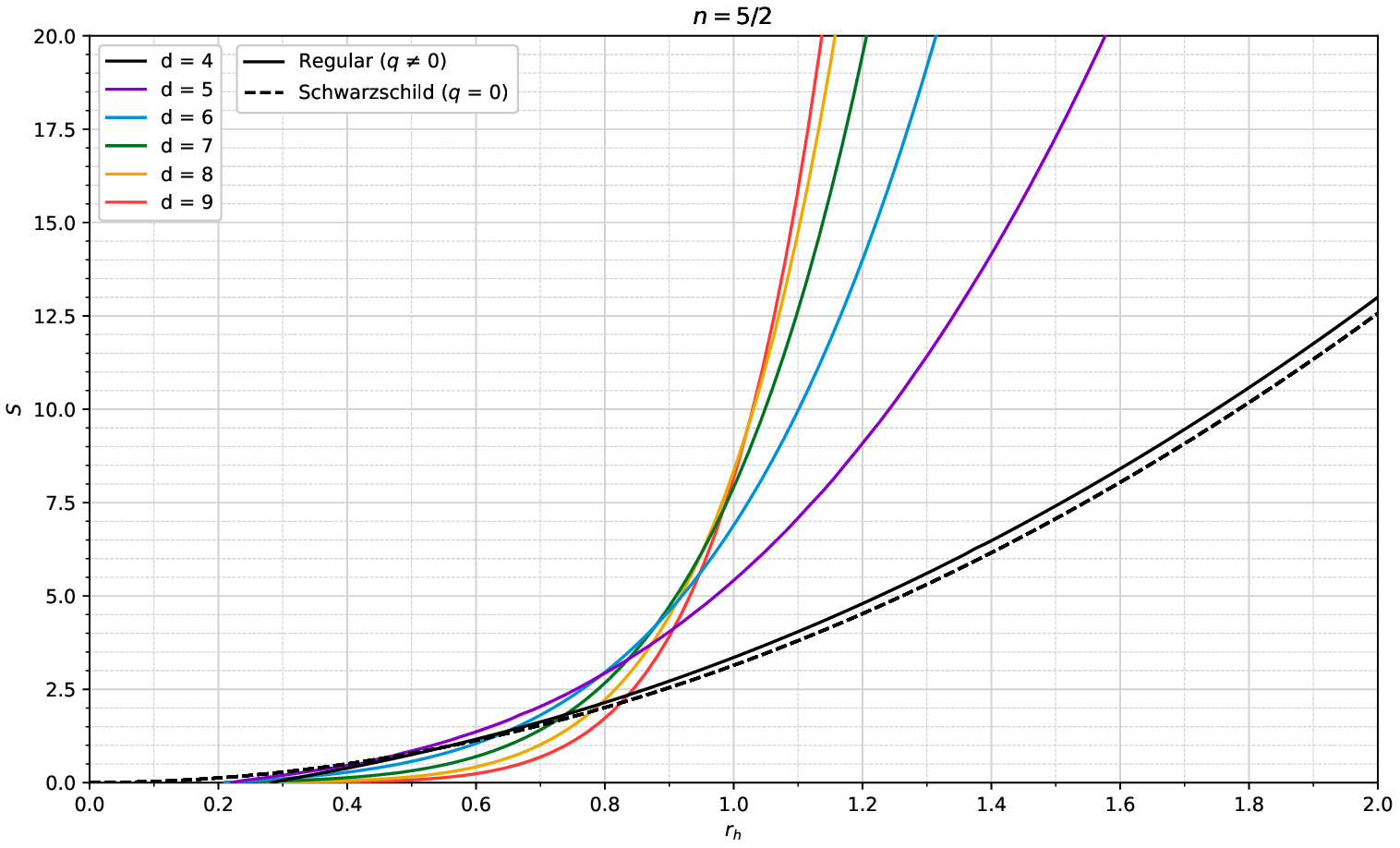}
	\end{minipage}
	\caption{The numerically calculated BH entropy in terms of the outer horizon radius for $q = 0.3$, as well as for different values of the polytropic index and dimensions. The dashed lines represent the Schwarzschild BH entropy in $4$ dimensions.}
			\label{figent} 
\end{figure}

To determine the entropy for any $n$, the horizon radius must be known. However, it is not always possible to solve the horizon equation exactly. Therefore, a numerical approach must be used.
Entropy \eqref{firstS} is plotted numerically with respect to the horizon radius in Fig. \ref{figent} \footnote{The Python code implementing the numerical methods used throughout this article is available from the authors upon request.}. As can be seen, the area law for entropy ($S = \pi r_h^2$)  does not hold in regular spacetimes such as \eqref{ent4H} and \eqref{ent4B} \cite{Waldent, ghosh1}.  Moreover, as the polytropic index $n$ increases, the deviation from the area law decreases. Conversely, as the dimension increases, the  deviation becomes significant. 

\subsection{Heat capacity}

The goal of this section is to study the local thermodynamic response of the regular BH \eqref{ad1} in the ensemble with fixed magnetic charge. The heat capacity $C_q$ provides a local thermodynamic criterion: $C_q>0$ corresponds to a locally stable thermal branch in this fixed-$q$ description, whereas $C_q<0$ signals local canonical thermodynamic instability. This criterion should not be confused with linear dynamical stability of metric and NLED perturbations.

\begin{figure}[t]
	\begin{minipage}[b]{0.5\linewidth}
		\includegraphics[width=1\linewidth]{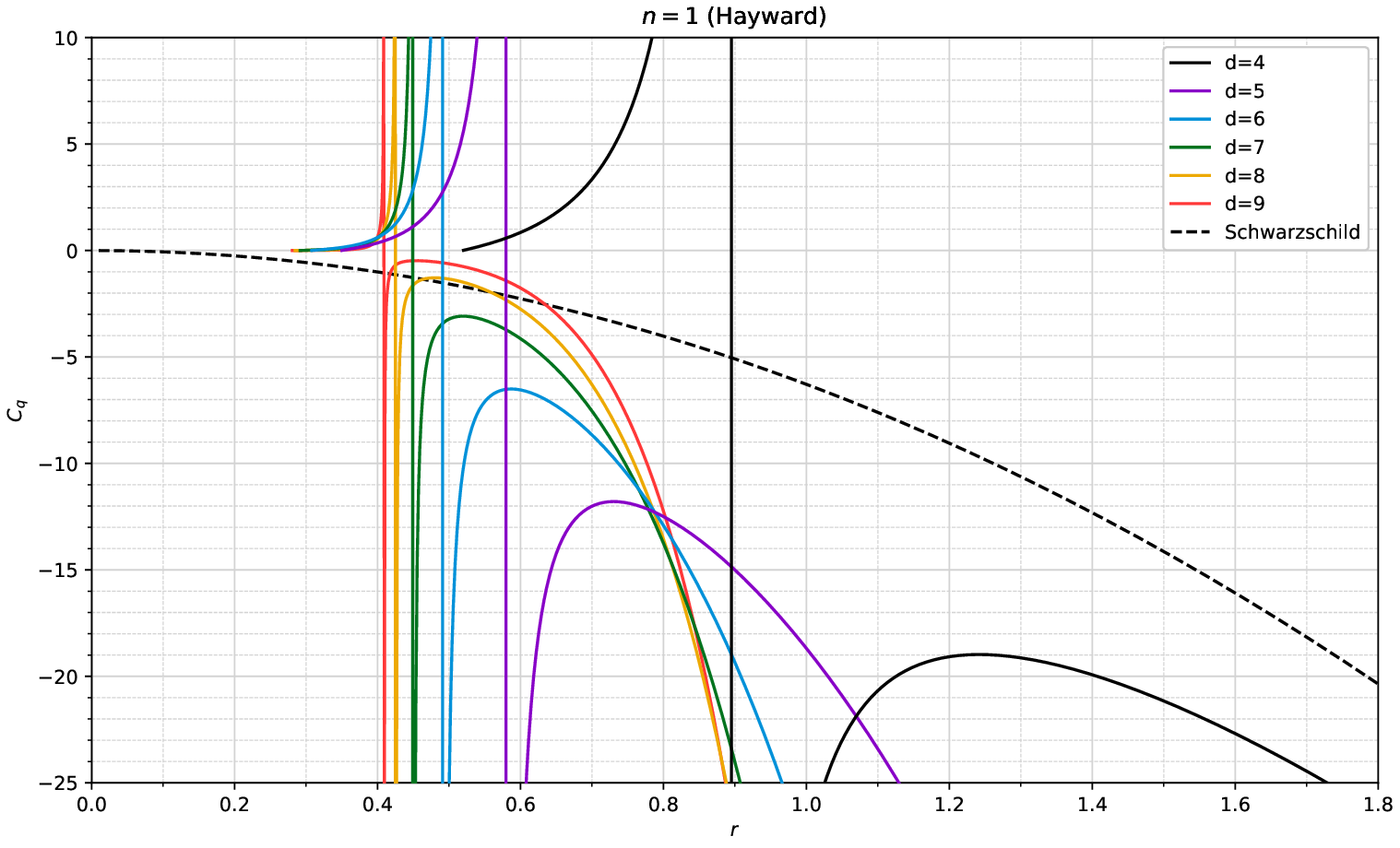}
	\end{minipage}  \vspace{0.5cm}
	\begin{minipage}[b]{0.5\linewidth}
		\includegraphics[width=1\linewidth]{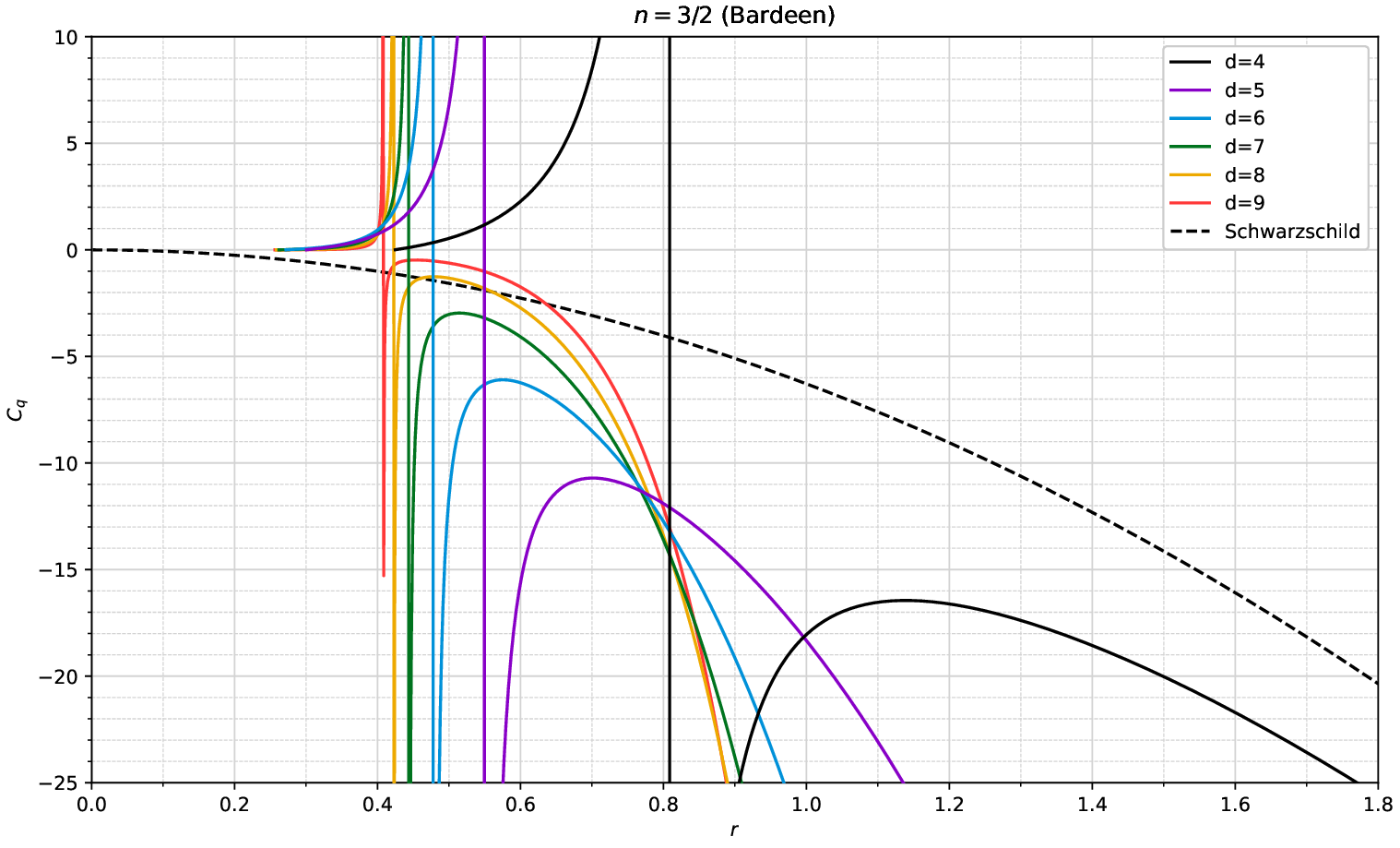}
	\end{minipage} 
	\begin{minipage}[b]{0.5\linewidth}
		\includegraphics[width=1\linewidth]{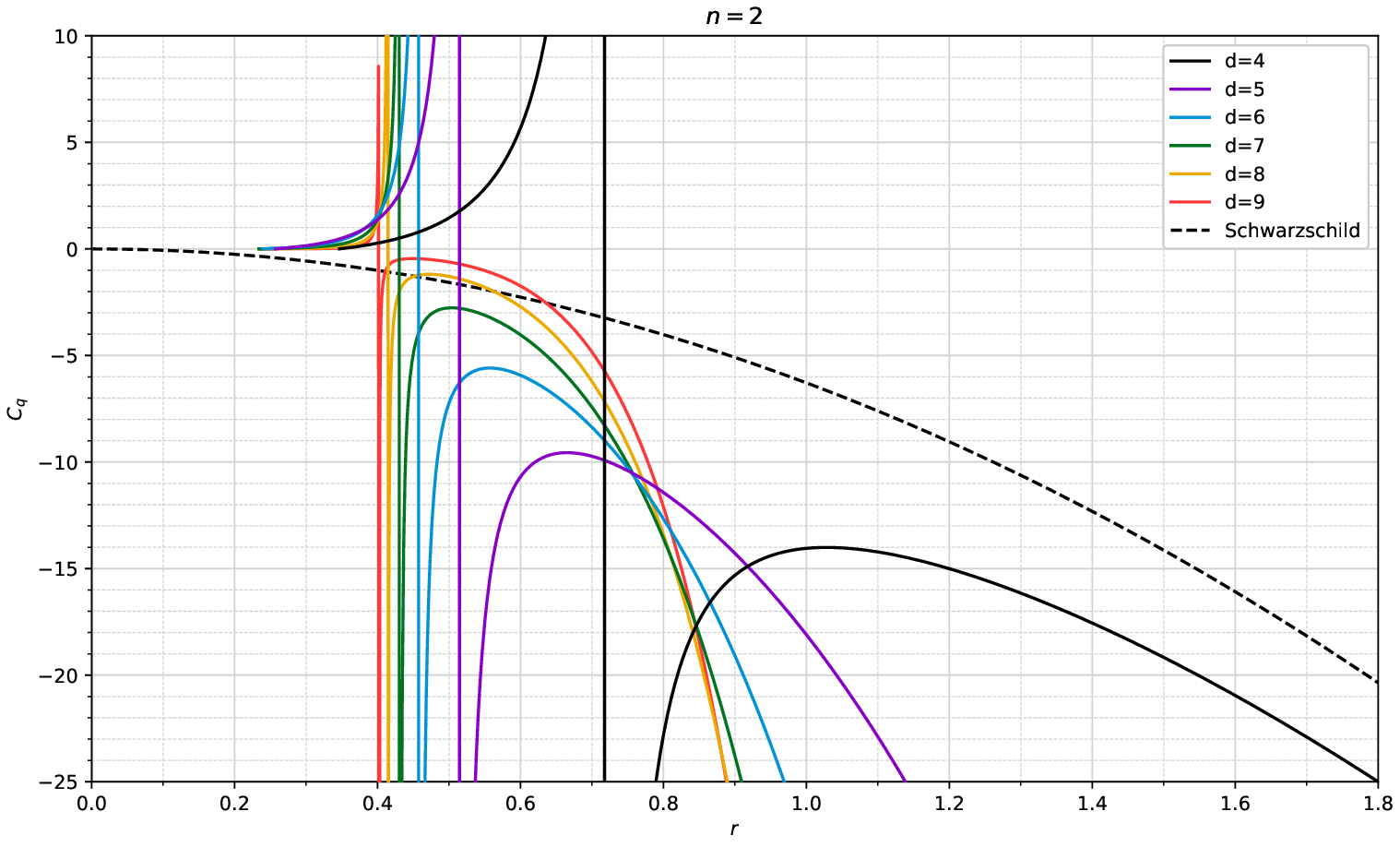}
	\end{minipage}
	\hfill
	\begin{minipage}[b]{0.5\linewidth}
		\includegraphics[width=1\linewidth]{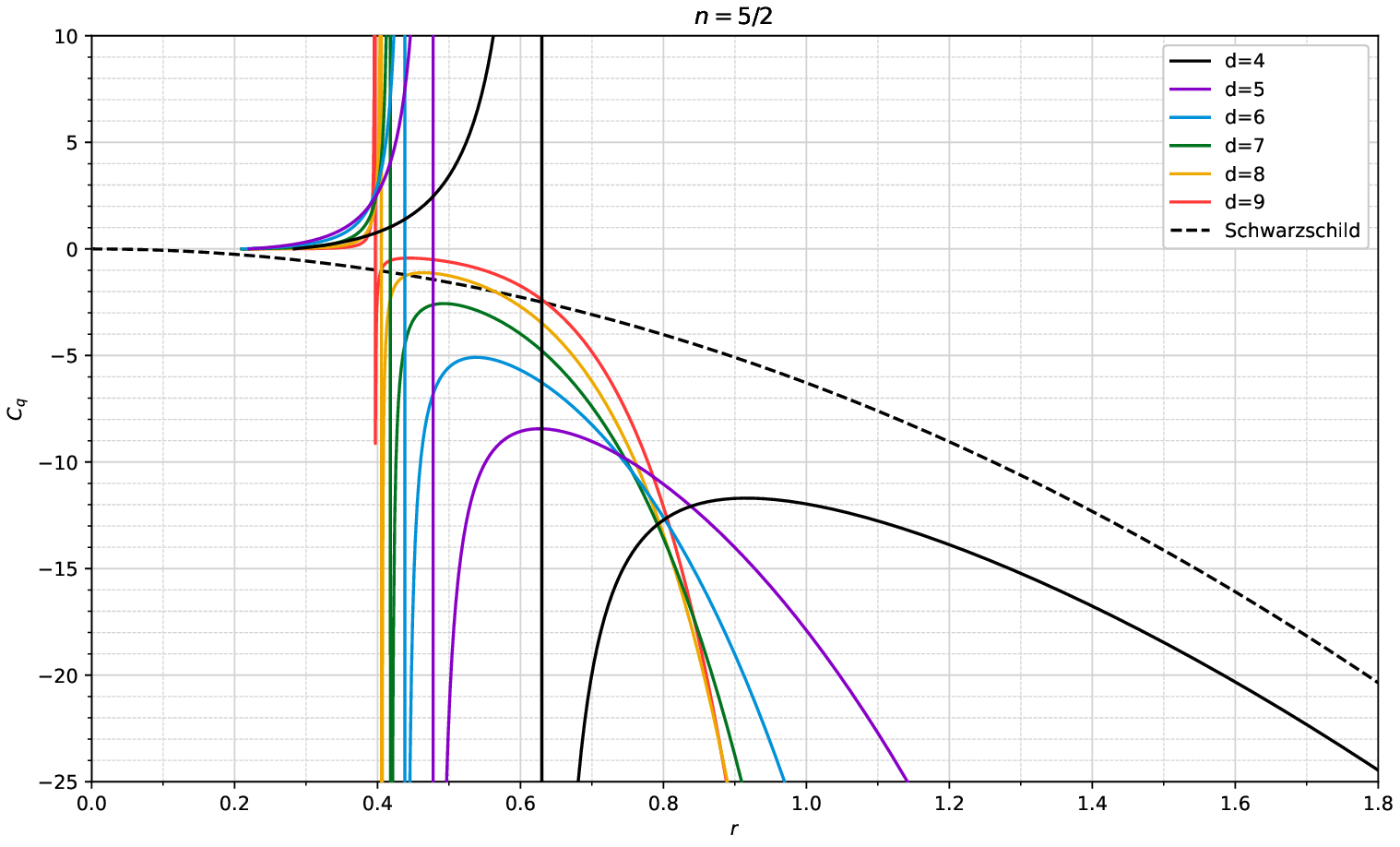}
	\end{minipage} 
	\caption{Heat capacity with respect to the horizon radius for  different polytropic indices and dimensions, with q set to $0.3$. The vertical line indicates the radius at which the heat capacity and its derivative diverge, suggesting a possible phase transition. The dashed line represents a Schwarzschild BH in $4$ dimensions.} 
	\label{figCq} 
\end{figure}

The heat capacity can be expressed as:
\begin{align}\label{Cq}
 C_{ q}= \left( \frac{\partial M}{\partial  T} \right)_{ q}= \left(\frac{\partial M}{\partial  r_\text{h}}\right)_{ q} \left( \frac{\partial  r_\text{h}}{\partial  T}\right)_{ q}.
\end{align}
Therefore, the heat capacity can be obtained using Eqs.  \eqref{qhorzionntilde}, \eqref{mfirst}, and \eqref{SG1}. Using  a numerical approach,  the heat capacity  can be computed for various values of $d$ and $n$. The results are plotted in Fig. \ref{figCq}. For $d$-dimensional Bardeen  and $4$-dimensional Hayward spacetimes, an analytical approach is feasible. For $d$-dimensional Bardeen BH

\begin{align}
	C_{q}\big|_{\text{Bardeen,}\,d}= \frac{(d-2) \Omega_{d-2} }{4 r_{\text{h}}}\frac{\left(q^{d-2}+r_{\text{h}}^{d-2}\right)^{\frac{2 d-3}{d-2}} \left((d-3) r_{\text{h}}^{d-2}-2 q^{d-2}\right)}{\left(d^2-4 d+7\right) q^{d-2} r_{\text{h}}^{d-2}+2 q^{2( d-2)}-(d-3) r_{\text{h}}^{2( d-2)}}.
\end{align}
as well as for $4$-dimensional Hayward and Bardeen BHs
\begin{align}
	C_{q}\big|_{\text{Hayward, $d=4$}}&=\frac{2 \pi  r_{\text{h}}^6 \left(3 q^2-r_{\text{h}}^2\right)}{\left(q^2-r_{\text{h}}^2\right)^2 \left(r_{\text{h}}^2-9 q^2\right)},\\
	C_{q}\big|_{\text{Bardeen, $d=4$}}&=\frac{2 \pi  \left( q^2+ r_h^2\right)^{5/2} \left( r_h^2-2  q^2\right)}{ r_h \left(2  q^4+7  q^2  r_h^2- r_h^4\right)},
\end{align}
These results are the same as those in \cite{man1, akbar, Myung}. Furthermore, in the limit of $ q\to 0$
\begin{align}
 C_{\text{ST}}=- \frac{(d-2) \Omega _{d-2}}{4}  r_{\text{h}}^{d-2},
\end{align}
we recover the heat capacity of the $d$-dimensional ST BH \cite{Ghosh:2014pga}. The negative value across all radii indicates that the BH is thermodynamically unstable. Unlike the ST BH, the regular BH \eqref{ad1} has a singularity in its heat capacity, which signals a possible phase transition. For the $d$-dimensional Bardeen BH, this occurs at
\begin{align}
	r_{\text{PT}}=q \left(4 (-7+4 d-d^2+\sqrt{(d-1) (d ((d-7) d+23)-25)}-7)\right)^{\frac{1}{2-d}}.
\end{align}
As Fig. \ref{figCq} shows, the heat capacity sign is negative and therefore the BH is thermodynamically unstable for horizon radii greater than $r_{\text{PT}}$ (indicated by a vertical line), unlike the ST BH. For $r<r_{\text{PT}}$, the BH is stable. The minimum horizon radius is again $r_{\text{ext}}$ below which temperature becomes negative and is therefore unacceptable.

Within the fixed-$q$ thermodynamic description, the sign of $C_q=(\partial M/\partial T)_q$ characterizes the local response of the BH branch to thermal fluctuations. A positive $C_q$ identifies a locally stable branch in this ensemble, whereas $C_q<0$
identifies a locally unstable one. The divergence of $C_q$ occurs when $(\partial T/\partial r_\text{h})_q=0$ and separates branches with different local thermal responses. We therefore regard this divergence as a candidate continuous phase-transition point. Establishing a genuine global phase transition would require, in addition, an analysis of the appropriate free energy
and competing thermodynamic branches, which is beyond the present scope. \\
This thermodynamic analysis does not remove the perturbative stability problem. The authors of \cite{DeFelice:2024seu} derived the linear perturbation dynamics of four-dimensional nonsingular NLED BHs and found an angular Laplacian instability in a finite region around the regular center. The instability is associated with a negative squared angular propagation speed of one vector mode; the growing vector perturbation then excites a dynamical even-parity gravitational perturbation, so that the static background cannot in general remain a steady configuration. Importantly, this unstable region should not be identified a priori with the region strictly inside the Cauchy horizon; in the representative solutions studied in Ref. \cite{DeFelice:2024seu}, it can extend beyond the inner horizon while remaining in the BH interior. Consequently, the $d=4$ BH sector of the present NLED construction should not be described as linearly dynamically stable, even where $C_q>0$.\\
For $d>4$, the four-dimensional result of Ref. \cite{DeFelice:2024seu} is not by itself a proof of the same instability because the perturbation decomposition and the dimension-dependent coefficients of the quadratic action change. However, no suppression of the instability can be inferred from the background solution alone. A definitive statement requires deriving the full $d$-dimensional quadratic action for coupled metric-NLED perturbations and checking the kinetic matrix, no-ghost
conditions, and radial and angular characteristic speeds. Since that analysis is not performed here, we make no claim of linear dynamical stability for $d>4$.

\section{OSD collapse in $d$ dimensions}
\label{SecCol}

Considering the spacetime \eqref{genmet}, we examine the gravitational collapse of a star in $d$ dimensions according to the OSD model \cite{Oppenheimer:1939ue, Datt:1938uwc}. In the classical OSD scenario, the stellar interior is filled with a homogeneous and isotropic perfect fluid, while the exterior is described by the Schwarzschild solution. The two regions can be smoothly joined without the need for a thin-shell layer.

In our analysis, we embed the star in regular BH spacetime \eqref{genmet} with metric component \eqref{ad1}. This generalizes the OSD collapse to $d$ dimensions.   The stellar region forms a high-density region relative to its surroundings, leading to gravitational collapse. Although the EMT of the stellar matter is not known a priori, it can be determined by requiring smooth matching of the interior and exterior metrics at the stellar surface. For the interior, we adopt a spatially flat, $d$-dimensional FLRW metric and choose the time coordinate so that it coincides with the proper time of comoving observers. Meanwhile, the exterior time coordinate is identified with the proper time of freely falling observers. 

Based on the assumption we made in section \ref{sec4}, we restrict ourselves to $\theta_i = \pi/2$ (except for $\theta_{d-2}$), and assume that  the stellar surface moves radially,  equivalent to setting $L=0$ in Eq. \eqref{Lagpho2}. Additionally, we assume that the stellar surface is timelike (i.e., $\epsilon=-1$). Under these conditions, Eq. \eqref{LagEep} reduces to
\begin{align}
	E^2 = \dot r^2 +A_d(r).
\end{align}
If we set $r_i$ as the initial radius of the collapsing star and take it to be sufficiently large, then $E\to1$. Thus, it is convenient to write the metric \eqref{genmet} in the Painlevé-Gullstrand (PG) coordinates  \cite{Painleve:1921,Gullstrand:1922,Martel:2000rn}, which describe a freely falling radial observer starting from rest at infinity. The four-velocity of such an observer takes the following form
$
u^{\alpha}\partial_{\alpha}	= A_d(r)^{-1/2} \partial_t	- \sqrt{1-A_d(r)}\,\partial_r .
$
This gives the observer’s proper time $\tau$  as
$
d\tau = dt	+ A_d(r)^{-1}	\sqrt{1-A_d(r)}\, dr
$
. Substituting this expression for $dt$ into the metric \eqref{genmet} leads directly to its PG representation
\begin{align}
	ds^2 = -d\tau^2 + \left(dr + \sqrt{1-A_d(r)} d\tau\right)^2+r^2 d \Omega_{d-2}^2.
\end{align}
The interior metric in PG coordinates  is also expressed as
\begin{align}
	\label{PGFRW}
	ds^2 = - d\tau^2 + \left(d r - r H(\tau) d \tau\right)^2 + r^2d\Omega_{d-2}^2,
\end{align}
where $r(\tau)=a(\tau)r_\text{c}$, $H(\tau) = \dot{r}(\tau)/r(\tau)$, $a(\tau)$ is the scale factor and  $r_\text{c}$ is the comoving radial coordinate. 

The parametric equations on the surface of the star are described by $r = R(\tau)$ and $t = T(\tau)$ where $\tau$ is the proper time of observers comoving with the surface \cite{Poisson:2009pwt}. On the surface of the star $\Sigma$
\begin{align}
	ds^2_\Sigma &= -d \tau^2 + a^2(\tau) R_\text{c}^2 d\Omega_{d-2}^2 \hspace{3.6cm} \text{ from inside}\\
	ds^2_\Sigma &= -(A_d(R) \dot T^2 - A_d(R)^{-1} \dot R^2)d \tau^2 + R^2 d\Omega_{d-2}^2 \hspace{1cm} \text{ from outside}
\end{align}
According to the Israel junction conditions \cite{Israel}, the induced metric and the extrinsic curvature should be continuous across the boundary of the two geometries to get a smooth transition. The continuity of the induced metric implies $R(\tau)= a(\tau) R_\text{c}$ and $A(R) \dot T^2 - A(R)^{-1} \dot R^2=1$. To impose the second junction condition, the extrinsic curvature must be identical on both sides of the stellar surface
\begin{align}
{}^{\text{(in)}}K^{\tau}_{\tau}&=0,  \hspace{1.65cm} {}^{\text{(in)}}K^{\theta_i}_{\theta_i}=1/R,\\
	{}^{\text{(out)}}K^{\tau}_{\tau}&=\dot{\gamma}/\dot R, \hspace{1cm} {}^{\text{(out)}}K^{\theta_i}_{\theta_i}=\gamma/R,
\end{align}
where $\gamma = \sqrt{\dot R^2 +A_d(r)}$. To smoothly join two geometries, we require $\gamma=1$. This yields the evolutionary equation of the stellar surface as 
\begin{align}
	\label{dotR}
	\dot R(\tau) = - \sqrt{1-A_d(R)}.
\end{align}
The negative sign is chosen due to the contraction of the star. To obtain the Friedmann equation in $d$ dimensions, we must compute the components of the Einstein tensor:
\begin{align}
	\label{errFRW}
	G^{t }{}_{t }&= -\frac{(d-1)(d-2)}{2}\frac{\dot a^2}{a^2} = - 8\pi \rho_{\text{star}}, \\
	\label{eppFRW}
	G^{r}{}_{r } &=G^{\theta_i}{}_{\theta_i }= -\frac{(d-2)(d-3)}{2}\frac{\dot a^2}{a^2}-(d-2)\frac{\ddot a}{a} = 8\pi p_{\text{star}}.
\end{align}
The first is the Friedmann equation in $d$ dimensions.  
Substituting $R(\tau)=a(\tau) R_\text{c}$ and Eq. \eqref{dotR} into it yields
\begin{align}
	\label{rhocol}
	\rho_{\text{star}} = 	\frac{(d-1)(d-2)}{2}\frac{1-A_d(R(\tau))}{8\pi R(\tau)^2}.
\end{align}
Using Eq. \eqref{errFRW} and \eqref{eppFRW}, the continuity equation becomes $\dot \rho + (d-1)  (\rho +p){\dot a}/{a}=0$. Combining the continuity equation with the density function \eqref{rhocol}, the stellar pressure can be calculated as
\begin{align}
	\label{pcol}
	p_{\text{star}} =\frac{d-2}{16 \pi  R(\tau)^2} \left(R(\tau) A_d'(R(\tau))+(d-3) (A_d(R(\tau))-1)\right).
\end{align}
The isotropic density \eqref{rhocol} and pressure \eqref{pcol} provide the total energy density and the isotropic pressure evaluated at the stellar surface. These quantities are a combination of the background contribution from \eqref{rhop} and  an additional source associated with the star itself. It should be mentioned that the radial pressure of the exterior metric \eqref{err} is equal to the pressure at the surface of the star \eqref{pcol} and as a result the pure radial pressure on the stellar surface is $p_{\text{star}} - p_\text{r}  |_{r=R}=0$. hus, the radial pressure jump across the stellar surface vanishes, and no thin shell is required. The stellar surface can therefore be	consistently chosen to follow the radial geodesic.

By dividing  \eqref{rhocol} by \eqref{err}, using the fact that $p_{\text{star}} = p_\text{r}  |_{r=R}$ and substituting the resulting radius into \eqref{pcol}, straightforward calculations show that the EOS takes the form
	\begin{align}
		\label{EOS}
		p_{\text{star}}  =-\left(\frac{16 \pi  q^{\frac{1}{3} (d-1) (2 n)}}{(d-2) (d-1) m^{\frac{2}{3} (d-1) n+\frac{3-4 n}{n}}}\right)^{\frac{3 (d-2)}{2 (d-1) n}}\rho_{\text{star}}^{\frac{3 (d-2)}{2 (d-1) n}+1} 
	\end{align}
	In four dimensions, Eq. \eqref{EOS} reduces to the usual polytropic form $p\propto\rho^{1+1/n}$. In higher dimensions, it can be	written as a generalized polytropic EOS, 
	\begin{align}
	p\propto\rho^{1+1/n_{\text{eff}}}, \hspace{1cm}n_{\text{eff}}=\frac{2n(d-1)}{3(d-2)}.
	\end{align}
	Thus, the effective polytropic index depends both on the original index $n$ and on the spacetime dimension $d$, and it reduces to $n_{\text{eff}}=n$ for $d=4$.
	
	Using the energy conditions studied in subsection \ref{SSecEC} and substituting the metric component \eqref{ad1} into \eqref{rhocol} and \eqref{rhop}, we find that all energy conditions are satisfied for all radii, except the SEC, which is only satisfied for 
	\begin{align}
		R_{\text{SEC}}>\left(\frac{(d-3) m^{(d-2) \left(1-\frac{3}{2 n}\right)}}{2 q^{d-2}}\right)^{-\frac{2 n}{3 (d-2)}},
	\end{align}
	This means that for other radii, the stellar matter contains exotic matter. 
	
\subsection{OSD collapse into $d$-dimensional regular BH}	

	By substituting $A_d(r)$ from \eqref{ad1} into  \eqref{dotR} and performing the integration, we obtain the evolution of the stellar surface 
\begin{align}
	\label{SOTS}
	\tau &=\frac{2}{d-1} \bigg(-\frac{ R^{\frac{d-1}{2}} }{m^{\frac{d-3}{2}}}\, {}_2F_1\left[-\frac{n(d-1)}{3 (d-2)} ,-\frac{n(d-1)}{3 (d-2)} ;1+\frac{n(d-1)}{3 (d-2)} ;-m^{\frac{1}{2} (d-2) \left(\frac{3}{n}-2\right)} q^{d-2} R^{-\frac{3 (d-2)}{2 n}}\right]\nonumber\\
	&+\frac{  R_0^{\frac{d-1}{2}} }{m^{\frac{d-3}{2}}}\, _2F_1\left[-\frac{n(d-1)}{3 (d-2)} ,-\frac{n(d-1)}{3 (d-2)} ;1+\frac{n(d-1)}{3 (d-2)} ;-m^{\frac{1}{2} (d-2) \left(\frac{3}{n}-2\right)} q^{d-2} R_0^{-\frac{3 (d-2)}{2 n}}\right]\bigg),
\end{align}
where $_2F_1$ is the hypergeometric function, and the integration constant $R_0$ is chosen such that  $R=R_0$ at $\tau = 0$. For $n=1$ and $n=3/2$ in $d=4$ we will recover the OSD‌ collapse into Hayward and Bardeen BHs, respectively, as studied in \cite{our1}. For $q=0$, the  stellar surface evolution reduces to that of the ST spacetime
\begin{align}
		\tau_{\text{ST}} =\frac{2 m^{\frac{3-d}{2}}}{d-1} (R^{\frac{d-1}{2}}-R_0^{\frac{d-1}{2}}),
\end{align}
and for $d=4$, we recover the result of standard OS collapse $\tau_{\text{Sch}}  =\frac{2}{3 \sqrt{m}}(R^{3/2}-R_0^{3/2})$ \cite{Blau, Rezzolla:2004}. Using the dimensionless parameters, Eq. \eqref{SOTS} is plotted in Fig. \ref{figcol} for various polytropic indices and spacetime dimensions, with the magnetic charge held fixed.
As it is evident from Eq. \eqref{SOTS}, the collapsing stellar surface takes infinite proper time to reach the center (it diverges as $R\to 0$), so the surface never reaches the center. The duration of the collapse increases with the dimension and polytropic index.

\subsubsection{Horizons}

\begin{figure}[t]
	\begin{minipage}[b]{0.5\linewidth}
		\includegraphics[width=1\linewidth]{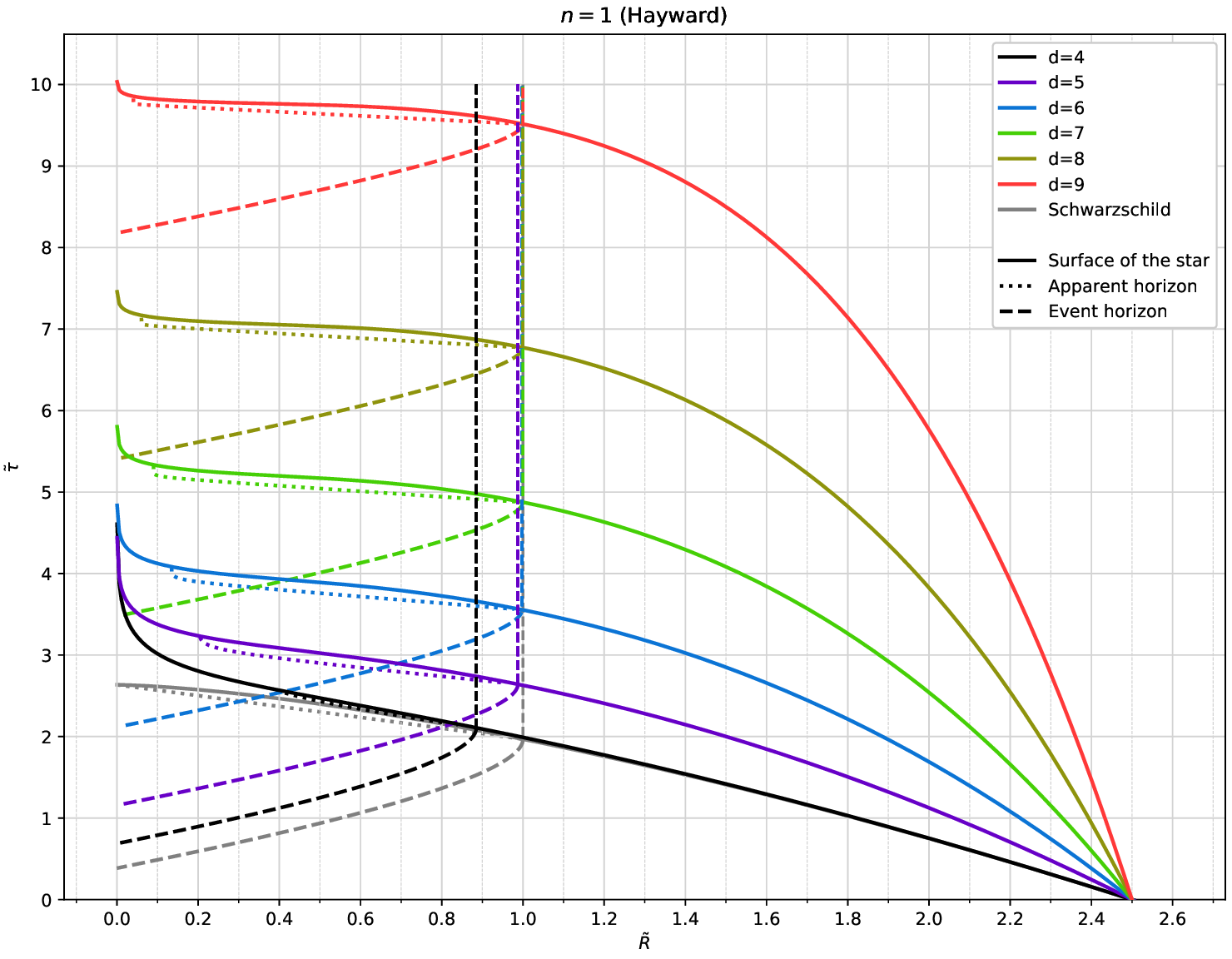} 
	\end{minipage} \vspace{0.2cm}
	\begin{minipage}[b]{0.5\linewidth}
		\includegraphics[width=1\linewidth]{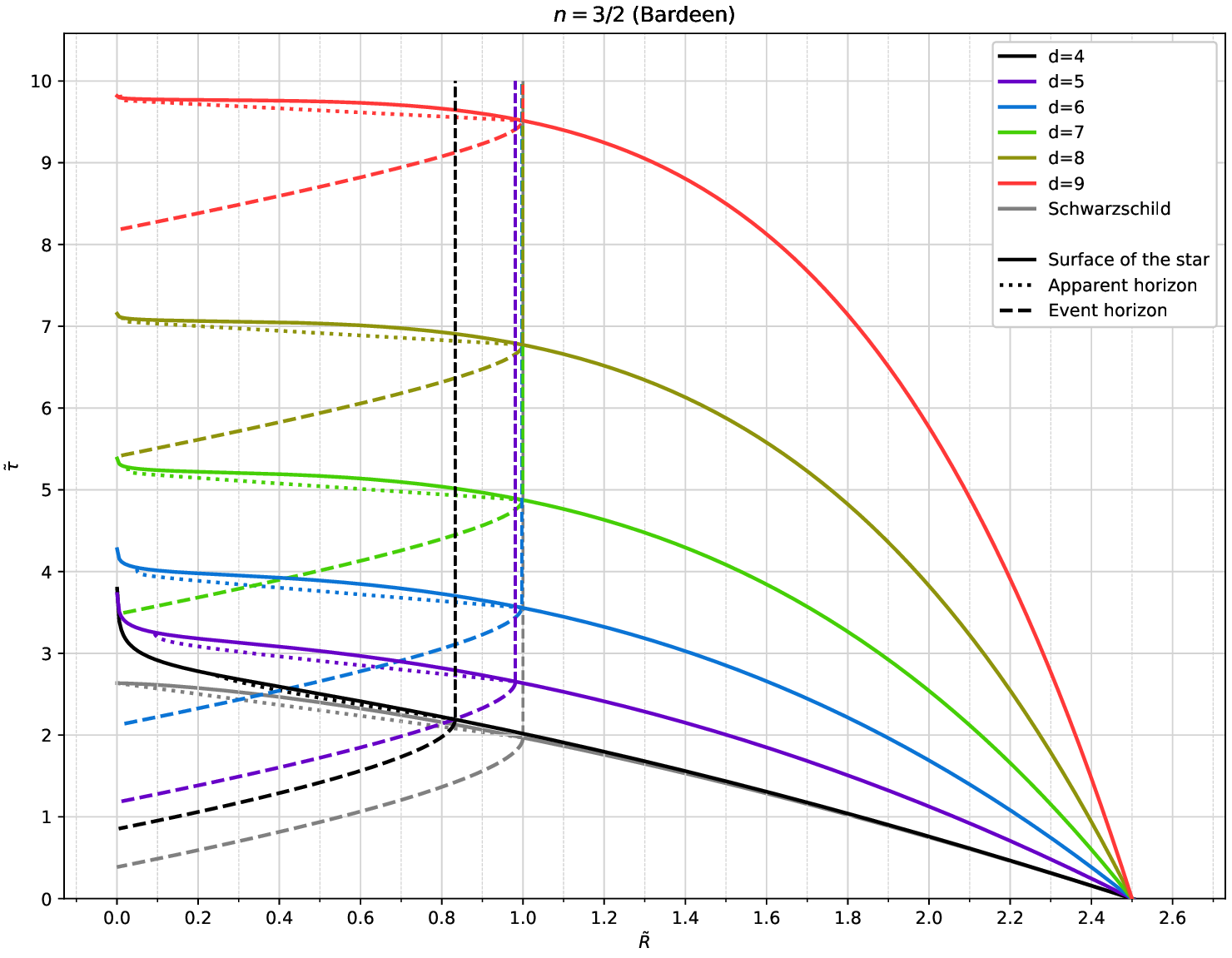} 
	\end{minipage} \\
	\begin{minipage}[b]{0.5\linewidth}
		\includegraphics[width=1\linewidth]{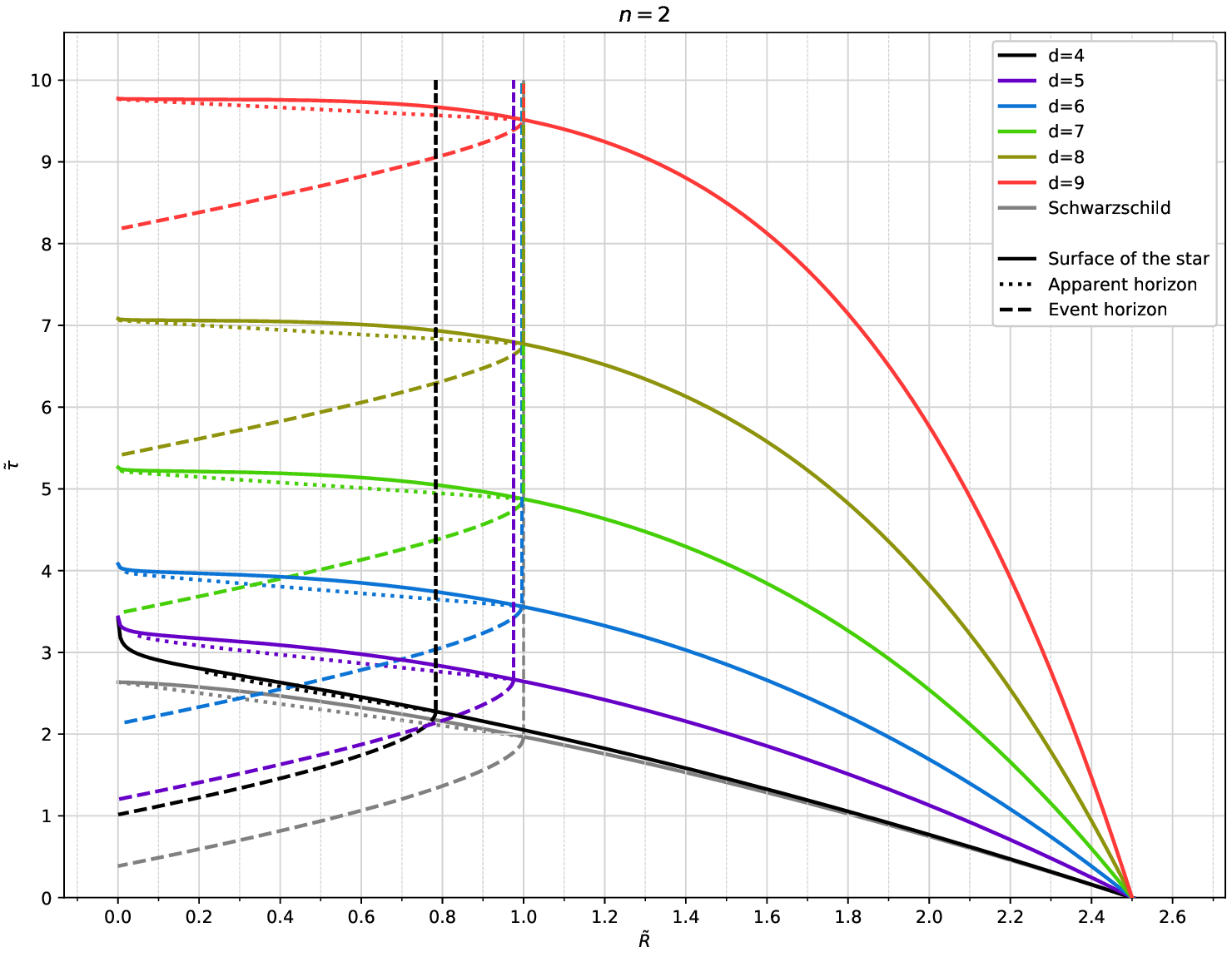} 
	\end{minipage}
	\begin{minipage}[b]{0.5\linewidth}
		\includegraphics[width=1\linewidth]{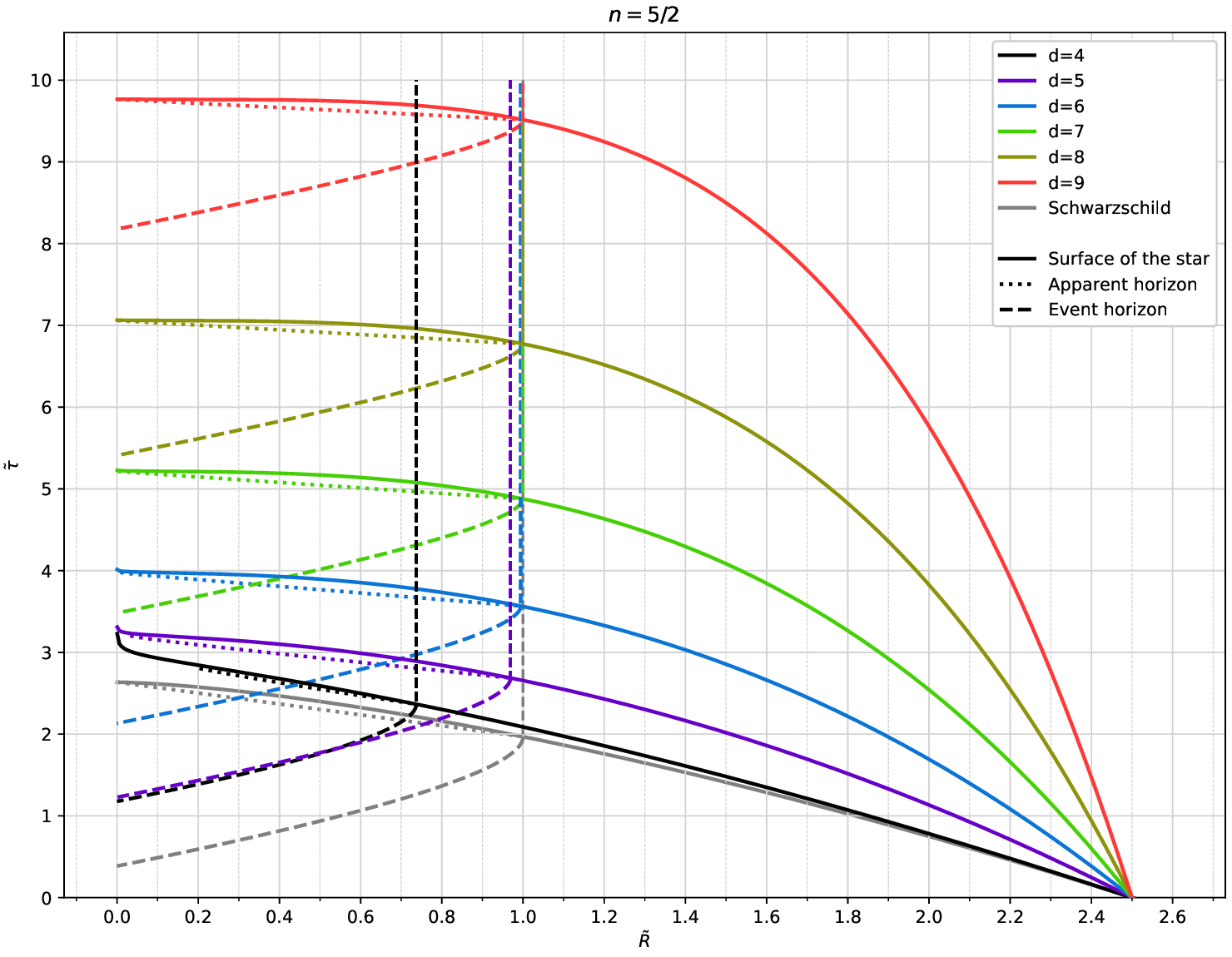} 
	\end{minipage} 
	\caption{ The evolution of the star surface and horizons in
		OSD collapsing scenario into $d$-dimensional regular BH \eqref{ad1} for different polytropic indices and dimensions and with the fixed values of $\tilde q=0.3$ and $R_0=2.5$. The quantities are normalized to $m$, which are denoted by tildes. The solid, dotted, and dashed curves represent the surface of the star, the apparent horizon, and the event horizon, respectively. The Schwarzschild case is plotted as a grey line.}
	\label{figcol}
\end{figure}

Using the metric \eqref{PGFRW}, inside the star, the Hubble parameter can be expressed as $H(\tau) = \dot R(\tau)/ R(\tau)$ and the radial null trajectories satisfy
\begin{align}
	\frac{dr}{d\tau}= -1 + r H,\hspace{0.5cm} \text{or}\hspace{0.5cm}\frac{dr}{d\tau} = 1 + r H.
\end{align}
Since $H<0$, the left expression corresponds to ingoing null geodesics. The right expression represents ingoing null geodesics for $r>-1/H$ and outgoing null geodesics for $r<-1/H$. This means that there are trapped surfaces with radii greater than $-1/H$. Therefore, using Eqs. \eqref{dotR} and \eqref{ad1}, the apparent horizon as a function of radius of star can be calculated as
\begin{align}
	\label{AHCol}
	r_\text{AH}= - \frac{1}{H} =  \frac{R^{\frac{d-1}{2}}}{m^{\frac{d-3}{2}}  }\left(1+ \frac{m^{\left(\frac{d-2}{2} \right) \left(\frac{3}{n}-2\right)} q^{d-2} }{ R^{\frac{3 (d-2)}{2 n}}}\right)^{\frac{n(d-1)}{3 (d-2)}}.
\end{align}
Eq. \eqref{SOTS} together with Eq. \eqref{AHCol} describe the evolution of the apparent horizon  during the OSD collapse into the $d$-dimensional regular BH \eqref{ad1}. 
The apparent horizon is plotted as a dotted line in Fig. \ref{figcol}. It forms when the stellar surface crosses the exterior event horizon and vanishes when the surface reaches the exterior Cauchy horizon. 

By applying \eqref{dotR}, the equation describing the apparent horizon surface can be written as
	\begin{align}
		F(r,\tau) = r - \frac{R(\tau)}{\sqrt{1-A_d(R(\tau))}}.
	\end{align}
	On the horizon $R H/\sqrt{1- A_d(R)}=-1$,  so the normal vector to the apparent horizon is 
	\begin{align}
		n_a=\partial_a F=-\left(1+ \frac{R  A^\prime_d(R)}{2 (1- A_d(R))},1,0,0\right).
	\end{align}
	Thus,
	\begin{align}
		n_a n^a &= - \left(1+ \frac{R  A^\prime_d(R)}{2 (1- A_d(R))}\right)^2 + 2\left(1+ \frac{R  A^\prime_d(R)}{2 (1- A_d(R))}\right)\\ \nonumber
		&=\frac{4 (d-1) q^{d-2} m^{\frac{d-2}{2}  \left(\frac{3}{n}-2\right)} R^{-\frac{3 (d-2)}{2 n}}-(d-5) (d-1)}{4 \left(q^{d-2}  m^{\frac{1}{2} (d-2) \left(\frac{3}{n}-2\right)} R^{\frac{-3 (d-2)}{2 n}}+1\right)^2}
	\end{align}
	For $d=4$ and $d=5$, the above expression is positive. Therefore, the apparent horizon is timelike, but for $d>5$, the apparent horizon will be null for
	\begin{align}
		R^{\text{null}}_{\text{AH}}= \left(\left(\frac{4}{d-5}\right)^{1/3}q^{\frac{d-2}{3}} m^{-\frac{(d-2) (2 n-3)}{6 n}}\right)^{\frac{2 n}{d-2}}
	\end{align}
	and for $R>R^{\text{null}}_{\text{AH}}$ the apparent horizon is spacelike.
	
	It is well-known that the union of all apparent horizons is called a trapping horizon (TH) \cite{Hayward1994} which is a hypersurface that admits a foliation by spacelike $d-2$-surfaces on which the expansion of the outgoing null normal vanishes, $\theta_\ell\big|_\text{h}=0$, while the expansion of the ingoing null normal is nonzero, $\theta_n\big|_\text{h}\neq0$, and $n^a \nabla_a \theta_\ell\big|_\text{h} \neq 0$. A TH is called outer if $n^a \nabla_a \theta_\ell\big|_\text{h} < 0$, inner if $n^a \nabla_a \theta_\ell\big|_\text{h} > 0$ , future if $\theta_n\big|_\text{h}<0$ , and past if $\theta_n\big|_\text{h}>0$.  To demonstrate the THs in the collapsing scenario described above, we begin with the expansion parameter for the exterior metric \cite{Sadeghi:2024rar}.
	\begin{align}
		\theta_{\ell,\,\text{out}} &= \frac{2}{r}\left(1 - \sqrt{1- A_d(r)}\right),\\
		\theta_{n,\,\text{out}} &=- \frac{1}{r}\left(1 + \sqrt{1- A_d(r)}\right).
	\end{align}
At the horizon, i.e., where $A_d(r_{\text{h}})=0$,
 	\begin{align}
	 \theta_{\ell,\,\text{out}}\big|_{r_{\text{h}}} =0, \hspace{1cm} \theta_{n,\,\text{out}}\big|_{r_{\text{h}}} = -\frac{2}{r} <0.
 \end{align}
This  means horizons are marginally trapped tubes\footnote{A marginally trapped tube \cite{10JFT} is a hypersurface foliated by $d-2$-dimensional surfaces, called marginally trapped surfaces (MTS), on which the expansion of the outgoing null normal vanishes.}. Furthermore,
	\begin{align}
		n^b \nabla_b  \theta_{\ell,\,\text{out}} \big|_{r_{\text{h}}} = -\frac{1}{2r_{\text{h}^2}} ( 4 + r A'(r_{\text{h}})).
	\end{align}
In Section \ref{sec5}, we saw that for the larger value of $r_{\text{h}}$, i.e., the event horizon, the derivative of the metric function is always positive. Consequently, the expression above is always negative, identifying the event horizon as a future outer TH. For the interior, similar calculations lead to
\begin{align}
	\ell^a_{\text{in}} &= (1, 1+ r H ,0 ,0,\dots),\\
	n^a_{\text{in}} &= \frac{1}{2} (1, -1 + r H,0,0,\dots)\label{nnll},
\end{align}
and the expansion parameter can be calculated as
\begin{align}
	\theta_{\ell,\text{in}} &= \frac{2}{r}\left( 1+ r H\right),\\
	\theta_{n,\text{in}} &= -\frac{1}{r}\left( 1-r H\right).
\end{align}
For $r=-1/H$,	$\theta_{\ell,\text{in}}\big|_\text{h} =0$, and $\theta_{n,\text{in}}\big|_\text{h}<0$. The Lie derivative of the expansion parameter becomes
\begin{align}
	n^a \nabla_a \theta_\ell\big|_{\text{h},\text{in}} =\dot{H}+2H^2.
\end{align}
	From Friedmann's equations $H^2 = \frac{16\pi}{(d-1)(d-2)} \rho_{\text{star}}$ and $\dot H = - \frac{8\pi}{d-2}(\rho_{\text{star}} +p_{\text{star}})$, we obtain
\begin{align}
	\label{hprho}
	\dot{H}+2H^2 = \frac{8\pi}{(d-1)(d-2)}\left[(5-d) \rho_{\text{star}}- (d-1)p_{\text{star}} \right].
\end{align}

Substituting Eqs. \eqref{rhocol} and \eqref{EOS} into Eq. \eqref{hprho}, it is straightforward to  verify that the resulting expression is always positive, which classifies the TH as a future inner TH. This result demonstrates that the  characteristics of the TH remain unchanged in higher dimensions.
As Fig. \ref{figcol} shows, when the surface of the star reaches the smaller value of $r_{\text{h}}$, i.e., the Cauchy horizon, the TH disappears. Thereafter, there are no trapped surfaces inside the star. This indicates that, for a non-singular collapse scenario, the disappearance of trapped surfaces in the final stages of collapse is essential.

To determine the radius of the interior event horizon,
$r_{\text{{EH}}}$,
we consider the outgoing null geodesics that reach the stellar surface precisely when it intersects the exterior event horizon.
As mentioned earlier, outgoing null geodesics satisfy
\begin{align}\label{rehgen}
	\dot{ r}_{\text{{EH}}} = 1 + {r}_{\text{{EH}}}{H}.
\end{align}
Using Eq. \eqref{dotR}
and the chain rule 
$\dot{{r}}_{\text{{EH}}} = (d{r}_{\text{{EH}}}/d{R})\dot{{R}}$,
Eq. \eqref{rehgen} becomes
\begin{align}\label{109}
	\frac{d{r}_{\text{{EH}}}}{d{R}}
	- \frac{{r}_{\text{{EH}}}}{{R}} &= - \left(1-A_d(R)\right)^{-1/2}\\
	&= -\frac{R^{\frac{d-3}{2}}}{m^{\frac{d-3}{2}}} \left(\frac{q^{d-2} +m^{\left(\frac{d-2}{2}\right) \left(\frac{3}{n}-2\right)}}{R^{\frac{3 (d-2)}{2 n}}}+1\right)^{\frac{n(d-1) }{3 (d-2)}}.
\end{align}
Given the boundary condition
${r}_{\text{{EH}}}({r}_\text{h}) = {r}_\text{h}$, this equation can be solved analytically as 
\begin{align}
	\label{EHCol}
	{r}_{\text{{EH}}}&= R -\frac{2 R^{\frac{d-1}{2}}}{(d-3) m^{\frac{d-3}{2}}}\, {}_2F_1\left[-\frac{n(d-3) }{3 (d-2)},-\frac{n(d-1) }{3 (d-2)};1-\frac{n(d-3) }{3 (d-2)}; -\frac{m^{\left(\frac{d-2}{2}\right) \left(\frac{3}{n}-2\right)} q^{d-2}}{R^{\frac{3 (d-2)}{2 n}}}\right]\nonumber\\
	&+ \frac{2 R {r}_\text{h}^{\frac{d-3}{2}}}{(d-3) m^{\frac{d-3}{2}}} \, {}_2F_1\left[-\frac{n(d-3) }{3 (d-2)},-\frac{n(d-1) }{3 (d-2)};1-\frac{n(d-3) }{3 (d-2)};-\frac{m^{\left(\frac{d-2}{2}\right) \left(\frac{3}{n}-2\right)}q^{d-2}}{ {r}_\text{h}^{\frac{3 (d-2)}{2 n}}}\right].
\end{align}
Eqs. \eqref{SOTS} and \eqref{EHCol} describe the evolution of the event horizon as a function of proper time. See Fig. \ref{figcol}. As the number of dimensions increases, so does the initial radius of the event horizon. In Fig. \ref{figcol},  the value of the parameter $\tilde q=0.3$ is fixed for all plots. With increasing dimensionality, the resulting event horizon approaches that of the Schwarzschild BH, a consequence of the fixed choice of $q$ (see Fig. \ref{figqr}).
For the ST BH limit ($q=0$), Eq. \eqref{EHCol} reduces to
\begin{align}
	{r}_{\text{{EH,\,ST}}}=\frac{R}{3-d}\left(\frac{2 R^{\frac{d-3}{2}}}{m^{\frac{d-3}{2}}}-d+1\right),
\end{align}
 and for Schwarzschild BH, it becomes ${r}_{\text{{EH,\,Sch}}}=3 R-\frac{2 R^{3/2}}{\sqrt{m}}$. 
 
A crucial point concerns the choice of the initial radius $R_{0}$.
In addition to the boundary condition ${r}_{\text{{EH}}}({r}_\text{h}) = {r}_\text{h}$, the initial radius of the star must be chosen such that the event horizon starts to grow from a  non-negative value. This sets a lower limit for the allowed initial radius of the star. If the initial radius is smaller than this limit, the event horizon would form before the OSD collapse begins, which is unphysical. For collapse to a ST BH, the event horizon radius vanishes at $ R_0 /m= \left(\frac{d-1}{2}\right)^{\frac{2}{d-3}}$. This reduces to $R_{0,\text{min}}/m=9/4$ for the Schwarzschild BH in $4$ dimensions. Since Schwarzschild BHs naturally impose the most restrictive lower limit on $R_{0}/m$, we chose $R_0/m=2.5$ in Fig. \ref{figcol} to satisfy this condition.

		As shown in Fig. \ref{figcol}, let $\tau_\text{B}$ denote the proper time at which the event horizon begins to expand, and let the corresponding radius be $R_\text{B}$. By definition, $\tau_\text{B} = \tau(R_\text{B})$ and $r_{\text{EH}}(R_\text{B}) = 0$. In view of Eq. \eqref{SOTS}, we have
	\begin{align}
		R_{\text{B}}&= \frac{2 R_{\text{B}}^{\frac{d-1}{2}}}{(d-3) m^{\frac{d-3}{2}}}\, {}_2F_1\left[-\frac{n(d-3) }{3 (d-2)},-\frac{n(d-1) }{3 (d-2)};1-\frac{n(d-3) }{3 (d-2)}; -\frac{m^{\left(\frac{d-2}{2}\right) \left(\frac{3}{n}-2\right)} q^{d-2}}{R_{\text{B}}^{\frac{3 (d-2)}{2 n}}}\right]\nonumber\\
		&- \frac{2 R_{\text{B}} {r}_\text{h}^{\frac{d-3}{2}}}{(d-3) m^{\frac{d-3}{2}}} \, {}_2F_1\left[-\frac{n(d-3) }{3 (d-2)},-\frac{n(d-1) }{3 (d-2)};1-\frac{n(d-3) }{3 (d-2)};-\frac{m^{\left(\frac{d-2}{2}\right) \left(\frac{3}{n}-2\right)}q^{d-2}}{ {r}_\text{h}^{\frac{3 (d-2)}{2 n}}}\right].
	\end{align}

       The above equation clearly demonstrates that the radius $R_{\text{B}}$ is independent of the initial stellar surface radius $R_0$; rather, it depends only on the outer event horizon radius and the metric parameters. By employing Eq. \eqref{SOTS} to determine $\tau_\text{B}$ and differentiating it with respect to $R_0$,  we obtain, after straightforward simplification (see Appendix \ref{A2}),
	\begin{align}
		\label{dtauB}
		\frac{ d\tau_\text{B}}{d R_0} =\frac{R_0^{\frac{d-3}{2}}}{m^{\frac{d-3}{2}}} \, _2F_1\left[-\frac{n(d-1)}{3 (d-2)},1-\frac{n(d-1)}{3 (d-2)};\frac{n(d-1) }{3 (d-2)}+1;-m^{\frac{d-2}{2}\left(\frac{3}{n}-2\right)} q^{d-2} R_0^{-\frac{3 (d-2)}{2 n}}\right].
		\end{align}
	For given values of $q$, $n$ and $d$, and for $R_0/m > \left(\frac{d-1}{2}\right)^{\frac{2}{d-3}}$, one can easily show that the expression above is always positive. This implies that increasing the initial stellar radius $R_0$ delays the formation of the event horizon, i.e., the horizon forms at a later proper time.

By solving Eq. \eqref{109} with the initial condition ${r}_{\text{{EH}}}=r_h$ at $R=r_h$, we obtain:
\begin{align}
	\label{1091}
	r_{\text{EH}} = R\left(1-\int_{r_{\text{h}}}^{R_{0,\text{min}}} \frac{dR}{R \sqrt{1 - A_d(R)}}\right).
\end{align}
Defining the smallest possible initial radius for the star, $R_{0, \text{min}}$, as the value where $r_{\text{EH}}=0$ , we find that\footnote{For further details and the extension to the spatially closed case see \cite{Khodabakhshi:2025fmf}.} 
\begin{align}
	\label{rEHSC1}
	1 = \int_{r_{\text{h}}}^{R_{0,\text{min}}} \frac{dR}{R \sqrt{1 - A_d(R)}}.
\end{align}
Substituting Eq. \eqref{ad1} into Eq. \eqref{rEHSC1} and then expanding it in terms of the small dimensionless magnetic charge,  $\tilde q\ll1$, we obtain
	\begin{align}
		\label{rEHSCpre}
		1 &\approx \int_{\tilde r_{\text{h}}}^{\tilde R_{0,\text{min}}} \tilde R^{\frac{d-5}{2}} d\tilde R+ \frac{n \tilde q^{d-2} (d-1)}{3(d-2)} \int_{\tilde r_{\text{h}}}^{\tilde R_{0,\text{min}}} \tilde R^{\frac{d (n-3)-5 n+6}{2 n}}d \tilde R +\mathcal{O}(\tilde q^{2(d-2)})\\
		\label{rEHSC2}
		&= \frac{2}{d-3}\left(\tilde R_{0,\text{min}}^{\frac{d-3}{2}}- \tilde r_{\text{h}}^{\frac{d-3}{2}}\right)+  \frac{2 n q^{d-2}}{d (n-3)-3( n-2)} \left(\tilde R_{0,\text{min}}^{\frac{d (n-3)-3( n-2)}{2 n}} - \tilde r_{\text{h}}^{\frac{d (n-3)-3( n-2)}{2 n}}\right)+\mathcal{O}(\tilde q^{2(d-2)}).
	\end{align}
	On the other hand, in the limit  $\tilde q\ll1$,  the horizon radius $\tilde r_\text{h}$ can be obtained from Eq. \eqref{qhorizon} 
	\begin{align}
		\label{rhexpanded}
		\tilde r_\text{h}\approx 1 - \frac{2n}{3}\frac{d-1}{d-2} \tilde q^{d-2} +\mathcal{O}(\tilde q^{2(d-2)}).
	\end{align}
	In this limit, it is expected that the minimum initial radius can be written as $\tilde R_{0,\text{min}}  =\left(\frac{d-1}{2}\right)^{\frac{2}{d-3}} + \delta$ where $\left(\frac{d-1}{2}\right)^{\frac{2}{d-3}}$  is the ST limit (e.g., $9/4$ for $d=4$) and $\delta$ is a small correction representing the deviation of $\tilde R_{0,\text{min}}$ in the regular model from the ST case. To determine  $\delta$, it is sufficient to substitute $\tilde R_{0,\text{min}} $ and  \eqref{rhexpanded} into Eq. \eqref{rEHSC2}:
	\begin{align}
		1 &\approx \frac{2}{d-3}\left(\left(\frac{3}{2}\right)^{d-3} \left(\frac{2 (d-3)}{9} \delta +1\right)-\left(1-\frac{(d-1) n q^{d-2}}{3 (d-2)}\right)\right)\nonumber\\
		&+\frac{2 (d-1)}{3 (d-2)}\frac{n^2 q^{d-2}}{n(d-3) -3 (d-2)}\left(\left(\frac{3}{2}\right)^{d-3 -\frac{3(d-2)}{n}}-1\right) +\mathcal{O}(\tilde q^{2(d-2)}),
	\end{align}
	which leads us to
	\begin{align}
		\label{R0min}
		\tilde R_{0,\text{min}} &\approx \left(\frac{d-1}{2}\right)^{\frac{2}{d-3}}-\left(\frac{2}{3}\right)^{d-5} \bigg[\frac{3 \left(\frac{2}{3}\right)^{4-d}-(d-1)}{d-3} \nonumber\\
		&+\frac{2 (d-1) n q^{d-2}}{3 (d-2)(d-3)}+\frac{2 (d-1) n^2 q^{d-2}}{n(d-3) -3 (d-2) }\left(\left(\frac{3}{2}\right)^{\frac{n(d-3)-3(d-2)}{n}}-1\right) \bigg] +\mathcal{O}(\tilde q^{2(d-2)}).
	\end{align}
	The first term in the bracket vanishes at $d=4$. Moreover, its derivative with respect to $d$ is strictly positive, hence this term is positive for all $d>4$.  The second term is positive for any value of $q>0$, $n>0$ and $d>4$. For the third term, consider the function $f(x) = (y^x - 1)/x$ where $y>1$. Since $y^x-1$ has the same sign as $x$, the third term ($y=3/2$ and $x={(n(d-3)-3(d-2))}/{n} $) is positive for $n>3(d-2)/(d-3)$. If $n<3(d-2)/(d-3)$, then both the numerator and denominator are negative. Therefore, the third term is positive in all admissible cases\footnote{For the polytropic index $n=3(d-2)/(d-3)$, one must refer back to Eq. \eqref{rEHSCpre} and perform the calculation directly.}. Thus, the second term in Eq. \eqref{R0min} is always negative, which implies that  small values of the magnetic charge always decrease the minimum initial radius required for the event horizon formation in the collapse scenario. In $4$ dimensions, for the Hayward case ($n=1$), $\tilde R_{0,\text{min}}\approx {9}/{4}-{713 \tilde q^2}/{405} +\mathcal{O}(\tilde q^{4})$,  while for the Bardeen case ($n=3/2$), $\tilde R_{0,\text{min}} \approx {9}/{4}-{25 \tilde q^2}/{9}+\mathcal{O}(\tilde q^{4})$.

\section{Conclusion}
\label{seccon}

In this work, we have conducted a comprehensive investigation into a general class of $d$-dimensional regular BHs characterized by a de Sitter core, sourced by NLED. This framework serves as a higher-dimensional generalization of the well-known 4-dimensional Bardeen and Hayward BHs, which are recovered for specific values of the polytropic index $n$ \cite{our1}. By extending the analysis beyond the standard four-dimensional spacetime, we have elucidated the intricate interplay between the spacetime dimensionality, the magnetic charge parameter, and the polytropic index in shaping the geometric, optical, thermodynamic, and dynamic properties of regular BHs.

Our analysis of the metric structure reveals that the presence of magnetic charge ensures a regular center with finite curvature, thereby resolving the central singularity inherent in classical solutions. In the absence of charge, the solution reduces to the singular ST BH. We derived exact conditions for the existence of horizons and demonstrated that while the extremality condition for the charge is independent of the polytropic index, the extremal radius depends on it. A key geometric finding is that as the number of dimensions increases, the distinctions between the radii of the event horizons, photon spheres, and shadows diminish, suggesting that higher-dimensional regular BHs converge behaviorally towards their singular counterparts in terms of geometric scales. Furthermore, we analyzed the ECs satisfied by the source matter. While the WEC and NEC are satisfied everywhere, the SEC is violated at certain radii, and the DEC may also be violated depending on the specific values of $n$ and $d$, indicating the presence of exotic matter regions necessary to sustain the regular core. Our solutions are curvature-regular in the sense that the curvature invariants remain finite at the center. However, we do not identify curvature regularity with full geodesic completeness. The maximal extension may be geodesically incomplete for some choices of $d$ and $n$, as in the Hayward case in 4 dimensions.

Regarding the optical properties, we identified the existence of both stable and unstable photon spheres. For BH configurations, an unstable outer photon sphere exists, whereas horizonless compact objects (formed when $q > q_{\text{ext}}$) can possess both stable and unstable photon spheres. We established that photon spheres and shadows persist only up to a critical magnetic charge limit, denoted as $q_{\text{deg}}$. Beyond this threshold, the object transitions into a horizonless compact object devoid of a shadow. Our numerical results indicate that as the dimension $d$, the magnetic charge $q$, or the polytropic index $n$ increases, the shadow's size decreases. Crucially, we observed that as the number of dimensions increases, the differences between the radius of the photon sphere and the radius of the shadow become negligible, implying that detailed studies of photon spheres in very high dimensions may be less critical for observational distinctions.

Thermodynamically, the proposed regular BHs exhibit rich behavior distinct from the ST BH. Unlike the ST BH, which is always thermodynamically unstable, our model possesses regions of local stability characterized by a positive heat capacity. We observed that the temperature vanishes at the extremal horizon radius,  in contrast to the ST BH, for which the temperature diverges as the horizon radius approaches zero. Therefore, the extremal horizon radius must be imposed as the lower limit in the calculations of entropy and heat capacity, as it has not been considered in previous studies. The heat capacity exhibits singularities signaling possible phase transitions, and the entropy deviates from the standard area law. Notably, this deviation from the area law becomes more significant as the dimension increases, while increasing the polytropic index reduces the deviation. These features suggest that curvature-regular BHs provide a useful effective setting for exploring thermodynamic modifications often associated with quantum-gravity-motivated corrections, particularly in higher dimensions where the entropy-area relation is modified.

While the exterior optical properties of regular BHs are expected to remain close to those of their singular counterparts within current observational uncertainties, the central novelty of the present work lies in the collapse sector. In particular, we generalized the OSD collapse scenario to the $d$-dimensional regular geometry introduced in this work. By matching an interior FLRW metric to the exterior regular solution, we tracked the evolution of the stellar surface, apparent horizon, and event horizon analytically. Our results indicate that the formation of the BH takes a longer proper time as the number of dimensions and the polytropic index increase. We also derived a critical lower bound for the initial stellar radius ($R_{0,\text{min}}$) required for physical BH formation. Through a perturbative analysis for small magnetic charges, we demonstrated that small values of the magnetic charge decrease this minimum initial radius, thereby facilitating the collapse scenario compared to the neutral case. Specifically, in 4 dimensions, we found explicit corrections for the Hayward and Bardeen cases, showing that the reduction in $R_{0,\text{min}}$ is sensitive to the polytropic index.

We acknowledge a recent analysis by De Felice and Tsujikawa \cite{DeFelice:2024seu} demonstrating that non-singular BHs sourced by NLED in four dimensions are generically prone to angular Laplacian instability around the center due to a negative squared propagation speed in the angular direction. This result applies directly as a serious viability constraint on the $d=4$ NLED sector considered here. For $d>4$, the four-dimensional result cannot simply be assumed to remain unchanged, because the perturbation structure is dimension dependent. The structure of perturbation equations and stability criteria in higher dimensions ($d > 4$) can exhibit qualitative differences due to the explicit dimensional dependence in the Einstein-NLED field equations, and whether the four-dimensional instability mechanism persists, or is modified,  remains an open question. A dedicated linear stability analysis of the $d$-dimensional regular BHs introduced here constitutes an important and necessary direction for future research to fully assess their physical viability.\\
The present work focuses on establishing the background-level geometric, thermodynamic, and collapse properties of the $d$-dimensional generalization. As mentioned above, our thermodynamic-stability statements, should be understood strictly as local statements in the fixed-charge ensemble, not as evidence of full dynamical stability.

Finally, we emphasize that while our analysis is formulated in arbitrary dimensions to uncover universal scaling laws and dimensional dependencies, all observational implications are derived from the well-defined four-dimensional limit ($d=4$) of the theory. The magnetic charge $q$ is treated as a free parameter subject to empirical constraints; our shadow analysis in Section \ref{SecShad} demonstrates that values $\tilde{q} \lesssim 0.3$ are compatible with current EHT observations of M87* and Sgr A*. The dimensional generalization serves as a tool to identify which qualitative features, such as curvature singularity avoidance, the existence of stable photon spheres, and thermodynamic phase transitions—are robust across dimensions and which are specific to $d=4$.

Collectively, these results provide new insights into the viability of curvature-regular BHs as possible curvature-singularity-free endpoints of gravitational collapse in higher dimensions. The dependence of stability, shadow size, and collapse dynamics on the polytropic index and spacetime dimension offers potential observational signatures that could distinguish these regular solutions from classical BHs in future astrophysical surveys. Future work could explore the observational signatures of these $d$-dimensional objects in the context of gravitational wave astronomy and higher-dimensional holographic models.

\appendix
\section{Analytic solutions}
\renewcommand{\theequation}{A.\arabic{equation}}
\setcounter{equation}{0}

This appendix explores analytical solutions to the event horizon equation \eqref{firsthor}. For spacetime dimensions $d > 4$, an exact analytical solution is generally unattainable. This is because the horizon equation \eqref{firsthor} becomes a  quintic equation, which lacks a solution in closed form without introducing simplifying assumptions. However, exact solutions can be obtained for odd dimensions up to $d = 9$. We introduce  the scaled parameters $\tilde r = r/m$ and $\tilde q = q/m$.
\label{appa}
\subsection{Horizons in $5$ dimensions}
At $d=5$, Eq. \eqref{firsthor} becomes quadratic and can be easily solved as
\begin{align}
    \tilde r_{\pm}^{(5)}=\left(\frac{1}{2} \left(1 \pm\sqrt{1-4 \tilde q^3}\right)\right)^{\frac{4 n}{9}},
\end{align}
where $\tilde r_{+}$ is the outer (event) horizon, $\tilde r_{-}$ is the inner (Cauchy) horizon and $n$ is the polytropic index. "It is clear that for $q=(1/4)^{1/3}$, the horizon is extremal, while for $q>(1/4)^{1/3}$
	no real solution, and therefore no horizon exists.
	\subsection{Horizons in $7$ dimensions}
At $d=7$, Eq. \eqref{firsthor} becomes  a cubic equation with the following form:
\begin{align}
	\tilde{\mathfrak{R}}^2 - \tilde q^5 \tilde{\mathfrak{R}}^3 -1 = 0,
\end{align}
To solve this equation, we convert it into a depressed cubic equation by introducing a new variable $x$ defined as
\begin{align}
    x=\tilde{\mathfrak{R}}- \frac{1}{3\tilde q^{5}}.
\end{align}
To solve the horizon equation \eqref{firsthor} using this substitution, we must solve the resulting depressed cubic equation for $x$.
\begin{align}
   x^3 + px + s=0,
\end{align}
where
 \begin{align}
	p&= -\frac{1}{3 \tilde q^{10}},\hspace{2cm} s=\frac{1}{\tilde q^5} -\frac{2}{27 \tilde q^{15}},
\end{align}
The solution to this equation is
\begin{align}
   x_k = 2 \sqrt{\frac{-p}{3}} \cos\left[ \frac{1}{3}\arccos \left(\frac{3s}{2p}\sqrt{\frac{-3}{p}}\right)-\frac{2\pi k}{3}\right],
\end{align}
where $k=0,1,2$. The final form of the solution is given by
\begin{align}
	\tilde r = \tilde q^{2 n} \left(\frac{1}{3}+\frac{2}{3} \cos  \left(\frac{2 \pi  k}{3}-\frac{1}{3} \cos ^{-1}\big(\frac{1}{2} \left(2-27 \tilde q^{10}\right)\big)\right)\right)^{-\frac{2 n}{5}}.
	\end{align}
	If the expression $\left(2-27 \tilde q^{10}\right) =0$ the horizon is extremal and if $\left(2-27 \tilde q^{10}\right)>1$ the no horizon exists.

\subsection{Horizons in $9$ dimensions}
For $d=9$,  Eq.\eqref{firsthor} becomes a quartic
\begin{align}
\tilde q^7 \tilde{\mathfrak{R}}^4 - \tilde{\mathfrak{R}}^3 + 1 = 0,
\end{align}
where $\tilde{\mathfrak{R}}=\tilde r^{-\frac{21}{8 n}}$. Expressing the quartic equation in its depressed form yields the coefficients of the first and the second degrees in the associated depressed quartics are 
\begin{align}
p = - \frac{3}{8 \tilde q^{14}},\hspace{1cm}q=\frac{1}{8 \tilde q^{21}},
\end{align}
Substituting back for $\tilde r$, the solutions are
\begin{align}
	\label{R12}
	\tilde r_{1,2}&=\left(\frac{1}{4 \tilde  q^7}-S \pm \frac{1}{4} \sqrt{-\frac{1}{2 \tilde  q^{21} S}+\frac{3}{\tilde  q^{14}}-16 S^2}\right)^{-\frac{8 n}{21}},\\
		\label{R34}
	\tilde r_{3,4}&=\left(\frac{1}{4 \tilde  q^7}+S \pm \frac{1}{4} \sqrt{-\frac{1}{2 \tilde  q^{21} S}+\frac{3}{\tilde  q^{14}}-16 S^2}\right)^{-\frac{8 n}{21}},\\
\end{align}
where
\begin{align}
   S= \frac{1}{4} \sqrt{\frac{4 Q}{3 \tilde q^7}-\frac{16}{Q}+\frac{1}{\tilde q^{14}}}, \hspace{2cm}   Q=\left(\frac{3}{2} \sqrt{192 \tilde q^{21}+81}-\frac{27}{2}\right)^{1/3}.
\end{align}
To determine the real roots among the four roots, one should calculate the discriminant. The discriminant is  $\Delta =- \left(256 \tilde q^{21}+27\right)<0$ which indicates that the quartic equation has two real positive roots and two complex conjugate roots. The complex roots are physically irrelevant.  Therefore, among the solutions in \eqref{R12} and \eqref{R34}, two are real and positive, corresponding to the inner and outer horizons. This holds only when the expression under the square root is non-negative.

\section{A hypergeometric identity}
\renewcommand{\theequation}{B.\arabic{equation}}
\setcounter{equation}{0}
\label{A2}
Here we detail the simplification leading to Eq. \eqref{dtauB}.
The starting point is the differentiation of $\tau_B$ with respect to $R_0$ using Eq. \eqref{SOTS}
\begin{align}
	\frac{d\tau_B}{d R_0}=\frac{d}{d R_0} \left[ \frac{2 m^{\frac{3-d}{2}} }{d-1}  \left( R_0^{\frac{d-1}{2}} \, {}_2F_1\!\left[a,a;1-a; -c R_0^{-b}\right] -  R_B^{\frac{d-1}{2}} \, {}_2F_1\left[a,a;1-a; -c R_B^{-b}\right] \right) \right],
\end{align}
where
\begin{align}
	a \equiv -\frac{(d-1)n}{3(d-2)}, \quad 
	b\equiv \frac{3(d-2)}{2n}, \quad 
	c \equiv m^{\frac{d-2}{2} (\frac{3}{n}-2)}q^{d-2}.
\end{align}
Applying the hypergeometric derivative identity
\begin{align}
	\frac{d}{dz} \, {}_2F_1[a,a;1-a;z] = \frac{a^2}{1-a} {}_2F_1[a+1,a+1;2-a;z], \quad z \equiv -c R_0^{-b}.
\end{align}
together with $dz/dR_0 = c b R_0^{-b-1}$, we obtain
\begin{align}
	\frac{d\tau_B}{d R_0}&= \frac{2m^{\frac{3-d}{2}}}{d-1} \left[ \frac{d-1}{2} R_0^{\frac{d-3}{2}} \, {}_2F_1[a,a;1-a;z]+ z b  R_0^{\frac{d-3}{2}} \frac{a^2}{1-a} \, {}_2F_1[a+1,a+1;2-a;z]   \right],\\
	&=  \frac{2m^{\frac{3-d}{2}}}{d-1} R_0^{\frac{d-3}{2}}\left[ \frac{d-1}{2} \left( {}_2F_1[a,a;1-a;z] + \frac{2 z b}{d-1}   \frac{a^2}{1-a} \, {}_2F_1[a+1,a+1;2-a;z]\right)  \right],\\
	\label{B6}
	&= \frac{ R_0^{\frac{d-3}{2}}}{m^{\frac{d-3}{2}}}\left[  {}_2F_1[a,a;1-a;z] + \frac{a z}{1-a} \, {}_2F_1[a+1,a+1;2-a;z] \right].
\end{align}
where in the last step we used $\frac{2b}{d-1} \frac{a^2}{1-a}= -\frac{a}{1-a}$. To further simplify the expression, we use the series representations. For the first term we have
\begin{align}
	{}_2F_1[a,a;1-a;z]= \sum_{k=0}^{\infty} \frac{(a)_k (a)_k}{(1-a)_k} \frac{z^k}{k!},
\end{align}
where $(a)_k$ is Pochhammer symbol. The second term in \eqref{B6} can be written as
\begin{align}
	z \, {}_2F_1[a+1,a+1;2-a;z]= z \sum_{k=0}^{\infty} \frac{(a+1)_k (a+1)_k}{(2-a)_k} \frac{z^k}{k!}.
\end{align}
Shifting the summation index $k \to k-1$ in the second term and using Pochhammer identities,  $(a+1)_{k-1} = \frac{(a)_k}{a}$ and $ (2-a)_{k-1} = \frac{(1-a)_k}{1-a}$ yields
\begin{align}
	\frac{a z}{1-a} \, {}_2F_1[a+1,a+1;2-a;z] = \sum_{k=1}^{\infty} \frac{(a)_k (a+1)_{k-1}}{(1-a)_k} \frac{z^k}{(k-1)!}.
\end{align}
Thus,
\begin{align}
	{}_2F_1[a,a;1-a;z] + \frac{a}{1-a} \, z \, {}_2F_1[a+1,a+1;2-a;z]
	&= \sum_{k=0}^{\infty} \frac{(a)_k^2}{(1-a)_k} \frac{z^k}{k!} + \sum_{k=1}^{\infty} \frac{(a)_k (a+1)_{k-1}}{(1-a)_k} \frac{k z^k}{k!},\\
	&= 1 + \sum_{k=1}^{\infty} \frac{(a)_k}{(1-a)_k} \frac{z^k}{k!} \left[ (a)_k + k (a+1)_{k-1} \right].
\end{align}
Using the identity $(a)_k + k(a+1)_{k-1} = (a+1)_k$ we obtain the compact hypergeometric form
\begin{align}
	{}_2F_1[a,a;1-a;z]+ \frac{a}{1-a} \, z \, {}_2F_1[a+1,a+1;2-a;z]&= \sum_{k=0}^{\infty} \frac{(a)_k (a+1)_k}{(1-a)_k} \frac{z^k}{k!}\\
	&={}_2F_1[a,a+1;1-a;z]
\end{align}
Substituting back $a$, $z$ (and subsequently $c$ and $b$) yields \eqref{dtauB}.


\begin{thebibliography}{99}
\bibitem{34-35Mal}
 V.P. Frolov, G.A. Vilkovisky, in Proceedings of the Second Marcel Grossmann Meeting on General Relativity, Trieste, Italy, 5–11 July (1979); V.P. Frolov, G.A. Vilkovisky, Phys. Lett. B, \textbf{106}, 307–313 (1981).
\bibitem{kh4-5}
R. Penrose, Riv. Nuovo, Cim. \textbf{1}  252–276 (1969);  ibid., Gen. Rel. Grav. \textbf{34}, 1141 (2002).
\bibitem{Mukh}
V. Mukhanov, S. Winitzki, \textit{"Introduction to quantum effects in gravity"} ,
Cambridge University Press, (2007).

\bibitem{23-25Fro13}
M. Markov, JETP Letters \textbf{36}, 265 (1982); M. Markov, Ann. Phys. \textbf{155}, 333 (1984); J. Polchinski, Nucl.Phys. B \textbf{325}, 619 (1989).

\bibitem{Fro16}
V. P. Frolov, Phys. Rev. D. \textbf{94}, 104056 (2016).

\bibitem{1DyCqg05}
 A.D. Sakharov, Sov. Phys. JETP \textbf{22} 241 (1966).

\bibitem{2DyCqg05}
E.B. Gliner, Sov. Phys. JETP \textbf{22} 378 (1966).

\bibitem{3RodJcap-9-21Kho} 
J.M. Bardeen, in Proceedings of the International Conference GR5, Tbilisi, U.S.S.R. (1968).

\bibitem{4-8DyCqg04}
 E. Ayón-Beato,  A.  Garcia, Gen. Rel. Grav. \textbf{31} 629 (2003);
N. Breton,  Phys. Rev. D \textbf{67}, 124004 (2003).

\bibitem{54Fro16}
S. A. Hayward, Phys. Rev. Lett. \textbf{96}, 031103 (2006).

\bibitem{Zhou:2022yio}
T.~Zhou and L.~Modesto,
Phys. Rev. D \textbf{107}, no.4, 044016 (2023).

\bibitem{our1}
F.~Shojai, A.~Sadeghi, and R.~Hassannejad,
Class. Quantum Grav. \textbf{39}, 8, 085003 (2022).

\bibitem{vafa}
 A. Strominger and C. Vafa, Phys. Lett. B 379, 99 (1996).

\bibitem{malda1}
O. Aharony, S. S. Gubser, J. M. Maldacena, H. Ooguri, and Y. Oz, Phys. Rep. 323, 183 (2000).


\bibitem{tang}
F.~R.~Tangherlini,
Nuovo Cim. \textbf{27}, 636-651 (1963).



\bibitem{Myers:1986un}
R.~C.~Myers and M.~J.~Perry,
Ann. Phys. \textbf{172}, 304 (1986).



\bibitem{Emparan:2008eg}
R.~Emparan and H.~S.~Reall,
Living Rev. Rel. \textbf{11}, 6 (2008).



\bibitem{ali}
M.~S.~Ali and S.~G.~Ghosh,
Phys. Rev. D \textbf{98}, 8, 084025 (2018).


\bibitem{EventHorizonTelescope:2019dse}
K.~Akiyama \textit{et al.} [Event Horizon Telescope],
Astrophys. J. Lett. \textbf{875}, L1 (2019).

\bibitem{EventHorizonTelescope:2022wkp}
K.~Akiyama \textit{et al.} [Event Horizon Telescope],
Astrophys. J. Lett. \textbf{930}, 2, L12 (2022).


\bibitem{Cardoso:2019rvt}
V.~Cardoso and P.~Pani,
Living Rev. Rel. \textbf{22}, 1, 4 (2019).


\bibitem{Shaikh:2022ivr}
R.~Shaikh,
Mon. Not. Roy. Astron. Soc. \textbf{523}, 1, 375-384 (2023).

\bibitem{Abdikamalov:2019ztb}
A.~B.~Abdikamalov, A.~A.~Abdujabbarov, D.~Ayzenberg, D.~Malafarina, C.~Bambi, and B.~Ahmedov,
Phys. Rev. D \textbf{100}, 2, 024014 (2019).


\bibitem{Singh:2017vfr}
B.~P.~Singh, and S.~G.~Ghosh,
Ann. Phys. \textbf{395}, 127-137 (2018).

\bibitem{Roy:2023ine}
S.~Roy, S.~Chatterjee, and R.~Koley,
Eur. Phys. J. C \textbf{84}, 1, 47 (2024).


\bibitem{Papnoi:2014aaa}
U.~Papnoi, F.~Atamurotov, S.~G.~Ghosh, and B.~Ahmedov,
Phys. Rev. D \textbf{90}, 2, 024073 (2014).






\bibitem{bch}
J.~M.~Bardeen, B.~Carter, and S.~W.~Hawking,
Commun. Math. Phys. \textbf{31}, 161-170 (1973).

\bibitem{Dymnikova:2015yma}
I.~Dymnikova and M.~Khlopov,
Int. J. Mod. Phys. D \textbf{24}, no.13, 1545002 (2015).

\bibitem{Hawkingtemp}
S.~W.~Hawking,
Commun. Math. Phys. \textbf{43}, 199-220 (1975).



\bibitem{bonn1}
A.~Bonanno and M.~Reuter,
Phys. Rev. D \textbf{62}, 043008 (2000).


\bibitem{Hut}
P. Hut, 
Mon. Not. R. astr. Soc. 180, 379 - 389 (1977).



\bibitem{Pavn}
Diego Pavn,
Phys. Rev. D \textbf{43}, 2495 - 2497 (1991).

\bibitem{DeFelice:2024seu}
A.~De Felice and S.~Tsujikawa,
Phys. Rev. Lett. \textbf{134}, no.8, 081401 (2025).

\bibitem{Oppenheimer:1939ue}
J.~R.~Oppenheimer and H.~Snyder,
Phys. Rev. \textbf{56}, 455-459 (1939).

\bibitem{Datt:1938uwc}
B.~Datt,
Z. Phys. \textbf{108}, 5, 314-321 (1938).

\bibitem{Bambi:2014}
Y.~Liu, D.~Malafarina, L.~Modesto and C.~Bambi,
Phys. Rev. D \textbf{90}, 4, 044040 (2014).

\bibitem{Bambi:2013}
Y.~Liu, D.~Malafarina, L.~Modesto and C.~Bambi,
Phys. Rev. D \textbf{88}, 044009  (2013).

\bibitem{Zhang:2015}
Y.~Zhang, Y.~Zhu, L.~Modesto and C.~Bambi,
Eur. Phys. J. C \textbf{75}, 2, 96 (2015).

\bibitem{Ayon}
E.~Ayon-Beato and A.~Garcia,
Phys. Lett. B \textbf{493}, 149-152 (2000).


\bibitem{myers}
R. C. Myers and M. J. Perry, Ann. Phys. (N.Y.) \textbf{172}, 304 (1986).

\bibitem{dian}
X. Dianyan, Classical Quantum Gravity \textbf{5}, 871 (1988).

\bibitem{Tang}
Z.~Y.~Tang, B.~Wang and E.~Papantonopoulos,
Eur. Phys. J. C \textbf{81}, 4, 346 (2021).

\bibitem{Yin}
R.~Yin, J.~Liang and B.~Mu,
Phys. Dark Univ. \textbf{32}, 100831 (2021).


\bibitem{Rodrigues:2018bdc}
M.~E.~Rodrigues and M.~V.~de Sousa Silva,
JCAP \textbf{06}, 025 (2018).

\bibitem{Malaf}
D.~Malafarina and B.~Toshmatov,
Phys. Rev. D \textbf{105}, 12, L121502 (2022).

\bibitem{Malaf2}
D.~Malafarina,
\textit{"Semi-classical Dust Collapse and Regular Black Holes"}, Springer Series in Astrophysics and Cosmology. Springer, Singapore. (2023).


\bibitem{Guo:2022ghl}
G.~Guo, Y.~Lu, P.~Wang, H.~Wu and H.~Yang,
Phys. Rev. D \textbf{107}, 12, 124037 (2023).

\bibitem{Synge:1966okc}
J.~L.~Synge,
Mon. Not. Roy. Astron. Soc. \textbf{131}, , 3, 463-466 (1966).

\bibitem{Perlick:2021aok}
V.~Perlick and O.~Y.~Tsupko,
Phys. Rept. \textbf{947}, 1-39 (2022).

\bibitem{Bambi:2019tjh}
C.~Bambi, K.~Freese, S.~Vagnozzi and L.~Visinelli,
Phys. Rev. D \textbf{100}, 4, 044057 (2019).

\bibitem{Vagnozzi:2022moj}
S.~Vagnozzi, R.~Roy, Y.~D.~Tsai, L.~Visinelli, M.~Afrin, A.~Allahyari, P.~Bambhaniya, D.~Dey, S.~G.~Ghosh and P.~S.~Joshi, \textit{et al.}
Class. Quant. Grav. \textbf{40}, 16, 165007 (2023).

\bibitem{Beke}
J.~Bekenstein,
Scholarpedia \textbf{3}, 10, 7375 (2008).




\bibitem{Smarr}
L.~Smarr,
Phys. Rev. Lett. \textbf{30}, 71-73 (1973).

\bibitem{Man:2013hza}
J.~Man and H.~Cheng,
[arXiv:1312.6566 [hep-th]].

\bibitem{Singh:2019tgw}
D.~V.~Singh and S.~Siwach,
[arXiv:1909.11529 [hep-th]].

\bibitem{Sadeghi:2023aii}
J.~Sadeghi, S.~Noori Gashti, M.~R.~Alipour and M.~A.~S.~Afshar,
Annals Phys. \textbf{455}, 169391 (2023).

\bibitem{Sharif:2010pj}
M.~Sharif and W.~Javed,
J. Korean Phys. Soc. \textbf{57}, 217-222 (2010).


\bibitem{Waldent}
R.M.~Wald, Phys. Rev. D \textbf{48} R3427 (1993).


\bibitem{ghosh1}
D.~V.~Singh, S.~G.~Ghosh and S.~D.~Maharaj,
Nucl. Phys. B \textbf{981}, 115854 (2022).


\bibitem{man1}
J.~Man and H.~Cheng,
Gen. Rel. Grav. \textbf{46}, 1660 (2014).

\bibitem{akbar}
M. Akbar, N. Salem, and S. A. Hussein,
Chin. Phys. Lett. \textbf{29}, 070401 (2012).

\bibitem{Myung}
Y.~S.~Myung, Y.~W.~Kim and Y.~J.~Park,
Gen. Rel. Grav. \textbf{41}, 1051-1067 (2009).




\bibitem{Ghosh:2014pga}
S.~G.~Ghosh, U.~Papnoi and S.~D.~Maharaj,
Phys. Rev. D \textbf{90}, 4, 044068 (2014).



\bibitem{Painleve:1921}
P.~Painlevé, C. R. Acad. Sci. \textbf{173}, 677–680 (1921).

\bibitem{Gullstrand:1922}
A.~Gullstrand,  Arkiv för Matematik, Astronomi och Fysik. \textbf{16} (8): 1–15 (1922).

\bibitem{Martel:2000rn}
K.~Martel and E.~Poisson,
Am. J. Phys. \textbf{69}, 476-480 (2001).

\bibitem{Poisson:2009pwt}
E.~Poisson,
``A Relativist's Toolkit: The Mathematics of Black-Hole Mechanics,''
Cambridge University Press, (2009).

\bibitem{Israel}
W.~Israel, \textit{Nuovo Cimento}, \textbf{44} 4349 (1966).

\bibitem{Blau}
M.~Blau, "Lecture note on general Relativity",  http://www.blau.itp.unibe.ch/newlecturesGR.pdf (2020).


\bibitem{Rezzolla:2004}
L. Rezzolla,  "An Introduction to Gravitational Collapse to Black Holes", Lectures given at the Villa Mondragone International School of Gravitation and Cosmology, Frascati (Rome), Italy, Sept. 7th – 10th, (2004).

\bibitem{Hayward1994}
S.~A.~Hayward, Phys. Rev. D \textbf{49}, 6467 (1994).

\bibitem{Sadeghi:2024rar}
A.~Sadeghi, F.~Shojai and F.~Bahmani,
Eur. Phys. J. C \textbf{84}, no.2, 197 (2024).

\bibitem{10JFT}
A. Ashtekar and B. Krishnan,  Phys. Rev. Lett. \textbf{89}, 261101 (2002).

\bibitem{Khodabakhshi:2025fmf}
A. Sadeghi, "Black hole mimickers", Ph.D. thesis, University of Tehran, (2025); 
H.~Khodabakhshi, H.~Lu and F.~Shojai,
Phys. Rev. D \textbf{112}, no.12, 124057 (2025).
	
	\end{thebibliography}
\end{document}